\documentclass{article}
\usepackage{graphicx} 
\usepackage{graphicx}
\usepackage{tikz}
\usetikzlibrary{arrows.meta,positioning}
\usepackage{amssymb}
\usepackage{amsmath}
\usepackage{amsthm}
\usepackage{thmtools}
\usepackage{amsfonts}
\usepackage{mathtools}
\usepackage{enumerate}
\usepackage{nicefrac}
\usepackage{fullpage}
\usepackage[dvipsnames]{xcolor}
\usepackage{cite}
\usepackage[unicode=true]{hyperref}
\usepackage[linesnumbered,ruled,vlined,boxed,algo2e]{algorithm2e}
\usepackage[capitalise,nameinlink]{cleveref}
\crefname{claim}{Claim}{Claims}
\crefname{observation}{Observation}{Observations}
\crefname{algocf}{Algorithm}{Algorithms}
\crefname{problem}{Problem}{Problems}

\newtheorem{theorem}{Theorem}
\newtheorem{lemma}{Lemma}
\newtheorem{claim}{Claim}[lemma]

\newtheorem{definition}{Definition}

\newtheorem{problem}{Problem}

\newcommand{\avi}[1]{{\color{purple} \textbf{Avi}: #1}}

\newcommand{\eps}{\varepsilon}
\newcommand{\Ot}{\widetilde{O}}

\newcommand{\EE}{\mathbb{E}}

\DeclarePairedDelimiter{\pair}{\langle}{\rangle}

\newcommand{\wt}{\ell}
\newcommand{\codestyle}[1]{\texttt{#1}}
\newcommand{\RT}{\mbox{\codestyle{RT}}}
\renewcommand{\L}{\mbox{\codestyle{L}}}
\newcommand{\Route}{\codestyle{Route}}
\newcommand{\PickLevel}{\codestyle{PickLevel}}
\providecommand{\id}{\mathrm{id}}
\newcommand{\one}{\mathbf{1}} 

\title{Optimal (Randomized) Compact Routing Schemes}
\author{Shiri Chechik\thanks{Department of Computer Science, Tel Aviv University, Israel. E-mail {\tt shiri.chechik@gmail.com}.} \and
Avi Kadria\thanks{Department of Computer Science, Bar-Ilan University, Ramat Gan 5290002, Israel. E-mail {\tt avi.kadria3@gmail.com}.} \and
Gur Lifshitz\thanks{Department of Computer Science, Tel Aviv University, Israel. E-mail {\tt gur.lifshitz@gmail.com}.} \and Liam Roditty\thanks{Department of Computer Science, Bar-Ilan University, Ramat Gan 5290002, Israel. E-mail {\tt liam.roditty@biu.ac.il}. Supported in part by BSF grants 2016365 and 2020356.}}
\date{}

\begin{document}
\maketitle

\begin{abstract}
Compact routing schemes route messages along short paths while every vertex stores little information.
We study the trade-off between the stretch and the storage of compact routing schemes in undirected graphs, in four models: labeled and name-independent schemes, with average or worst-case local storage. We improve the best known stretch in all four models. We present here selected results.

Our main result is a labeled routing scheme with expected stretch $2k-1$ and $\widetilde O(n^{1/k})$ average local storage, in which only the routing is randomized.
Under the Erd\H{o}s girth conjecture, no scheme with this storage has stretch below $2k-1$, so our scheme is optimal, up to polylogarithmic factors in the storage.
With deterministic routing, we obtain stretch $\approx2.47k$, the best known stretch for deterministic routing with average storage.

Worst-case local storage is the classical setting of compact routing. The leading coefficient $4$ in the stretch $4k-5$ of Thorup and Zwick~[SPAA'01] was improved only by Chechik~[PODC'13], to $\approx3.68$ in unweighted graphs.
In unweighted graphs, we obtain expected stretch $\left( \frac83k+O(1), O(k) \right)$, and deterministic stretch $\approx3.05k$.

Name-independent routing is harder, since the labels cannot be chosen, and the source knows only the name of the destination. The best known stretch with $\widetilde O(n^{1/k})$ local storage in unweighted graphs was $96k$, by Abraham, Gavoille, and Malkhi~[DISC'04].
We improve it by more than a factor of $8$. With average storage, we obtain expected stretch $4k-5$, which matches the labeled scheme of Thorup and Zwick. With worst-case storage in unweighted graphs, we obtain expected stretch $\left( 8k+O(1), O(k) \right)$.

All our schemes for unweighted graphs extend to weighted graphs, with a small loss in the stretch in some models, and still improve the previous best unweighted bounds.
\end{abstract}

\pagebreak
\section{Introduction}\label{S-Intro}

Routing is a fundamental task in computer networks, studied since the 1970s~\cite{KleinrockKamoun77}. A \emph{routing scheme} delivers messages from a source vertex to a destination vertex, where every vertex forwards a message using only the information that it stores.
We study routing schemes in undirected graphs with $n$ vertices.

A routing scheme has two phases.
In the \emph{preprocessing phase}, the whole graph is known, and every vertex receives a \emph{routing table} and a short \emph{label}.
In the \emph{routing phase}, the source is given a message and the label of the destination, and attaches a short header to the message. Every vertex that holds the message forwards it to one of its neighbors, using only its own routing table, the label of the destination, and the header, which it may update. This repeats until the message reaches the destination.
In a \emph{labeled} routing scheme, the preprocessing phase chooses the labels. In a \emph{name-independent} routing scheme, the labels are the original names of the vertices, so they are fixed in advance and cannot be changed in the preprocessing phase.

Routing along shortest paths is possible by storing a full routing table at every vertex, but this takes $\Theta(n)$ words per vertex, which cannot be avoided.\footnote{Every routing scheme whose routes are shorter than three times the distance uses $\Omega(n)$ bits at some vertex on some graph~\cite{DBLP:journals/jpdc/GavoilleG01,DBLP:conf/wdag/Zinovyev22}.}
Compact routing schemes therefore use less storage, and route along paths that are longer than shortest paths.
Let $\hat d(u,v)$ be the length of the walk that a message traverses from $u$ to $v$. For randomized routing, it is a random variable.
A scheme has \emph{$(\alpha,\beta)$-stretch} if $\hat d(u,v)\le\alpha\, d(u,v)+\beta$ for every $u,v\in V$, where $d(u,v)$ is the distance between $u$ and $v$. For randomized routing, a scheme has \emph{$(\alpha,\beta)$-expected stretch} if $\EE[\hat d(u,v)]\le\alpha\, d(u,v)+\beta$ for every $u,v\in V$, with expectation over the choice of the route. When $\beta=0$, we write $\alpha$-stretch and $\alpha$-expected stretch, and call $\alpha$ the (expected) \emph{stretch} of the scheme.
The \emph{local storage} of a vertex is the number of words in its routing table. We bound either the local storage of every vertex (\emph{worst-case storage}) or the total storage divided by $n$ (\emph{average storage}). Our average-storage schemes work in weighted graphs. Our worst-case schemes are for unweighted graphs, and we extend them to weighted graphs in \Cref{S-Weighted}.

The design of compact routing schemes in undirected graphs has been studied extensively over the past several decades~\cite{DBLP:journals/jacm/PelegU89,DBLP:journals/jal/AwerbuchBLP90,DBLP:journals/siamdm/AwerbuchP92,DBLP:journals/jal/Cowen01,DBLP:journals/jal/EilamGP03,DBLP:conf/spaa/ThorupZ01,AbrahamGM04,DBLP:conf/spaa/AbrahamGM06,DBLP:journals/talg/AbrahamGMNT08,DBLP:conf/wdag/AbrahamG11,DBLP:conf/podc/Chechik13,DBLP:conf/podc/RodittyT15,kadria2025compact}. See \Cref{tab-results,tab-ni-results} for a summary of the previous relevant results.

Under the Erd\H{o}s girth conjecture,\footnote{The Erd\H{o}s girth conjecture~\cite{Erdos64} states that, for every $k\ge1$, there are graphs with $\Omega(n^{1+1/k})$ edges and girth at least $2k+2$.} every routing scheme with stretch below $2k+1$ uses $\Omega(n^{1+1/k})$ bits of total storage on some graph~\cite{DBLP:conf/spaa/ThorupZ01,DBLP:conf/wdag/Zinovyev22}. Applying this with $k-1$ in place of $k$, stretch $2k-1$ is optimal for $\Ot(n^{1/k})$ average local storage, also for expected stretch (\Cref{L-LB-random}).\footnote{$\Ot(\cdot)$ hides polylogarithmic factors in $n$, and a word has $O(\log n)$ bits.}

In this paper, we study four models of compact routing, defined by the choice between labeled and name-independent routing and between average and worst-case local storage.
In all of them, we consider an integer $k\ge2$ and $\Ot(n^{1/k})$ local storage, and we compare schemes by their stretch, mainly by the leading coefficient of $k$.
In all four models, we improve the best known stretch. 

\paragraph{Our techniques.}
Our results follow from new routing techniques: randomized routing, optimizing over all routing choices at once, and a cheaper search in trees for worst-case storage. In randomized routing, the routing tables and the labels are fixed in preprocessing, and only the source flips coins, to choose the route of the message. We bound the expected stretch for every source and destination. These techniques also apply in the harder name-independent model, and \Cref{S-Over} describes them in more detail.
We now present our results in the four models.

\subsection{Labeled routing with average storage}
Our main result is a labeled routing scheme with expected stretch $2k-1$ and $\Ot(n^{1/k})$ average local storage, which is optimal up to polylogarithmic factors in the storage, and improves upon the previous $\approx2.64k$-stretch, by Kadria and Roditty~\cite{kadria2025compact}.
\begin{restatable}{theorem}{thmAvgRandom}\label{T-avg-random}
For every integer $k\ge2$, there is a $(2k-1)$-expected stretch labeled compact routing scheme that uses average local storage of $\Ot(n^{1/k})$ words, labels of $\Ot(k)$ words, and headers of $\Ot(1)$ words.
\end{restatable}

Our second result is a deterministic scheme with stretch $\approx2.47k$. It improves the $\approx2.64k$ of Kadria and Roditty~\cite{kadria2025compact}.
\begin{restatable}{theorem}{thmAvgDet}\label{T-avg-det}
For every integer $k\ge2$, there is a $\bigl(1+\frac{k-1}{\ln(3/2)}\approx2.47k-1.47\bigr)$-stretch labeled compact routing scheme with deterministic routing that uses average local storage of $\Ot(n^{1/k})$ words, labels of $\Ot(k)$ words, and headers of $\Ot(1)$ words.
\end{restatable}
\begin{table}[t]
\centering
\small
\begin{tabular}{@{}llll@{}}
\hline
Stretch & Storage & Graphs & Reference \\
\hline
$O(k)$ & $\Ot(n^{1/k})$ average & unweighted & \cite{DBLP:journals/jacm/PelegU89} \\
$2^k-1$ & $\Ot(n^{1/k})$ average & weighted & \cite{DBLP:journals/jal/AwerbuchBLP90} \\
$\approx2.64k$, $k\ge3$ & $\Ot(n^{1/k})$ average & weighted & \cite{kadria2025compact} \\
$\approx2.47k-1.47$ & $\Ot(n^{1/k})$ average & weighted & \Cref{T-avg-det} \\
$2k-1$ & $\Ot(n^{1/k})$ average & weighted & \Cref{T-avg-random}$^\star$ \\
\hline
$O(k^2)$ & $\Ot(n^{1/k}\log\Delta)$ worst-case & weighted & \cite{DBLP:journals/siamdm/AwerbuchP92} \\
$4k-5$ & $\Ot(n^{1/k})$ worst-case & weighted & \cite{DBLP:conf/spaa/ThorupZ01} \\
$4k-7+\varepsilon$ & $\Ot(\varepsilon^{-1}n^{1/k}\log\Delta)$ worst-case & weighted & \cite{DBLP:conf/podc/RodittyT15} \\
$\approx3.68k$ & $\Ot(n^{1/k})$ worst-case & unweighted & \cite{DBLP:conf/podc/Chechik13} \\
$\approx3.05k+1$ & $\Ot(n^{1/k})$ worst-case & unweighted & \Cref{T-worst-det} \\
$(\frac{8}{3}k+6,\frac49k)\approx(2.67k+6,0.44k)$ & $\Ot(n^{1/k})$ worst-case & unweighted & \Cref{T-worst-random}$^\star$ \\
$\approx(1+\eps)3.42k+O(1)$ & $\Ot(\eps^{-1}n^{1/k}\log\Delta)$ worst-case & weighted & \Cref{T-weighted-det} \\
$\approx(1+\eps)3.07k+O(1)$ & $\Ot(\eps^{-1}n^{1/k}\log\Delta)$ worst-case & weighted & \Cref{T-weighted-rand}$^\star$ \\
\hline
\end{tabular}
\caption{Labeled routing schemes, where $\Delta$ is the aspect ratio of the graph and $0<\eps<1$. A pair $(\alpha,\beta)$ denotes $(\alpha,\beta)$-stretch, that is, $\hat d(u,v)\le\alpha\,d(u,v)+\beta$, or $\EE[\hat d(u,v)]\le\alpha\,d(u,v)+\beta$ for rows marked $^\star$. Under the Erd\H{o}s girth conjecture, no scheme with $\Ot(n^{1/k})$ average local storage has stretch below $2k-1$. $^\star$Expected stretch.}
\label{tab-results}
\end{table}

\subsection{Labeled routing with worst-case storage}
Thorup and Zwick~\cite{DBLP:conf/spaa/ThorupZ01} obtained stretch $4k-5$ with $\Ot(n^{1/k})$ local storage at every vertex.
Chechik~\cite{DBLP:conf/podc/Chechik13} obtained the first leading coefficient below $4$, with stretch approximately $3.68k$ and $\Ot(n^{1/k})$ worst-case local storage in unweighted graphs. For weighted graphs, she states stretch $(4-\alpha)k-\beta$, for absolute constants $\alpha,\beta>0$, with $\Ot(n^{1/k}\log D)$ local storage, where $D$ is the normalized diameter, and sketches the changes to her analysis for unweighted graphs.
Subsequently, Roditty and Tov~\cite{DBLP:conf/podc/RodittyT15} obtained stretch $4k-7+\varepsilon$, which improves the bounds for small values of $k$, for example to stretch $5+\varepsilon$ for $k=3$.

We present the first improvement over the $\approx3.68k$ of Chechik~\cite{DBLP:conf/podc/Chechik13}. With randomized routing, our leading coefficient is $\frac83\approx2.67$, the first below $3$, at the cost of an additive $\frac49k$.
Our randomized scheme has $(\frac83k+6,\frac49k)$-expected stretch.
\begin{restatable}{theorem}{thmWCrandom}\label{T-worst-random}
For every integer $k\ge2$, there is a $(\frac83k+6,\frac49k)$-expected stretch labeled compact routing scheme for unweighted graphs that uses worst-case local storage of $\Ot(n^{1/k})$ words, labels of $\Ot(k)$ words, and headers of $\Ot(1)$ words.
\end{restatable}

Without randomization, we obtain stretch $\approx3.05k$, which also improves the $\approx3.68k$ of Chechik~\cite{DBLP:conf/podc/Chechik13}.
\begin{restatable}{theorem}{thmWCDet}\label{T-worst-det}
For every integer $k\ge2$, there is a $\bigl(\frac{8k}{1+4\ln(3/2)}+1\approx3.05k+1\bigr)$-stretch labeled compact routing scheme for unweighted graphs with deterministic routing that uses worst-case local storage of $\Ot(n^{1/k})$ words, labels of $\Ot(k)$ words, and headers of $\Ot(1)$ words.
\end{restatable}

\subsection{Name-independent routing with average storage}
Name-independent routing has been studied extensively~\cite{DBLP:journals/jal/AwerbuchBLP90,DBLP:journals/siamdm/AwerbuchP92,AriasCLRT06,DBLP:journals/talg/AbrahamGMNT08,AbrahamGM04,AbrahamM05,AbrahamGM06upper,DBLP:conf/spaa/AbrahamGM06,AbrahamGGM06,DBLP:conf/podc/KonjevodRX06,KonjevodRX16,DBLP:journals/ipl/Laing07,LaingR07,DBLP:conf/wdag/GavoilleGHI13}, and yet the best known stretch with $\Ot(n^{1/k})$ local storage at every vertex in unweighted graphs is $96k$~\cite{AbrahamGM04}, far from the lower bound of $2k-1$~\cite{DBLP:conf/spaa/AbrahamGM06}.

We present new name-independent schemes, randomized and deterministic, with average and with worst-case local storage. \Cref{T-NI-avg,T-NI-det,T-NI-worst-random,T-NI-worst-det} below improve the previous $96k$ by a factor of more than $8$.
With $\Ot(n^{1/k})$ average local storage, we obtain expected stretch $4k-5$ with randomized routing, and stretch $\approx5.77k$ with deterministic routing.
\begin{restatable}{theorem}{thmNIavg}\label{T-NI-avg}
For every integer $k\ge2$, there is a $(4k-5)$-expected stretch name-independent compact routing scheme that uses average local storage of $\Ot(n^{1/k})$ words and headers of $\Ot(1)$ words.
\end{restatable}
\begin{restatable}{theorem}{thmNIdet}\label{T-NI-det}
For every integer $k\ge2$, there is a $\bigl(1+\frac4{2^{1/(k-1)}-1}\approx5.77k-4.77\bigr)$-stretch name-independent compact routing scheme with deterministic routing that uses average local storage of $\Ot(n^{1/k})$ words and headers of $\Ot(1)$ words.
\end{restatable}
The expected stretch $4k-5$ matches the stretch of the labeled scheme of Thorup and Zwick~\cite{DBLP:conf/spaa/ThorupZ01}. This is the best bound that an analysis based only on the distances from the source to the hierarchy of Thorup and Zwick can give, even if the source knows $d(u,v)$ (\Cref{S-NI-avg}).

\subsection{Name-independent routing with worst-case storage}
With $\Ot(n^{1/k})$ local storage at every vertex in unweighted graphs, we obtain $(8k+2e-7,2k-2)$-expected stretch with randomized routing, where $8k+2e-7\approx8k-1.56$, and stretch $\approx11.54k$ with deterministic routing.
\begin{restatable}{theorem}{thmNIworstRandom}\label{T-NI-worst-random}
For every integer $k\ge2$, there is a $(8k+2e-7,2k-2)$-expected stretch name-independent compact routing scheme for unweighted graphs that uses worst-case local storage of $\Ot(n^{1/k})$ words and headers of $\Ot(1)$ words.
\end{restatable}
\begin{restatable}{theorem}{thmNIworstDet}\label{T-NI-worst-det}
For every integer $k\ge2$, there is a $\bigl(\frac{8(k-1)}{\ln2}+9\approx11.54k-2.54\bigr)$-stretch name-independent compact routing scheme for unweighted graphs with deterministic routing that uses worst-case local storage of $\Ot(n^{1/k})$ words and headers of $\Ot(1)$ words.
\end{restatable}
\Cref{tab-ni-results} compares our name-independent schemes with the previous ones.
\begin{table}[t]
\centering
\small
\begin{tabular}{@{}llll@{}}
\hline
Stretch & Storage & Graphs & Reference \\
\hline
$\approx5.77k-4.77$ & $\Ot(n^{1/k})$ average & weighted & \Cref{T-NI-det} \\
$4k-5$ & $\Ot(n^{1/k})$ average & weighted & \Cref{T-NI-avg}$^\star$ \\
\hline
$O(k^29^k)$ & $\Ot(n^{1/k})$ worst-case & weighted & \cite{DBLP:journals/jal/AwerbuchBLP90} \\
$O(k^22^k)$ & $\Ot(n^{1/k})$ worst-case & weighted & \cite{AriasCLRT06} \\
$O(k^2)$ & $\Ot(n^{1/k}\log\Delta)$ worst-case & weighted & \cite{DBLP:journals/siamdm/AwerbuchP92} \\
$96k$ & $\Ot(n^{1/k}\log\Delta)$ worst-case & weighted & \cite{AbrahamGM04} \\
$\approx11.54k-2.54$ & $\Ot(n^{1/k})$ worst-case & unweighted & \Cref{T-NI-worst-det} \\
$(8k+2e-7,2k-2)$ & $\Ot(n^{1/k})$ worst-case & unweighted & \Cref{T-NI-worst-random}$^\star$ \\
$25.6(1+\eps)^2k+O(1)$ & $\Ot(\eps^{-1}n^{1/k}\log\Delta)$ worst-case & weighted & \Cref{T-weighted-ni-det} \\
$\approx14.62(1+\eps)^2k+O(1)$ & $\Ot(\eps^{-1}n^{1/k}\log\Delta)$ worst-case & weighted & \Cref{T-weighted-ni-rand}$^\star$ \\
\hline
\end{tabular}
\caption{Name-independent routing schemes, where $\Delta$ is the aspect ratio of the graph and $0<\eps<1$. A pair $(\alpha,\beta)$ denotes $(\alpha,\beta)$-stretch, that is, $\hat d(u,v)\le\alpha\,d(u,v)+\beta$, or $\EE[\hat d(u,v)]\le\alpha\,d(u,v)+\beta$ for rows marked $^\star$. Schemes with worst-case storage also bound the average storage. Under the Erd\H{o}s girth conjecture, no scheme with $\Ot(n^{1/k})$ local storage has stretch below $2k-1$~\cite{DBLP:conf/spaa/AbrahamGM06}. $^\star$Expected stretch.}
\label{tab-ni-results}
\end{table}

\subsection{Related work}
Further work on compact routing studies sparse partitions and neighborhood covers~\cite{DBLP:conf/stoc/AwerbuchBLP89,AwerbuchP90,ABCP93}, sparse graphs~\cite{DBLP:conf/infocom/AgarwalGH11,DBLP:journals/corr/abs-1201-2703}, near-linear total storage~\cite{RodittyT16}, and vertex priorities~\cite{ElkinFN18}.
Average stretch over vertex pairs~\cite{EilamGP07} and small stretch for most pairs in networks of low doubling dimension~\cite{KonjevodRXY07} differ from our expected stretch guarantees for each pair.

Routing has also been studied for trees~\cite{DBLP:conf/icalp/FraigniaudG01}, planar and bounded-genus graphs~\cite{DBLP:conf/icalp/GavoilleH99,DBLP:conf/cocoon/Lu02,thorup2004compact}, graphs excluding a fixed minor~\cite{DBLP:conf/wdag/AbrahamGM05}, chordal graphs and related classes~\cite{DBLP:conf/wdag/DourisboureG02,DBLP:conf/isaac/DraganL04}, doubling and Euclidean metrics~\cite{DBLP:conf/podc/AbrahamM04,Talwar04,Slivkins07,ChanGMZ16}, and geometric and unit disk graphs~\cite{DBLP:journals/dc/HassinP01,DBLP:conf/isaac/MulzerW20}.
Other graph families include interconnection networks~\cite{DBLP:journals/algorithmica/NarayananO99,DBLP:journals/jpdc/CalamoneriI03} and random and Internet-like graphs~\cite{DBLP:conf/podc/IwamaO02,DBLP:conf/wdag/IwamaO03,DBLP:conf/infocom/KrioukovFY04,DBLP:conf/dasc/TangYZ09}.

Other settings include implicit and interval routing~\cite{SantoroK85,vanLeeuwenT87,Gavoille00survey}, directed roundtrip routing~\cite{DBLP:journals/jal/CowenW04,DBLP:journals/talg/RodittyTZ08}, roundtrip spanners~\cite{ZhuLam17,ZhuLam18,DBLP:conf/soda/PachockiRSTW18,CenDG20,ZhuHanLam21,DBLP:conf/cocoon/StaffordZ23,DBLP:conf/soda/HarbuzovaJWX24}, fault-tolerant routing~\cite{DBLP:journals/algorithmica/ChechikLPR12,Chechik13ft,DoryP21}, and distributed construction~\cite{LenzenP13,SarmaDP15,ElkinN18,ElkinN18podc,LenzenPP19}.
Further lower bounds concern routing memory~\cite{DBLP:conf/podc/GavoilleP96} and routing in trees~\cite{DBLP:conf/stacs/FraigniaudG02}.

Related distance structures include approximate distance oracles~\cite{DBLP:journals/jacm/ThorupZ05}, path-reporting oracles~\cite{ElkinP16,ElkinNW16,DBLP:conf/focs/ElkinS23,DBLP:conf/icalp/Chechik024a,NeimanShabat24}, distance labeling schemes~\cite{Peleg00labeling,GavoilleKKPP01,DBLP:journals/jal/GavoillePPR04}, and spanners~\cite{Althofer93,DBLP:journals/rsa/BaswanaS07,DBLP:conf/icalp/RodittyTZ05}.
Distance oracles also admit trade-offs for sparse graphs~\cite{PatrascuRT12,PatrascuR14,RodittyT23,KadriaR25new}, with lower bounds depending on query time~\cite{DBLP:conf/focs/SommerVY09} or the pivot-based model~\cite{KadriaR26pivot}.
Further work concerns query time~\cite{DBLP:conf/focs/MendelN06,DBLP:conf/soda/Wulff-Nilsen13,DBLP:conf/stoc/Chechik14,DBLP:conf/stoc/Chechik15} and construction time~\cite{DBLP:conf/soda/Wulff-Nilsen12,KadriaR25faster}.

Other uses of randomization include congestion minimization~\cite{ValiantB81,Racke02}, probabilistic tree embeddings~\cite{AlonKPW95,Bartal96,FakcharoenpholRT04}, randomized search~\cite{KaoRT96}, and online bidding~\cite{ChrobakKNY08}.
For broader background on routing and distance structures, see~\cite{Peleg00book,Gavoille01,GavoilleP03,DBLP:reference/algo/Elkin08a,Sommer14,DBLP:reference/algo/Roditty16a}.

\subsection{Organization}
\Cref{S-Prel} gives the preliminaries, including the lower bound for randomized routing (\Cref{L-LB-random}), and \Cref{S-Over} gives an overview of our techniques.
\Cref{S-avg} presents our labeled schemes with average local storage, and \Cref{S-worst} presents our labeled schemes with worst-case local storage.
\Cref{S-Name-Independent,S-NI-worst} present our name-independent schemes with average and with worst-case local storage.
\Cref{S-Discussion} discusses open problems.
The appendices prove the optimization bounds (\Cref{S-avg-opt,S-worst-opt,S-NI-opt,S-NI-worst-opt}) and extend our worst-case schemes to weighted graphs (\Cref{S-Weighted}). \Cref{S-summary} summarizes all our results.

\section{Preliminaries}\label{S-Prel}
Let $G=(V,E,\wt)$ be a connected undirected graph with $n$ vertices and positive edge weights $\wt:E\to\mathbb{R}^+$, where every vertex $x$ has a unique identifier $\id(x)\in\{1,\ldots,n\}$.
Let $d(u,v)$ be the distance between $u$ and $v$, and let $P(u,v)$ be a shortest path between them.\footnote{We choose the shortest paths consistently, so that they are closed under reversal and under taking subpaths, e.g., by an infinitesimal perturbation of the weights. Distances are with respect to the original weights.}
For $X\subseteq V$, let $d(u,X)=\min_{x\in X}d(u,x)$, where $d(u,\emptyset)=\infty$, and let $p(u,X)$ be the vertex of $X$ closest to $u$.
We write $[k]=\{0,\ldots,k-1\}$.

Throughout the paper, all sizes are in words of $O(\log n)$ bits.\footnote{For the word bounds, the weights are polynomially bounded integers, possibly after rescaling, so every distance fits in $O(1)$ words.}
We write $\RT(u)$ for the routing table of a vertex $u$ and $\L(u)$ for its label. The routing table of $u$ contains $\L(u)$. A vertex that holds a message forwards it based on its routing table, the label of the destination, and a header that it may modify.
The ports $1,\ldots,\deg(x)$ of every vertex $x$ are fixed in advance (the \emph{fixed-port} model).

\subsection{Bunches and clusters}\label{S-Prel-TZ}
Following Thorup and Zwick~\cite{DBLP:journals/jacm/ThorupZ05}, for $X,Y\subseteq V$, the bunch of $u$ is $B(u,X,Y)=\{v\in X\mid d(u,v)<d(u,Y)\}$, and the cluster of $w$ is $C(w,Y)=\{v\in V\mid d(w,v)<d(v,Y)\}$.
For an integer $k\ge2$, let $V=A_0\supseteq A_1\supseteq\cdots\supseteq A_{k-1}\supseteq A_k=\emptyset$, and let $h_i(u)=d(u,A_i)$ be the \emph{height} of $u$ at level $i$.
The pivots of $u$ are defined by $p_{k-1}(u)=p(u,A_{k-1})$ and, for $i<k-1$, $p_i(u)=p_{i+1}(u)$ if $h_i(u)=h_{i+1}(u)$ and $p_i(u)=p(u,A_i)$ otherwise. Thus $d(u,p_i(u))=h_i(u)$.
The $i$-th bunch of $u$ is $B_i(u)=B(u,A_i,A_{i+1})$, and $B(u)=\bigcup_{i\in[k]}B_i(u)$.
For $w\in V$ and $i\in[k]$, let $C_i(w)=C(w,A_{i+1})$. For $w\in A_i\setminus A_{i+1}$, we write $C(w)=C_i(w)$.
For every $u,w\in V$, $w\in B(u)$ if and only if $u\in C(w)$.

We use the deterministic construction of Roditty, Thorup, and Zwick~\cite{DBLP:conf/icalp/RodittyTZ05}.
\begin{lemma}[Lemma~1 of~\cite{kadria2025compact}]\label{L-TZ-Size}
    For every integer $k\ge2$, we can deterministically compute sets $V=A_0\supseteq\cdots\supseteq A_{k-1}\supseteq A_k=\emptyset$ with $|A_i|=\Ot(n^{1-i/k})$ and $|B_i(u)|=\Ot(n^{1/k})$, for every $i\in[k]$ and $u\in V$.
\end{lemma}
Hence $|B(u)|=\Ot(n^{1/k})$, and double counting gives $\sum_{w\in V}|C(w)|=\sum_{u\in V}|B(u)|=\Ot(n^{1+1/k})$.
Our average-storage schemes use \Cref{L-TZ-Size}. A single cluster may be large, so our worst-case schemes use the following hierarchy, in which every cluster is small.
\begin{lemma}[Theorem~3.1 of~\cite{DBLP:conf/spaa/ThorupZ01}, see also Theorem~A.2 of~\cite{DBLP:conf/podc/Chechik13}]\label{L-A-center}
    For every integer $k\ge2$, we can compute sets $V=A_0\supseteq\cdots\supseteq A_{k-1}\supseteq A_k=\emptyset$ with $|A_i|=\Ot(n^{1-i/k})$ and $|B_i(u)|=\Ot(n^{1/k})$, such that $|C(w)|\le4n^{(i+1)/k}$ for every $i\in[k]$ and $w\in A_i\setminus A_{i+1}$.
\end{lemma}

Our routing schemes send the message from the source to one of its own pivots. This is possible since every vertex is in the cluster of each of its pivots.
\begin{lemma}[Lemma~4.1 of~\cite{DBLP:journals/jacm/ThorupZ05}]\label{L-Pivot-In-Bunch}
    For every $u\in V$ and $i\in[k]$, $p_i(u)\in B(u)$ and $u\in C(p_i(u))$.
\end{lemma}

Thorup and Zwick~\cite{DBLP:journals/jacm/ThorupZ05} showed that a cluster is closed under shortest paths from its root (proof of their Lemma~4.2). Hence $T(C(w))=\bigcup_{v\in C(w)}P(w,v)$ is a shortest-path tree rooted at $w$ with vertex set $C(w)$, which we call the \emph{cluster tree} of $w$.
All our schemes route along cluster trees, and the main difficulty is to choose a good tree to route on. Once the tree is chosen, the labeled tree routing of Thorup and Zwick~\cite{DBLP:conf/spaa/ThorupZ01} routes along the tree path.
\begin{lemma}[Corollary~2.7 of~\cite{DBLP:conf/spaa/ThorupZ01}]\label{T-Route-In-Trees}
    Let $T$ be a tree on at most $n$ vertices with fixed $O(\log n)$-bit ports.
    We can assign every vertex $v$ a label $\L(v,T)$ of $O(\log^2n/\log\log n)$ bits such that, given $\L(x,T)$ and $\L(v,T)$ for $x\neq v$, the port of the first edge on the $x$--$v$ path in $T$ can be computed in constant time.
\end{lemma}
We apply~\Cref{T-Route-In-Trees} to every cluster tree $T(C(w))$, with the ports of $G$.
\begin{lemma}[Lemma~16 of~\cite{kadria2025compact}]\label{L-Pivot-Cluster}
    If $h_{i+1}(v)>d(u,v)+h_i(u)$, then $v\in C(p_i(u))$.
\end{lemma}

\paragraph{A lower bound for randomized routing.}
The lower bound of Thorup and Zwick~\cite{DBLP:conf/spaa/ThorupZ01} and Zinovyev~\cite{DBLP:conf/wdag/Zinovyev22} is stated for deterministic routing. We adapt it to randomized routing, in which the routing tables and the labels are deterministic and only the routing uses coins.
\begin{lemma}\label{L-LB-random}
Assume the Erd\H{o}s girth conjecture. For every integer $k\ge2$ and every $s<2k-1$, every labeled routing scheme with randomized routing, expected stretch $s$, and labels of $o(n^{1/(k-1)})$ bits uses $\Omega(n^{1+1/(k-1)})$ bits of total storage on some graph with $n$ vertices.
\end{lemma}
\begin{proof}
Zinovyev~\cite[Theorem~8]{DBLP:conf/wdag/Zinovyev22} proves the bound for deterministic routing, with fixed ports, by recovering the port assignment of a fixed graph of girth $2k$ with $\Omega(n^{1+1/(k-1)})$ edges from the routing tables.
For every edge $(x,y)$, a route from $x$ to $y$ shorter than $2k-1$ traverses the edge itself, so simulating it reveals the port of $x$ towards $y$.
With expected stretch $s$, such a route has probability at least $p=\frac{2k-1-s}{2k-2}>0$, since $\hat d(x,y)\ge1$.
The simulation does not send a message, so it may try any outcome of the coins.
Among $O(\log n/p)$ random outcomes, fixed in advance, with high probability every edge has one that gives such a route, and naming it takes $O(\log\log n)$ bits per edge, which is of lower order than the bound.
So the simulation recovers the whole port assignment, and the bound holds.
\end{proof}

\section{Overview}\label{S-Over}

In this section, we first describe the routing framework common to all of our schemes. We then present the key new ideas behind each of our results and explain the difficulties they overcome.


\paragraph{The framework.}
All our schemes use the hierarchy of Thorup and Zwick~\cite{DBLP:journals/jacm/ThorupZ05} (\Cref{S-Prel-TZ}), and route along its cluster trees.
In the labeled schemes, the label of a vertex $v$ is
\[
\L(v) = \pair{\id(v),\ \{\pair{i, h_i(v), \id(p_i(v)), \L(v,T(C(p_i(v)))), \L(p_i(v),T(C(p_i(v))))} \mid i\in[k]\}}.
\]
Given $\L(v)$, the source $u$ first checks whether $v\in B(u)\cup C(u)$, and if so, routes to $v$ along a shortest path.
Otherwise, let $\tau=\min\{i\in[k]\mid p_i(v)\in B(u)\}$ be the first level at which the pivot of $v$ is in $B(u)$.
Then both $u$ and $v$ are in the cluster of $p_\tau(v)$, and Thorup and Zwick~\cite{DBLP:conf/spaa/ThorupZ01} route along $T(C(p_\tau(v)))$, with stretch $4\tau-1\le4k-5$. We call $p_\tau(v)$ the \emph{fallback pivot}, and this route the \emph{fallback route}.

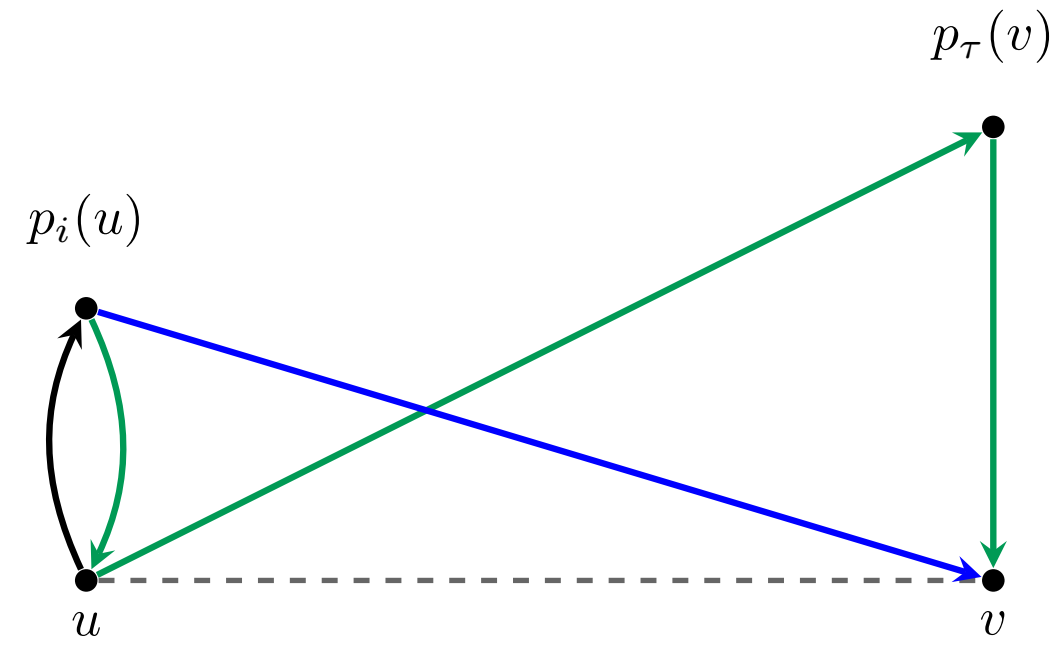
\begin{figure}[t]
\centering
\begin{tikzpicture}[>=stealth, every node/.style={circle, fill=black, inner sep=1.5pt}]
\node[label=below:$u$] (u) at (0,0) {};
\node[label=below:$v$] (v) at (6,0) {};
\node[label=above:$p_i(u)$] (pi) at (0,1.8) {};
\node[label=above:$p_\tau(v)$] (pt) at (6,3) {};
\draw[dashed, black!60, line width=1pt] (u) -- (v);
\draw[->, line width=1.2pt, ForestGreen] (pi) to[bend left=25] (u);
\draw[->, line width=1.2pt, ForestGreen] (u) -- (pt);
\draw[->, line width=1.2pt, ForestGreen] (pt) -- (v);
\draw[->, line width=1.2pt, black] (u) to[bend left=25] (pi);
\draw[->, line width=1.2pt, blue] (pi) -- (v);
\end{tikzpicture}
\caption{The message always goes from $u$ to $p_i(u)$ (black). It then takes the blue route if $v\in C(p_i(u))$, and the green route otherwise. The dashed line is $P(u,v)$.}
\label{F-avg-framework}
\end{figure}

Chechik~\cite{DBLP:conf/podc/Chechik13} and Kadria and Roditty~\cite{kadria2025compact} improve this stretch by first \emph{visiting} pivots of the source.
In a visit at level $i$, $u$ routes to $p_i(u)$, and $p_i(u)$ checks whether $v\in C(p_i(u))$. If so, the message continues to $v$ along $T(C(p_i(u)))$.
Otherwise, $d(u,v)\ge h_{i+1}(v)-h_i(u)$ by~\Cref{L-Pivot-Cluster}, and we call $h_{i+1}(v)-h_i(u)$ the \emph{threshold} of level $i$. The message then returns to $u$ and takes the fallback route.
So on failure, the message follows $u\to p_i(u)\to u\to p_\tau(v)\to v$ (\Cref{F-avg-framework}).

\paragraph{The difficulty.}
Kadria and Roditty~\cite{kadria2025compact} observed that if the source $u$ knew $d(u,v)$ in advance, it could get stretch $2k-1$. It would visit $p_i(u)$ for the first level $i$ with $h_{i+1}(v)-h_i(u)>d(u,v)$, since then $v\in C(p_i(u))$ by~\Cref{L-Pivot-Cluster}, or route through the fallback pivot if $\tau\le i$. As in the distance oracle of Thorup and Zwick~\cite{DBLP:journals/jacm/ThorupZ05}, every level below the chosen one raises the heights by at most $d(u,v)$, so the chosen level $i$ has $h_i(u)\le i\,d(u,v)$.
The source does not know $d(u,v)$. It knows only lower bounds on $d(u,v)$, from the labels and from failed visits, so it must choose which pivots to visit without knowing which visit succeeds.

\paragraph{New routing techniques.}
We overcome this difficulty with the following techniques.
\begin{itemize}
\item \textbf{Randomized routing.}
Only the routing is randomized. The routing tables and the labels are fixed before the routing starts.
We may view $d(u,v)$ as chosen by an adversary that knows the tables, the labels, and the distribution of the route, but not the coin flips of the source, since $d(u,v)$ is fixed with the graph before the source flips its coins.
Against a deterministic choice, the adversary picks a distance at which the chosen route is long.
Different routes are long at different distances, so a random choice prevents the adversary from picking a distance that is bad for all of them at once.

For example, let $k=3$, and suppose that the source knows that $d(u,v)\ge2$, $h_1(u)=2$, and that the fallback pivot is $p_2(v)$ with $h_2(v)=5$.
The fallback route has length up to $d(u,v)+2h_2(v)$, so its stretch can be $6$ at $d(u,v)=2$.
The visit at level $1$ has threshold $h_2(v)-h_1(u)=3$, so it can fail at $d(u,v)=3$, and then the route has length up to $3+2h_1(u)+2h_2(v)=17$, a stretch of $17/3$.
Both are above $2k-1=5$.
A fair coin between the two has expected stretch at most $\frac12(6+3)=4.5$ at $d(u,v)=2$, where the visit succeeds, and at most $\frac12(13/3+17/3)=5$ at $d(u,v)=3$, where the visit fails.

\item \textbf{Optimizing over all choices at once.}
In the labeled schemes and the deterministic name-independent schemes, the source bounds the cost of every possible choice before it sends the message, and solves an optimization problem over all levels at once, instead of comparing the levels one by one (\Cref{S-Over-avg}).
\item \textbf{Cheaper search in trees.}
With worst-case storage, a pivot cannot store its cluster, so a visit searches the cluster tree. We give a better search on trees, and search only a truncated tree (\Cref{S-Over-worst}).
\end{itemize}
We now describe these techniques for each model.

\subsection{Labeled routing with average storage}\label{S-Over-avg}
We start with this model, which is the simplest. Every vertex $u$ stores its bunch $B(u)$ and its cluster $C(u)$. This takes $\Ot(n^{1+1/k})$ total storage, that is, $\Ot(n^{1/k})$ on average. Its routing table is
\begin{align*}
\RT(u) = \ & \L(u) \\
\cup\ & \{\pair{\id(w), \L(u,T(C(w)))} \mid w\in B(u)\} \\
\cup\ & \{\pair{\id(x), \L(x,T(C(u)))} \mid x\in C(u)\}.
\end{align*}
We use the preprocessing of Kadria and Roditty~\cite{kadria2025compact}, with $\Ot(k)$ additional words per vertex, and change only the routing phase.
Our schemes visit a single pivot $p_i(u)$, for a level $i\in[\tau]$, before the fallback route. Since $p_0(u)=u$, the level $i=0$ means taking the fallback route directly. The key is to choose $i$, deterministically or at random, and to bound the resulting stretch.

\paragraph{Choosing the level.}
From $\RT(u)$ and $\L(v)$, the source immediately computes the lower bound
\[
M=\max\Big\{h_1(v),\ \max_{0\le i<\tau}\big(h_{i+1}(u)-h_i(v)\big),\ \max_{1\le i\le\tau}|h_i(u)-h_i(v)|\Big\}\le d(u,v).
\]
The three terms are at most $d(u,v)$ since $v\notin C(u)$, since $p_i(v)\notin B(u)$ for $i<\tau$, and by the triangle inequality (\Cref{S-avg-selection}).

We call $\hat d(u,v)-d(u,v)$ the \emph{extra cost}, and the extra cost divided by $d(u,v)$ the \emph{extra stretch}. If the visit succeeds, the route has length $h_i(u)+d(p_i(u),v)\le d(u,v)+2h_i(u)$. So the extra cost is at most $2h_i(u)$, and the extra stretch is at most $2h_i(u)/M$. If the visit fails, the fallback route has length at most $d(u,p_\tau(v))+h_\tau(v)\le d(u,v)+2h_\tau(v)$. So the extra cost of $u\to p_i(u)\to u\to p_\tau(v)\to v$ is at most $2h_i(u)+2h_\tau(v)$, and since $d(u,v)$ is at least the threshold, the extra stretch is at most
\[
\frac{2h_i(u)+2h_\tau(v)}{\max\{M,h_{i+1}(v)-h_i(u)\}}.
\]

Thus, for every level $i$, the extra stretch is at most $\beta_i$:
\[
\beta_i=\max\left\{\frac{2h_i(u)}M,\frac{2h_i(u)+2h_\tau(v)}{\max\{M,h_{i+1}(v)-h_i(u)\}}\right\}.
\]

The source knows every term of $\beta_i$, so it can compute $\beta_i$ for every level before visiting any pivot. In the deterministic case, we choose a level that attains $\beta^*=\min_{i\in[\tau]}\beta_i$, and show that $\beta^*\le\tau/\ln(3/2)\le(k-1)/\ln(3/2)$ in~\Cref{L-avg-det-bound}. This gives stretch $1+(k-1)/\ln(3/2)$ (\Cref{T-avg-det}).

\paragraph{The randomized choice.}
Our randomized scheme flips coins only to choose the level $i$.
We prove that, in expectation over the choice of $i$,
\[
\EE[\hat d(u,v)]  \le (2\tau + 1) d(u,v) \le (2k-1)d(u,v),
\]
which matches the lower bound that follows from the Erd\H{o}s girth conjecture (\Cref{L-LB-random}).

The source chooses the distribution over $[\tau]$ that minimizes its largest expected extra stretch over all distances $d\ge M$. 
The bound on the expected extra cost changes only when $d$ crosses a threshold, and between thresholds the bound on the expected extra stretch decreases.
Hence the source needs to consider only $d=M$ and the $O(k)$ thresholds.
Thus it solves a linear program with $O(k)$ variables and constraints, and samples $i$ from an optimal distribution (\Cref{S-avg-rand}).

\paragraph{Why the value is at most $2\tau$.}
Fix a distance guess $D\ge M$, and let $i$ be the first
level in $\{0,\ldots,\tau-2\}$ with
$h_{i+2}(u)-h_i(u)>M+D$, if such a level exists.
By the definition of $M$, we have
$h_{i+2}(u)-h_{i+1}(v)\le M$, and hence
\[
h_{i+1}(v)-h_i(u)
\ge h_{i+2}(u)-h_i(u)-M
>D.
\]
Thus the visit at level $i$ succeeds whenever $d(u,v)\le D$.
Moreover, by the choice of $i$, every earlier increase over
two levels is at most $M+D$. Summing these increases from
level $0$ or $1$, according to the parity of $i$, and using
$h_0(u)=0$ and $h_1(u)\le M$, gives
$h_i(u)\le\frac i2(M+D)\le iD$.

If no such level exists, the same summation gives
$h_{\tau-1}(u)\le(\tau-1)D$. If
$h_\tau(v)-h_{\tau-1}(u)>D$, visiting level $\tau-1$
succeeds whenever $d(u,v)\le D$, with extra cost at most
$2(\tau-1)D$. Otherwise, $h_\tau(v)\le\tau D$, so the
fallback route has extra cost at most $2\tau D$.
Thus, if the source knew $d(u,v)$, setting $D=d(u,v)$
would give extra cost at most $2\tau d(u,v)$.

Now consider a distribution of distances $d\ge M$, and
a level $i$ whose visit succeeds whenever $d\le D$.
The visit contributes at most $2h_i(u)$ to the extra cost,
and it can require the fallback route only when $d>D$.
Its expected extra cost is therefore at most
\(
2h_i(u)+2h_\tau(v)\Pr[d>D].
\)
To obtain the bound $2\tau\EE[d]$, after paying
$2h_i(u)$ for the visit, we have
$2\tau\EE[d]-2h_i(u)$ left for the expected fallback cost.
Since the fallback contributes at most $2h_\tau(v)$,
it suffices that
\[
\Pr[d>D]
\le
\frac{\tau\EE[d]-h_i(u)}{h_\tau(v)}.
\]
We prove that, for every distribution of distances, some
level has expected extra cost at most $2\tau\EE[d]$,
using the growth bounds above and treating the last level
and the direct fallback route separately.
By LP duality, this gives a distribution over levels whose
expected extra cost is at most $2\tau d$ for every $d\ge M$,
without knowing $d(u,v)$ (\Cref{S-avg-opt-rand}).

\subsection{Labeled routing with worst-case storage}\label{S-Over-worst}
In this model, as in the scheme of Thorup and Zwick~\cite{DBLP:conf/spaa/ThorupZ01}, every vertex $v$ stores only its bunch $B(v)$ and its level-$0$ cluster $C_0(v)=C(v,A_1)$, which has $\Ot(n^{1/k})$ vertices by \Cref{L-A-center}. A vertex can no longer store its cluster. So in a visit, $p_i(u)$ cannot check whether $v\in C(p_i(u))$. 
To solve this problem, Chechik~\cite{DBLP:conf/podc/Chechik13} distributed the cluster among the vertices of its tree $T(C(p_i(u)))$. Roughly speaking, the visit then solves name-independent routing on this tree: it reaches $v$ if $v\in C(p_i(u))$, and returns to $p_i(u)$ otherwise.

We improve upon the routing scheme of Chechik~\cite{DBLP:conf/podc/Chechik13} with three ideas:
\begin{enumerate}
\item We improve the name-independent routing on trees, and search only in a truncated cluster tree.
\item We choose the level of the visit by considering all the levels at once, rather than one by one.
\item We choose this level at random, as with average storage.
\end{enumerate}

\paragraph{Better name-independent routing on trees.}
Let $T$ be an unweighted tree on $N$ vertices, with root $w$ and radius $R$, and let $S\ge2$. We give a name-independent routing scheme on $T$ that finds the record of $v$, or determines that $v\notin T$, after a walk of length at most $R+2\lceil\log_S N\rceil-2$ from $w$ (\Cref{T-S-NI-Tree}). Every vertex stores $\Ot(S)$ words, and $\Ot(1)$ words in expectation.

Our idea is to randomize the construction of Abraham, Gavoille, and Malkhi~\cite{AbrahamGM04}. They store the $N$ pairs $\pair{\id(x),\L(x,T)}$, sorted by name, at the first $N/S$ vertices in a preorder of $T$, with an index that guides the search from $w$. They showed that the search for $\id(v)$ makes at most $\log_S(N/S)$ excursions of two edges each. So reaching the pair of $v$ takes at most $R+2\log_S(N/S)$ steps, but the first $N/S$ vertices of the preorder store $\Ot(S)$ words each.

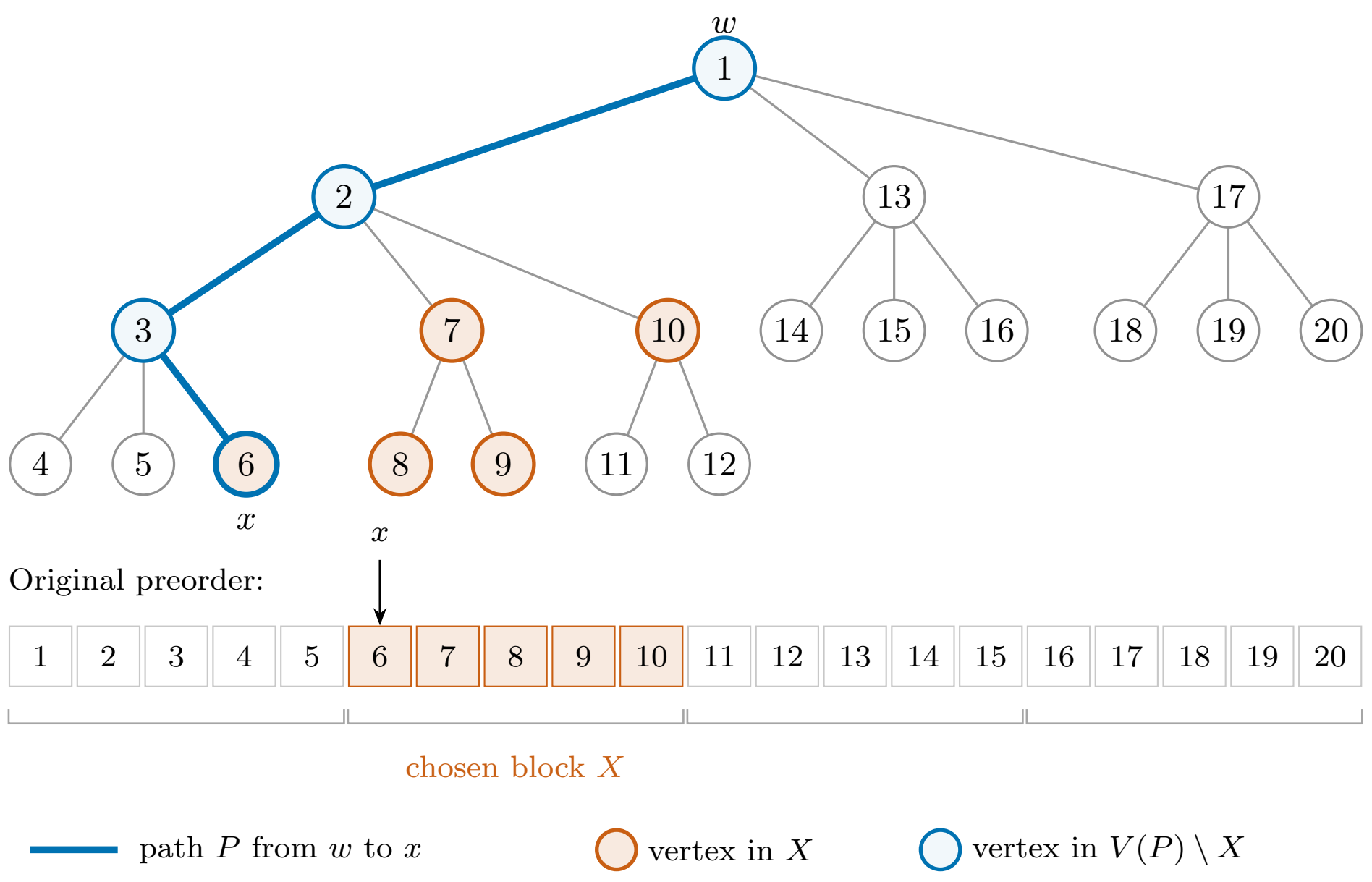
\begin{figure}[!ht]
\centering
\definecolor{pathblue}{HTML}{0072B2}
\definecolor{blockorange}{HTML}{C95F13}
\definecolor{treegray}{HTML}{757575}
\begin{tikzpicture}[
  x=.94cm,
  y=.94cm,
  tree edge/.style={
    draw=treegray!75,
    line width=.6pt
  },
  root path/.style={
    draw=pathblue,
    line width=1.8pt
  },
  tree vertex/.style={
    circle,
    draw=treegray!80,
    fill=white,
    line width=.6pt,
    minimum size=5.6mm,
    inner sep=0pt,
    font=\small
  },
  connector vertex/.style={
    tree vertex,
    draw=pathblue,
    line width=1pt,
    fill=pathblue!5
  },
  block vertex/.style={
    tree vertex,
    draw=blockorange,
    line width=1pt,
    fill=blockorange!13
  },
  first vertex/.style={
    block vertex,
    draw=pathblue,
    line width=1.5pt
  },
  order cell/.style={
    rectangle,
    draw=treegray!40,
    fill=white,
    minimum width=5.7mm,
    minimum height=5.5mm,
    inner sep=0pt,
    font=\footnotesize
  },
  selected cell/.style={
    order cell,
    draw=blockorange,
    fill=blockorange!13
  },
  note/.style={
    font=\footnotesize,
    align=left,
    inner sep=0pt
  },
  legend/.style={
    font=\footnotesize,
    inner sep=0pt
  }
]

  \coordinate (p1) at (1.0,0);
  \coordinate (p2) at (-2.7,-1.25);
  \coordinate (p3) at (-4.65,-2.55);
  \coordinate (p4) at (-5.65,-3.85);
  \coordinate (p5) at (-4.65,-3.85);
  \coordinate (p6) at (-3.65,-3.85);
  \coordinate (p7) at (-1.65,-2.55);
  \coordinate (p8) at (-2.15,-3.85);
  \coordinate (p9) at (-1.15,-3.85);
  \coordinate (p10) at (.45,-2.55);
  \coordinate (p11) at (-.05,-3.85);
  \coordinate (p12) at (.95,-3.85);
  \coordinate (p13) at (2.65,-1.25);
  \coordinate (p14) at (1.65,-2.55);
  \coordinate (p15) at (2.65,-2.55);
  \coordinate (p16) at (3.65,-2.55);
  \coordinate (p17) at (5.9,-1.25);
  \coordinate (p18) at (4.9,-2.55);
  \coordinate (p19) at (5.9,-2.55);
  \coordinate (p20) at (6.9,-2.55);

  \foreach \u/\v in {
    1/2,1/13,1/17,2/3,2/7,2/10,3/4,3/5,3/6,
    7/8,7/9,10/11,10/12,13/14,13/15,13/16,
    17/18,17/19,17/20}
    \draw[tree edge] (p\u)--(p\v);

  \draw[root path] (p1)--(p2)--(p3)--(p6);

  \foreach \i in {4,5,11,12,13,14,15,16,17,18,19,20}
    \node[tree vertex] at (p\i) {\i};

  \foreach \i in {1,2,3}
    \node[connector vertex] at (p\i) {\i};

  \foreach \i in {7,8,9,10}
    \node[block vertex] at (p\i) {\i};

  \node[first vertex] at (p6) {6};
  \node[above=2mm,font=\small] at (p1) {\(w\)};
  \node[below=3.3mm,font=\small,text=black] at (p6) {\(x\)};

  \node[note,anchor=west] at (-5.97,-5.0)
    {Original preorder:};

  \foreach \i in {1,...,20}{
    \pgfmathsetmacro{\xx}{-5.65+.66*(\i-1)}
    \ifnum\i>5
      \ifnum\i<11
        \node[selected cell] (s\i) at (\xx,-5.72) {\i};
      \else
        \node[order cell] (s\i) at (\xx,-5.72) {\i};
      \fi
    \else
      \node[order cell] (s\i) at (\xx,-5.72) {\i};
    \fi
  }

  \foreach \i/\j in {1/5,6/10,11/15,16/20}{
    \draw[treegray!65,line width=.5pt]
      ([yshift=-2mm]s\i.south west)--++(0,-1.3mm)
      -| ([yshift=-2mm]s\j.south east);
  }

  \node[below=5mm of s8,font=\footnotesize,text=blockorange]
    {chosen block \(X\)};

  \draw[-{Stealth[length=1.6mm]},line width=.6pt]
    ([yshift=6mm]s6.north)--(s6.north);
  \node[above=6.5mm of s6,font=\footnotesize]
    {\(x\)};

  \draw[root path] (-5.75,-7.6)--(-4.9,-7.6);
  \node[legend,anchor=west] at (-4.7,-7.6)
    {path \(P\) from \(w\) to \(x\)};

  \node[block vertex,minimum size=3.6mm] at (-.05,-7.6) {};
  \node[legend,anchor=west] at (.25,-7.6)
    {vertex in \(X\)};

  \node[connector vertex,minimum size=3.6mm] at (3.1,-7.6) {};
  \node[legend,anchor=west] at (3.4,-7.6)
    {vertex in \(V(P)\setminus X\)};

\end{tikzpicture}

\caption{Selecting a consecutive preorder block for \(N=20\)
and \(S=4\). Numbers on vertices give their positions in the
original preorder. The chosen block is \(X=\{6,7,8,9,10\}\),
its first vertex is \(x=6\), and \(P=(1,2,3,6)\).}
\label{fig:ni-tree-preorder}
\end{figure}

We split the preorder into $S$ consecutive blocks of about $N/S$ vertices each, and store the pairs in a block chosen uniformly at random (see~\Cref{fig:ni-tree-preorder}). A vertex is in the chosen block with probability $1/S$, and then it stores $O(S)$ pairs, so its expected storage is $\Ot(1)$. We need a random block since every vertex is in many cluster trees. A fixed block, such as the first one, can contain the same vertex in many of these trees, and then this vertex stores $\Ot(S)$ words in each of them. A random block need not contain $w$. The ancestors of block vertices that lie outside the block lie on the path from $w$ to the first vertex of the block, and the vertices of this path guide the search into the block, so the search keeps the same length bound.

The main difficulty in adapting their construction is that a preorder prefix induces a connected subtree, whereas a random block may be disconnected. We must show that the path from $w$ to the block can guide the search while preserving the search-length and storage bounds.

\paragraph{Truncated cluster trees.}
We apply this scheme to every cluster tree, with $S=\Ot(n^{1/k})$
chosen so that the cluster of every vertex of $A_i\setminus A_{i+1}$ has at most $S^{i+1}$ vertices.
The source visits only levels with $h_{i+1}(u)>h_i(u)$, so $p_i(u)\in A_i\setminus A_{i+1}$.
Thus, in the cluster tree of a visited pivot, the search, including the return to the root, has
length at most $2R+2i$.

The problem is that the radius of $T(C(p_i(u)))$ can be much larger
than $h_i(u)+d(u,v)$, and a search pays twice the radius.
If $h_{i+1}(v)>d(u,v)+h_i(u)$, then $v\in C(p_i(u))$ and $d(p_i(u),v)<h_{i+1}(v)$,
so it suffices to search $T(C(p_i(u)))$ truncated at this radius.

We do this as follows.
For every $w\in A_i\setminus A_{i+1}$, let $T_j$ be the subtree
of $T(C(w))$ induced by the $\min\{2^j,|C(w)|\}$ vertices of $C(w)$ closest to $w$,
for $j=0,\ldots,\lceil\log_2|C(w)|\rceil$, breaking ties consistently.
For $j\ge1$, we store the records
$\pair{\id(v),\L(v,T(C(w)))}$, for $v\in V(T_j)$,
and their search indices in the smaller tree $T_{j-1}$, as before.
The search then stays inside $T_{j-1}$, which has a small radius, while still detecting every $v\in V(T_j)$.
The pivot uses the first $T_j$ that contains all cluster vertices
at distance less than $h_{i+1}(v)$ from it.
The radius of $T_{j-1}$ is then less than $h_{i+1}(v)$.

Since $T_j$ has at most twice as many vertices as $T_{j-1}$,
the expected storage contributed by each search structure
remains $\Ot(1)$ per vertex, with worst-case storage $\Ot(S)$.
At each level $i$, a vertex belongs to $\Ot(n^{1/k})$ clusters,
each with $O(\log n)$ such structures.
The blocks of different search structures are chosen independently,
so with high probability, every vertex stores $\Ot(n^{1/k})$ words, and we repeat the choice of the blocks until this holds.

\paragraph{The routing phase.}
The source computes $M$ as before.
Only the cost of a visit to $p_i(u)$ changes.

After routing to $p_i(u)$, the message searches in $T_{j-1}$, with $T_j$ chosen as above.
The search, including the return to $p_i(u)$, has length at most $2h_{i+1}(v)+2i$.
If the search succeeds, the message continues to $v$ along a path of length $d(p_i(u),v)$.
If it fails, then $d(u,v)\ge h_{i+1}(v)-h_i(u)$ as before, and the message returns to $u$ and takes the fallback route.

Thus, a visit at level $i$ has extra stretch at most
\[
\beta_i=\max\left\{
    \frac{2h_i(u)+2h_{i+1}(v)+2i}{M},\quad
    \frac{2h_i(u)+2h_{i+1}(v)+2i+2h_\tau(v)}
         {\max\{M,h_{i+1}(v)-h_i(u)\}}
\right\}.
\]
As with average storage, the source computes the bound $\beta_i$ for every level before visiting any pivot, and visits a single pivot, chosen deterministically or at random. The source also chooses the search radius, which may be smaller than $h_{i+1}(v)$. A search of radius $h_i(u)+t$, for $M<t\le h_{i+1}(v)-h_i(u)$, costs $4h_i(u)+2t+2i$, and a failure shows that $d(u,v)\ge t$. The coefficients $\frac{8}{1+4\ln(3/2)}$ and $\frac83$ are the values of the resulting optimizations (\Cref{L-worst-det-opt,L-worst-rand-opt}).

\subsection{Name-independent routing}\label{S-Over-NI}

In this model, $u$ gets only $\id(v)$, and there are no labels. This causes two difficulties. First, $u$ cannot route through the pivots of $v$, since it does not know them. Intuitively, a vertex that could tell $u$ the pivots of $v$ would store a map from $\id(v)$ to a label of $v$, and could then route to $v$ by itself. Second, $u$ cannot compute the lower bound $M$, since it does not know the heights of $v$. It knows only that $d(u,v)\ge h_1(u)$ if $v\notin B(u)$.

\paragraph{Random scales.}
We change the routing phase to visit a sequence of pivots $p_i(u)$, in increasing order of level, until a visit succeeds. The source guesses $d(u,v)$ at geometrically increasing scales $D\ge h_1(u)$, and for each scale, it visits the pivot of the first level $i$ whose \emph{gap} $h_{i+1}(u)-h_i(u)$ exceeds $2D$. If $d(u,v)\le D$, then $h_{i+1}(v)\ge h_{i+1}(u)-d(u,v)>h_i(u)+d(u,v)$, so this visit succeeds by~\Cref{L-Pivot-Cluster}. Hence each failure shows that $d(u,v)>D$. With randomization, the scales are $e^{j+X}$, for $j\in\mathbb{Z}$, where the shift $X$ is uniform in $[0,1)$ and drawn once per message. In expectation, the first gap $h_1(u)\le d(u,v)$ costs at most $2d(u,v)$, and every other gap costs at most $4d(u,v)$. With the final route of length $d(u,v)$, this gives expected stretch $1+2+4(k-2)=4k-5$ (\Cref{T-NI-avg}).

\paragraph{Deterministic scales.}
Deterministically, the source picks the sequence of levels with the smallest worst-case bound. Let $\lambda=1/(2^{1/(k-1)}-1)$ and let $i_1<\cdots<i_m$ be the levels of the sequence. If $h_{i_1}(u)+\cdots+h_{i_t}(u)\le\lambda\,(h_{i_{t-1}+1}(u)-h_{i_{t-1}}(u))$ for every $t\ge2$, then the extra stretch is at most $4\lambda$, since the source continues past a level only if $2d(u,v)$ is at least its gap. We show that such a sequence always reaches level $k-1$. This gives stretch $1+4\lambda\le1+4(k-1)/\ln2\approx5.77k$ (\Cref{T-NI-det}), and this is tight: when the gaps are $2^{i/(k-1)}$, no sequence does better.

\paragraph{Worst-case storage.}
With worst-case storage, a visit searches the cluster tree of $p_i(u)$ as in the labeled schemes. The source does not know $h_{i+1}(v)$, so the radius of the search is $h_i(u)+D$, given by the current guess $D$. A failed search costs $4h_i(u)+2D+2i$, so the guesses now also pay for the radius. This gives $(8k+2e-7,2k-2)$-expected stretch and stretch $\approx11.54k$ (\Cref{T-NI-worst-random,T-NI-worst-det}).

\subsection{Weighted graphs}
Our average-storage schemes work in weighted graphs with the same bounds. With worst-case storage, the search in a cluster tree makes up to $i+1$ excursions of two edges each, so in a tree with maximum edge weight $W$ it costs up to $2(i+1)W$ more than in an unweighted tree. Since $W$ can be as large as the radius of the tree, a search at level $i$ may cost $\Theta(i)$ times its radius.

To avoid this, for every $j\ge0$ with $W_{j-1}$ at most the diameter, we preprocess the graph $G_j$ that contains only the edges of weight at most $W_j=(1+\eps)^j$. This multiplies the storage by $O(\eps^{-1}\log\Delta)$. If the source knew $d(u,v)$, it could route in $G_j$ for the smallest $j$ with $W_j\ge d(u,v)$. This graph contains a shortest path from $u$ to $v$, and every excursion costs at most $2(1+\eps)d(u,v)$. The source does not know $d(u,v)$, and a small $W_j$ makes the search cheaper, but it may make the distance in $G_j$ larger, or disconnect $u$ from $v$. Our routing chooses the level and the graph jointly, using the heights in all the graphs $G_j$ (\Cref{S-Weighted}).

\section{Labeled routing with average storage}\label{S-avg}
In this section, we present our labeled routing schemes with $\Ot(n^{1/k})$ average local storage, and prove \Cref{T-avg-random,T-avg-det}.
Both schemes use the same preprocessing (\Cref{S-avg-prepr}) and the same routing framework (\Cref{S-avg-framework}).
The source either routes directly to the destination in $T(C(v))$ or $T(C(u))$, or first visits one of its own pivots. If the destination is not in that pivot's cluster, the message returns to the source and routes to the destination in the cluster tree of one of the destination's pivots.

The two schemes differ only in which pivot the source visits (\Cref{S-avg-selection}). We present the randomized choice first (\Cref{S-avg-rand}) and then the deterministic one (\Cref{S-avg-det}).
The proofs of the two optimization bounds are in~\Cref{S-avg-opt}.

\subsection{Preprocessing}\label{S-avg-prepr}
We use the preprocessing of Kadria and Roditty~\cite{kadria2025compact}, with $\Ot(k)$ additional words per vertex.
We compute the deterministic hierarchy of~\Cref{L-TZ-Size} and the tree labels of~\Cref{T-Route-In-Trees} for every cluster tree.

The label of a vertex $v\in V$ is
\[
\L(v) = \pair{\id(v),\ \{\pair{i, h_i(v), \id(p_i(v)), \L(v,T(C(p_i(v)))), \L(p_i(v),T(C(p_i(v))))} \mid i\in[k]\}}.
\]
The routing table of a vertex $u\in V$ contains its label and the following two separately tagged parts:
\begin{align*}
\RT(u) = \ & \L(u) \\
\cup\ & \{\pair{\id(w), \L(u,T(C(w)))} \mid w\in B(u)\} \\
\cup\ & \{\pair{\id(x), \L(x,T(C(u)))} \mid x\in C(u)\}.
\end{align*}
The label $\L(u)$ gives the pivot distances and the pivots of $u$, and the labels needed to route to the pivots.
The two parts are separate dictionaries keyed by identifiers.
The first part gives the bunch membership test at $u$ and the label of $u$ in every cluster tree that contains $u$.
The second part gives the cluster membership test at $u$ and the labels of the vertices of $C(u)$ in $T(C(u))$.

\begin{lemma}\label{L-avg-size}
    $\sum_{u\in V}|\RT(u)|=\Ot(n^{1+1/k})$, and $|\L(u)|=\Ot(k)$ for every $u\in V$.
\end{lemma}
\begin{proof}
Every tree label has $\Ot(1)$ words, so $|\RT(u)|=\Ot(|B(u)|+|C(u)|+k)$ and $|\L(u)|=\Ot(k)$.
The total bound follows from $\sum_u|C(u)|=\sum_u|B(u)|=\Ot(n^{1+1/k})$ and $k\le\log n$.
\end{proof}

\subsection{The routing framework}\label{S-avg-framework}
Fix a source $u$ and a destination $v\ne u$ for the rest of this section.

\paragraph{The naive case.}
If $v\in B(u)$, then $u$ routes to $v$ in $T(C(v))$, using the level-$0$ tree label in $\L(v)$.
If $v\in C(u)$, then $u$ routes to $v$ in $T(C(u))$, using the tree label in $\RT(u)$.
In both cases, the message follows a shortest path.
For the rest of the section, we assume that $v\notin B(u)\cup C(u)$.

\paragraph{The general case.}
Let $\tau=\min\{i\in[k]\mid p_i(v)\in B(u)\}$ be the fallback level.
Using $\RT(u)$ and $\L(v)$, the source $u$ computes a level $i=\PickLevel(u,v)\in[\tau]$.
We define $\PickLevel$ in \Cref{S-avg-rand,S-avg-det}.

Then $u$ routes to $p_i(u)$ in $T(C(p_i(u)))$, and $p_i(u)$ checks, using $\RT(p_i(u))$, whether $v\in C(p_i(u))$.
If so, $p_i(u)$ routes to $v$ in $T(C(p_i(u)))$, using the tree label of $v$ stored in $\RT(p_i(u))$.
Otherwise, the message returns to $u$ in $T(C(p_i(u)))$, and $u$ routes to $v$ in $T(C(p_\tau(v)))$. We call this last route the \emph{fallback route}.
Since $T(C(p_\tau(v)))$ is a shortest-path tree rooted at $p_\tau(v)$, the fallback route has length at most $d(u,p_\tau(v))+h_\tau(v)$.
For $i=0$, we have $p_0(u)=u$ and $v\notin C(u)$, so the message takes the fallback route directly.
See~\Cref{A-avg-route} and~\Cref{F-avg-framework}.

Next, we show that every vertex on the routing walk has the information it needs.
To route in a cluster tree $T(C(w))$ towards a target $y$, the header contains $\id(w)$ and $\L(y,T(C(w)))$.
\begin{lemma}\label{L-avg-routing-works}
For every $w\in V$ and every vertex $z$ of $T(C(w))$, $\L(z,T(C(w)))\in\RT(z)$.
\end{lemma}
\begin{proof}
The vertex set of $T(C(w))$ is $C(w)$ (\Cref{S-Prel-TZ}), so $z\in C(w)$ and $w\in B(z)$. Hence $\L(z,T(C(w)))$ is in the first part of $\RT(z)$, keyed by $\id(w)$.
\end{proof}
Every vertex on the walk can compute the next port by \Cref{T-Route-In-Trees}. The label of $p_i(u)$ in $T(C(p_i(u)))$ is in $\L(u)$, which is stored in $\RT(u)$, the label of $v$ in $T(C(p_i(u)))$ is in the second part of $\RT(p_i(u))$, and the label of $v$ in $T(C(p_\tau(v)))$ is in $\L(v)$. The return label is given by \Cref{L-avg-routing-works}.
The header contains, besides $\L(v)$, only $i$, the current tree identifier and target label, and the return label, that is, $\Ot(1)$ words.

\begin{algorithm2e}[t]
\caption{$\Route(u,v)$}\label{A-avg-route}
\SetKw{KwRoute}{Route}
\If{$v\in B(u)$}{\KwRoute $u\leadsto v$ in $T(C(v))$\;\KwRet\;}
\If{$v\in C(u)$}{\KwRoute $u\leadsto v$ in $T(C(u))$\;\KwRet\;}
$\tau\gets\min\{i\in[k]\mid p_i(v)\in B(u)\}$\;
$i\gets\PickLevel(u,v)$\;
\KwRoute $u\leadsto p_i(u)$ in $T(C(p_i(u)))$\;
\If(\tcp*[h]{tested at $p_i(u)$}){$v\in C(p_i(u))$}{\KwRoute $p_i(u)\leadsto v$ in $T(C(p_i(u)))$\;\KwRet\;}
\KwRoute $p_i(u)\leadsto u$ in $T(C(p_i(u)))$\;
\KwRoute $u\leadsto v$ in $T(C(p_\tau(v)))$\;
\end{algorithm2e}

\subsection{Choosing the level}\label{S-avg-selection}
We first describe what the source knows when it chooses $i$: a lower bound $M$ on $d(u,v)$, and a bound on the cost of each choice of $i$. Then we define and analyze the randomized $\PickLevel$, followed by the deterministic one.

The source does not know $d(u,v)$.
To choose $i$, it computes
\[
M=\max\left\{h_1(v),\ \max_{0\le i<\tau}\big(h_{i+1}(u)-h_i(v)\big),\ \max_{1\le i\le\tau}|h_i(u)-h_i(v)|\right\},
\]
which is a lower bound on $d(u,v)$.
Indeed, $h_1(v)\le d(u,v)$ since $v\notin C(u)$.
By the minimality of $\tau$, $p_i(v)\notin B(u)$ for $0\le i<\tau$, so $h_{i+1}(u)\le d(u,p_i(v))\le d(u,v)+h_i(v)$.
Finally, $|h_i(u)-h_i(v)|\le d(u,v)$ by the triangle inequality.
Summing $h_1(v)\le M$ and $h_{i+1}(v)-h_i(v)=\big(h_{i+1}(v)-h_{i+1}(u)\big)+\big(h_{i+1}(u)-h_i(v)\big)\le2M$ for $1\le i<\tau$ gives $h_\tau(v)\le(2\tau-1)M$.
Together,
\begin{equation}\label{e-avg-M}
0<h_\tau(v)\le(2\tau-1)M\qquad\text{and}\qquad M\le d(u,v).
\end{equation}

If the source chooses $i\in[\tau]$, then
\begin{equation}\label{e-avg-penalties}
\hat d(u,v)\le d(u,v)+2h_i(u)+2h_\tau(v)\cdot\one[d(u,v)\ge h_{i+1}(v)-h_i(u)],
\end{equation}
where $\one[\cdot]$ is the indicator function.
Indeed, the message first travels $h_i(u)$ to $p_i(u)$.
If $v\in C(p_i(u))$, it then travels $d(p_i(u),v)\le h_i(u)+d(u,v)$.
Otherwise, $d(u,v)\ge h_{i+1}(v)-h_i(u)$ by~\Cref{L-Pivot-Cluster}, and the message travels $h_i(u)$ back to $u$ and then at most $d(u,p_\tau(v))+h_\tau(v)\le d(u,v)+2h_\tau(v)$ along the fallback route.
The first extra term pays for the pivot visit, and the second for the fallback.
The source knows every term of this bound except $d(u,v)$.
Below, $d\ge M$ denotes a possible value of $d(u,v)$.

Both choices need only consider the possible values
\[
\Theta=\{M\}\cup\{h_{i+1}(v)-h_i(u)\mid i\in[\tau],\ h_{i+1}(v)-h_i(u)>M\}.
\]
Indeed, let $\theta<\theta'$ be consecutive values of $\Theta$, or $\theta=\max\Theta$ and $\theta'=\infty$.
Every indicator $\one[d\ge h_{i+1}(v)-h_i(u)]$ is constant for $d\in[\theta,\theta')$.
For each level, the bound $2h_i(u)+2h_\tau(v)\cdot\one[d\ge h_{i+1}(v)-h_i(u)]$ on the extra cost is constant on this interval, as is its expectation under any distribution over the levels.
After dividing by $d$, each resulting bound on the extra stretch, or on the expected extra stretch under a distribution, is maximized on the interval at $d=\theta$.
Since $M=\min\Theta$, these intervals cover all $d\ge M$, so each bound attains its maximum over $d\ge M$ in $\Theta$.

\subsection{Randomized \texorpdfstring{$\PickLevel$}{PickLevel}}\label{S-avg-rand}
We now define the randomized $\PickLevel$, which samples a level to minimize the worst-case expected extra stretch.
\thmAvgRandom*

Let $\mathcal P(X)$ denote the set of distributions over a finite set $X$.
The source solves
\begin{equation}\label{e-avg-LP}
z^*=\min\left\{z\;\middle|\;\sum_{i\in[\tau]}\pi_i\big(2h_i(u)+2h_\tau(v)\cdot\one[d\ge h_{i+1}(v)-h_i(u)]\big)\le zd\ \text{for every }d\in\Theta,\quad\pi\in\mathcal P([\tau])\right\}.
\end{equation}
This linear program has $O(k)$ variables and constraints.\footnote{The coefficients are rational under our weight convention, allowing an exact rational optimum and exact sampling with random bits. Source computation is unrestricted.}
Given an optimum $(\pi^*,z^*)$, $\PickLevel(u,v)$ returns $i$ with probability $\pi_i^*$.
This is the only random choice during routing.

\paragraph{Stretch analysis.}
\begin{lemma}\label{L-avg-rand-exp}
$\EE[\hat d(u,v)]\le(1+z^*)d(u,v)$.
\end{lemma}
\begin{proof}
By the choice of $\Theta$, the constraint of~\eqref{e-avg-LP} for $\pi^*$ and $z^*$ holds for every $d\ge M$, not only for $d\in\Theta$.
Since $d(u,v)\ge M$, the claim follows from~\eqref{e-avg-penalties} by taking the expectation over $i$.
\end{proof}

The following lemma bounds the value of the linear program and yields the desired expected stretch.
\begin{restatable}{lemma}{lemLPvalue}\label{L-avg-LP-value}
$z^*\le2\tau$.
\end{restatable}
\begin{proof}
The proof is deferred to~\Cref{S-avg-opt-rand}.
\end{proof}

We now prove~\Cref{T-avg-random}.
\begin{proof}[Proof of~\Cref{T-avg-random}]
\Cref{L-avg-size} bounds the storage and the labels.
By~\Cref{L-avg-routing-works}, every vertex on the walk computes the next port, so the message is always delivered, with headers of $\Ot(1)$ words besides $\L(v)$.
The naive case has stretch $1$.
Otherwise, by~\Cref{L-avg-rand-exp} and~\Cref{L-avg-LP-value},
\[
\EE[\hat d(u,v)]\le(1+z^*)d(u,v)\le(2\tau+1)d(u,v)\le(2k-1)d(u,v).
\qedhere
\]
\end{proof}

\subsection{Deterministic \texorpdfstring{$\PickLevel$}{PickLevel}}\label{S-avg-det}
We now define the deterministic $\PickLevel$, which chooses the level with the smallest worst-case extra stretch.
\thmAvgDet*

For $i\in[\tau]$, let
\[
\beta_i=\max\left\{\frac{2h_i(u)}M,\frac{2h_i(u)+2h_\tau(v)}{\max\{M,h_{i+1}(v)-h_i(u)\}}\right\},
\]
and let $\beta^*=\min_{i\in[\tau]}\beta_i$.
The deterministic $\PickLevel(u,v)$ returns a level attaining $\beta^*$.

\paragraph{Stretch analysis.}
By the choice of $\Theta$ and~\eqref{e-avg-penalties}, choosing such a level gives stretch at most $1+\beta^*$.
The following bound therefore suffices.
\begin{restatable}{lemma}{lemDetBound}\label{L-avg-det-bound}
$\beta^*\le\tau/\ln(3/2)$.
\end{restatable}
\begin{proof}
The proof is deferred to~\Cref{S-avg-opt-det}.
\end{proof}

\begin{proof}[Proof of~\Cref{T-avg-det}]
The storage, label, and header bounds follow as in the proof of~\Cref{T-avg-random}.
The naive case has stretch $1$.
Otherwise,~\Cref{L-avg-det-bound} gives stretch at most
\[
1+\beta^*\le1+\frac\tau{\ln(3/2)}\le1+\frac{k-1}{\ln(3/2)}.
\]
For the bound $4k-5$, level $0$ is the fallback route ($p_0(u)=u$). Since $h_0(u)=0$ and $h_1(v)-h_0(u)=h_1(v)\le M$, we have $\beta^*\le\beta_0=2h_\tau(v)/M\le2(2\tau-1)$ by~\eqref{e-avg-M}, giving stretch at most $4\tau-1\le4k-5$.
\end{proof}
For every $k\ge3$, this improves the previous bound of Kadria and Roditty~\cite{kadria2025compact}. The comparison is in~\Cref{S-avg-opt-compare}.
The profile in~\Cref{S-avg-opt-tight} shows that the coefficient $1/\ln(3/2)$ cannot be improved using only the inequalities of our analysis.

\section{Labeled routing with worst-case storage}\label{S-worst}

In this section, we present our labeled routing schemes with
$\Ot(n^{1/k})$ local storage at every vertex in unweighted graphs, and prove
\Cref{T-worst-det,T-worst-random}.
We use the routing framework of \Cref{S-avg}, and \Cref{S-Over-worst} describes the changes.
The main difference is that a pivot can no longer store its
cluster, so testing whether the destination is in
the cluster requires a search in the cluster tree.

We first give a name-independent routing scheme on trees
(\Cref{S-worst-trees}), and use it to construct the routing tables (\Cref{S-worst-prepr}).
Our two schemes use the same preprocessing and
differ only in the choice of the visited pivot.
We present the deterministic choice (\Cref{S-worst-det}) and then the
randomized one (\Cref{S-worst-rand}).

\subsection{Name-independent routing on trees}\label{S-worst-trees}

\Cref{T-Route-In-Trees} requires the label of the destination. In our schemes with worst-case storage, a message reaches the root of a cluster tree with only the identifier of the destination.
Abraham, Gavoille, and Malkhi~\cite{AbrahamGM04} give a name-independent routing scheme on trees that trades worst-case local storage for additive stretch.
For a parameter $S$, a vertex in their scheme stores up to $\Ot(S)$ words, and the vertices of the tree store $\Ot(1)$ words on average.
This average bound is not enough for us. Every vertex is in many cluster trees, and the same vertex could store $\Ot(S)$ words in many of them.
We therefore use a randomized variant, in which every vertex stores $\Ot(1)$ words in expectation. The expected storage of a vertex then adds up over the trees that contain it.
\begin{definition}\label{D-S-NI-Tree}
    Let $T$ be a tree rooted at $w$.
    A randomized \emph{$S$-name-independent routing scheme} on $T$
    receives only $\id(v)$ at $w$. If $v\in V(T)$, it finds $\L(v,T)$ and routes the message to $v$. Otherwise, it determines that $v\notin V(T)$ and returns the message to $w$.
    We denote by $\hat d_T(w,v)$ the total length of this walk.
    Each vertex stores at most $\Ot(S)$ words for every outcome of
    the preprocessing, and $\Ot(1)$ words in expectation over
    the preprocessing randomness.
\end{definition}
We use such a scheme on a cluster tree to test whether the destination is in the cluster, and if so, to route to it.

\begin{lemma}\label{T-S-NI-Tree}
Let \(T\) be an unweighted tree on \(N\ge 2\) vertices,
with root \(w\) and radius
\(R=\max_{x\in V(T)} d_T(w,x)\).
For every integer \(2\le S\le N\), there is an
\(\Ot(S)\)-name-independent routing scheme on \(T\)
such that, for every name \(\id(v)\),
\[
    \hat d_T(w,v)\le
    \begin{cases}
        3R+ 2\lceil \log_S N \rceil - 2,
            & v\in V(T),\\
        2R+2\lceil \log_S N \rceil - 2,
            & v\notin V(T).
    \end{cases}
\]
Moreover, the record of $v$ is found, or its absence is determined, after a walk of length at most $R+2\lceil\log_S N\rceil-2$ from $w$.
\end{lemma}

The search first locates the record \(\pair{\id(v),\L(v,T)}\)
associated with the destination name.
We distribute these records among a subset of the vertices
and store indices that guide the search from \(w\).
Once the record is found, the message uses \(\L(v,T)\) to reach \(v\).

\paragraph{Preprocessing.}
Let \(N=|V(T)|\), and let \(2\le S\le N\) be the parameter
from \Cref{T-S-NI-Tree}.
We first compute the tree-routing labels \(\L(v,T)\).
Each vertex stores its own label, and every nonroot vertex
stores the port leading to its parent.

Fix an ordering of the children of each vertex and compute
a preorder traversal of \(T\).
Partition this order into \(S\) consecutive blocks whose sizes
differ by at most one, and choose one block uniformly at random.
Let \(X\) be its set of vertices.
The set \(X\) contains \(\Theta(N/S)\) vertices,
which will each store \(O(S)\) records.
The random choice is used to balance the expected storage
among the vertices.

Let \(x\) be the first vertex of \(X\) in the preorder,
let \(P\) be the path from \(w\) to \(x\), and let
\(T'=T[X\cup V(P)]\).
Since \(X\) is a consecutive preorder block, every ancestor
of a vertex of \(X\) that lies outside \(X\) belongs to \(P\).
Thus, \(T'\) is a rooted subtree containing \(w\) and \(X\).

We now order the children in \(T'\) for the placement of records.
At each vertex of \(V(P)\setminus X\), place the child continuing
along \(P\) first, and order the remaining children by decreasing
number of vertices of \(X\) in their subtrees.
At each vertex of \(X\), order all children by the same quantity.
This second preorder determines the placement of records
on the already chosen set \(X\).

Sort the records \(\pair{\id(v),\L(v,T)}\), for \(v\in V(T)\),
by name and partition them into \(|X|\) consecutive groups
whose sizes differ by at most one.
Assign these groups, in order, to the vertices of \(X\)
encountered by the second preorder.
The records stored in each subtree of \(T'\) therefore form
a consecutive part of the sorted list.
Each vertex \(u\in V(T')\) stores the interval \(I(u)\)
from the smallest to the largest name stored in its subtree.

To choose among the children of a vertex, we use an index
whose tables are distributed among the vertex and its children.
This avoids storing the complete list of child intervals
at the parent.
Children with larger subtrees receive shorter words
in the following encoding, so that reaching them requires
fewer index lookups.

For an ordered list of children, encode the first \(S\) children
by one-digit words over \(\{0,\ldots,S-1\}\),
the next \(S^2\) by two-digit words, and so on.
Within each group, assign words in lexicographic order.
For each word length, the representative of a nonempty prefix is the first child with that prefix.
The prefix's interval spans the names stored in the subtrees
of all children with that prefix.
The root table lists the intervals, representatives,
and word lengths of all first-digit prefixes.
For each proper prefix of a word, store at its representative
a table listing its nonempty one-digit extensions.
Each entry includes the corresponding interval
and the port leading from the parent to its representative.

At a vertex of \(X\), build this index for all its children
in \(T'\), and store the root table locally.
At a vertex of \(V(P)\setminus X\), handle the child continuing
along \(P\) separately.
Build an index for the remaining children in \(T'\) and, if any exist,
store its root table at one of them.
These children belong to \(X\).
The vertex itself stores the interval and port of the child
continuing along \(P\), and the port leading to the
root-table holder, if one exists.

\begin{figure}[!ht]
\centering
\definecolor{pathblue}{HTML}{0072B2}
\definecolor{blockorange}{HTML}{C95F13}
\definecolor{treegray}{HTML}{757575}

\begin{tikzpicture}[
  tree edge/.style={draw=treegray!75,line width=.6pt},
  root path/.style={draw=pathblue,line width=1.8pt},
  tree vertex/.style={
    circle,draw=treegray!80,fill=white,
    line width=.6pt,minimum size=5.6mm,
    inner sep=0pt,font=\small
  },
  connector vertex/.style={
    tree vertex,draw=pathblue,
    line width=1pt,fill=pathblue!5
  },
  block vertex/.style={
    tree vertex,draw=blockorange,
    line width=1pt,fill=blockorange!13
  },
  first vertex/.style={
    block vertex,draw=pathblue,line width=1.5pt
  },
  storage box/.style={
    rectangle,rounded corners=2pt,
    line width=.8pt,align=center,
    inner sep=4pt,font=\footnotesize
  },
  path box/.style={
    storage box,draw=pathblue,fill=pathblue!5
  },
  block box/.style={
    storage box,draw=blockorange,fill=blockorange!8
  }
]
  \node[connector vertex] (v1) at (.7,0) {1};
  \node[connector vertex] (v3) at (-5,-4.55) {3};
  \node[first vertex] (v6) at (-4,-7.2) {6};
  \node[block vertex] (v8) at (-1.65,-7.2) {8};
  \node[block vertex] (v9) at (-.15,-7.2) {9};

  \node[above=2mm,font=\small] at (v1.north) {\(w\)};
  \node[below=1mm,font=\small,text=black] at (v6.south) {\(x\)};

  \node[path box,text width=3.4cm] (v2) at (-2,-1.7) {
    \textbf{2}\\[1mm]
    \(I(2)=[a_1,a_{20}]\)\\
    \([a_1,a_4]\to 3\)\\
    Otherwise: \(7\) (index)
  };

  \node[block box,text width=4.5cm] (v7) at (-.6,-4.55) {
    \textbf{7}\\[1mm]
    Records: \(a_5,\ldots,a_8\)\\
    \(I(7)=[a_5,a_{16}]\)\\[1mm]
    Index of \(2\):\\
    \([a_5,a_{16}]\to 7,\quad [a_{17},a_{20}]\to 10\)\\[1mm]
    Own index:\\
    \([a_9,a_{12}]\to 8,\quad [a_{13},a_{16}]\to 9\)
  };

  \node[block box,text width=3cm] (v10) at (4,-4.55) {
    \textbf{10}\\[1mm]
    Records: \(a_{17},\ldots,a_{20}\)\\
    \(I(10)=[a_{17},a_{20}]\)
  };

  \draw[root path] (v1.south)--(v2.north);
  \draw[root path] (v2.south west)--(v3.north);
  \draw[root path] (v3.south)--(v6.north);

  \draw[tree edge] (v2.south)--(v7.north);
  \draw[tree edge] (v2.east)--(v10.north);
  \draw[tree edge] (v7.south)--(v8.north);
  \draw[tree edge] (v7.south)--(v9.north);

\end{tikzpicture}

\caption{The subtree \(T'=T[X\cup V(P)]\) from
\Cref{fig:ni-tree-preorder}, with storage shown at vertices
\(2,7,10\). Here \(a_1<\cdots<a_{20}\) are the sorted names.
Each record includes its routing label.
Vertex \(7\) stores its own root table and that of vertex \(2\).}
\label{fig:ni-tree-storage}
\end{figure}
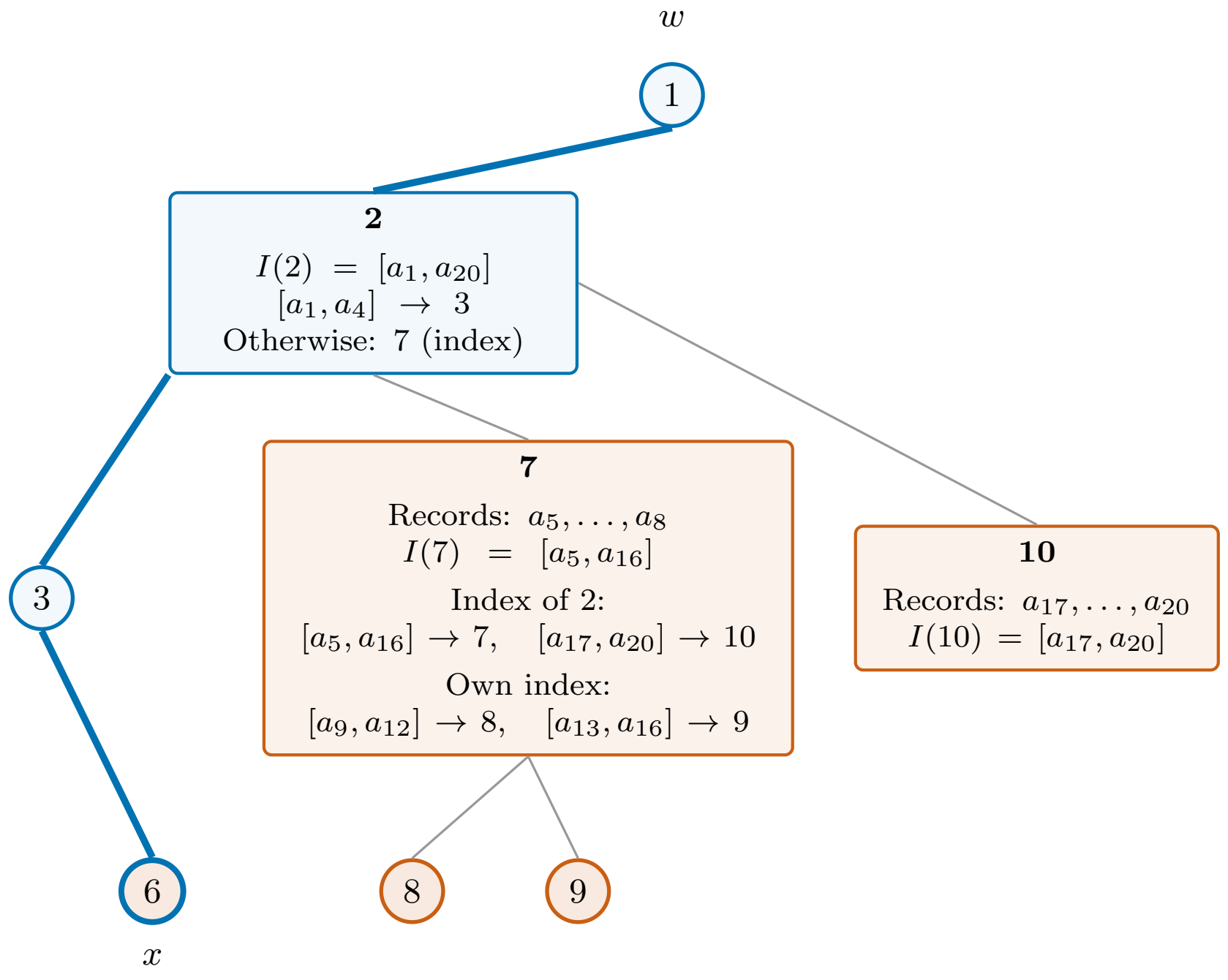

\paragraph{Routing.}
The search starts at \(w\) with the destination name \(\id(v)\).
During the search, the message either searches the current subtree
or reads a table in a child index.
In the latter case, it records the vertex whose children
are being indexed and whether it is reading the root table
or a table for a specified word length and prefix.

When the search is at \(u\) (and not reading an index table), it first checks whether \(\id(v)\in I(u)\).
If not, report failure.
If \(u\in X\), check its stored records.
If the destination's record is found, use \(\L(v,T)\)
to deliver the message to \(v\) along the shortest path in \(T\).
Otherwise, read the local root table to choose among
the children of \(u\).

If \(u\in V(P)\setminus X\), compare \(\id(v)\) with the interval
of the child continuing along \(P\).
If it belongs to that interval, forward the message to that child.
Otherwise, move to the designated child holding the root table
for the remaining children.
If no such child exists, report failure.

Within an index, use the destination name to select
a matching entry in the root table and then refine
the selected prefix using the corresponding tables.
Tables stored at the current vertex are read locally.
To reach a representative from its parent, traverse
the edge leading to that child.
To move between two children of the same parent,
follow the two-edge path through the parent.
The message carries the outgoing port to use upon returning
to the parent and remains in the index-reading mode
throughout this transition.
Once a complete word is identified, resume the subtree search
at the child encoded by that word.

If no interval matches the destination name in a consulted table,
report failure. Upon failure, return the message to \(w\) by following parent ports.

\begin{proof}[Proof of \Cref{T-S-NI-Tree}]
Every vertex stores \(\Ot(1)\) words outside \(X\) and
\(\Ot(S)\) words inside \(X\).
Since each vertex belongs to \(X\) with probability \(1/S\),
its expected storage is \(\Ot(1)\).
We now bound \(\hat d_T(w,v)\).

Let \(m(u)\) be the number of vertices of \(X\) in the
subtree of \(T'\) rooted at \(u\).
Thus \(m(w)=|X|\), and \(m(u)\ge 1\) throughout \(T'\).
Since the records are ordered by name, every interval test
preserves the destination's record if it exists.
The search therefore ends either at its holder or with
a correct report of absence.

The search follows a descending path, with excursions
to other children to read index tables.
Each excursion traverses one edge in each direction.
We bound their number as in~\cite{AbrahamGM04}, using
\(m(u)\) in place of the subtree size.

Consider an index at \(u\) that selects a child \(y\)
whose word has length \(\ell\).
Reading the successive prefixes requires at most
\(\ell-1\) excursions before reaching \(y\).
For \(\ell\ge 2\), the rank of \(y\) in the indexed list
is at least \(1+S+\cdots+S^{\ell-1}>S^{\ell-1}\).
Since these children are ordered by decreasing \(m\),
\[
    m(y)<m(u)/S^{\ell-1}.
\]
If the search fails inside the index, take \(y\) to be
the last representative visited.
Its word belongs to the same length group, and at most
\(\ell-1\) excursions precede it, so the same argument applies.
Failure at a locally stored root table adds no excursion.

Reading a root table stored at a child of a vertex in
\(V(P)\setminus X\) may cause one additional excursion.
This happens at most once: after leaving \(P\) through
an indexed child, the search stays in a subtree contained
in \(X\).

Let \(q\) count the excursions other than this possible
additional one.
Each index contributing a positive number of excursions
decreases \(m\) by a factor strictly greater than \(S\)
raised to the power of that number.
Along the remaining descending edges, \(m\) does not increase.
Since \(m\) starts at \(|X|\) and remains at least one,
\(S^q<|X|\) whenever \(q>0\).
It follows that
\(q\le\lceil\log_S|X|\rceil-1\).
Including the possible additional excursion gives at most
\(\lceil\log_S|X|\rceil\) excursions.

The balanced partition gives
\[
    |X|\le\lceil N/S\rceil
    \le S^{\lceil\log_S N\rceil-1},
\]
so there are at most \(\lceil\log_S N\rceil-1\) excursions, each of length \(2\).
The descending path has length at most \(R\), so the record of $v$ is found, or its absence is determined, after a walk of length at most \(R+2\lceil\log_S N\rceil-2\).
If \(v\in V(T)\), the message then reaches \(v\) in \(T\) after at most \(2R\) more steps.
Otherwise, it returns to \(w\) after at most \(R\) more steps.

\end{proof}

\subsection{Preprocessing}\label{S-worst-prepr}

We compute the hierarchy of \Cref{L-A-center}, and let \(S=4\lceil n^{1/k}\rceil\), so every cluster of a vertex of \(A_i\setminus A_{i+1}\) has at most \(S^{i+1}\) vertices.
Every vertex \(u\) stores \(\L(u)\) and the first part of \(\RT(u)\) from \Cref{S-avg-prepr}, with the labels as defined there.
Instead of the second part, \(u\) stores \(C_0(u)\), which has \(O(S)\) vertices, as a dictionary of tree labels in \(T(C_0(u))\).

For each cluster tree \(T(C(w))\), order its vertices by distance
from \(w\), breaking ties consistently.
Let \(T_j\) be the subtree induced by the first
\(\min\{2^j,|C(w)|\}\) vertices, for
\(0\le j\le\lceil\log_2|C(w)|\rceil\).
For each \(j\ge 1\), distribute the records
\[
    \{\pair{\id(v),\L(v,T(C(w)))}:v\in V(T_j)\}
\]
among the vertices of \(T_{j-1}\), using the construction of
\Cref{T-S-NI-Tree}.
Thus, the search takes place in \(T_{j-1}\), but determines
whether \(v\in V(T_j)\).
Since \(|V(T_j)|\le 2|V(T_{j-1})|\), each record holder stores
at most twice as many records as before.
The index construction and the search bound are unchanged.
All records use labels for the full cluster tree \(T(C(w))\).

If \(|V(T_{j-1})|\le S\), store the records of \(T_j\)
directly at \(w\).
The record for \(T_0=\{w\}\) is also stored at \(w\).
For each \(j\), the root stores the distance to the first
cluster vertex outside \(T_j\), or \(\infty\) if there is none.
Given a radius, it can therefore select the first \(T_j\)
containing every cluster vertex strictly within that radius.

Choose the preorder blocks independently for different
search structures.
Each vertex belongs to \(\Ot(n^{1/k})\) cluster trees,
each with \(O(\log n)\) such structures.
Each structure contributes \(\Ot(1)\) expected words and
at most \(\Ot(S)\) words to a vertex.
Independence and concentration give \(\Ot(n^{1/k})\) storage
at every vertex with high probability.
The explicit dictionaries and the information stored at
each root fit the same bound.
We repeat the block choices if this bound is exceeded.

\subsection{Deterministic routing}\label{S-worst-det}
\thmWCDet*

If \(v\in B(u)\cup C_0(u)\), route to \(v\) along a shortest path.
Otherwise, compute \(\tau\) and \(M\) as in
\Cref{S-avg-selection}, where \(v\notin C_0(u)\) gives \(h_1(v)\le d(u,v)\).
The source chooses either the fallback route or a single visit
to \(p_i(u)\), for \(1\le i<\tau\).

Let \(\delta_i=h_{i+1}(v)-h_i(u)\) be the threshold of level \(i\) and \(H=h_\tau(v)\).
We consider only levels with \(\delta_i>M\).
For a level with \(\delta_i\le M\), the visit may fail already when \(d(u,v)=M\), so its bound exceeds that of the fallback route.
For every considered level,
\[
    h_{i+1}(u)\ge h_{i+1}(v)-M>h_i(u).
\]
Hence \(p_i(u)\in A_i\setminus A_{i+1}\), so its cluster
has at most \(S^{i+1}\) vertices.

\paragraph{The cost of a visit.}
The message first routes to \(p_i(u)\).
The pivot selects the first \(T_j\) containing every vertex
of its cluster at distance less than \(h_{i+1}(v)\).
If \(j=0\), it checks the record locally.
Otherwise, it searches for \(\id(v)\) using the structure
stored in \(T_{j-1}\).

By the minimality of \(j\), every vertex of \(T_{j-1}\)
is at distance less than \(h_{i+1}(v)\) from the pivot.
Since distances are integers, the radius of \(T_{j-1}\) is at most \(h_{i+1}(v)-1\).
By \Cref{T-S-NI-Tree} with \(R\le h_{i+1}(v)-1\) and \(\lceil\log_S|V(T_{j-1})|\rceil\le i+1\), the walk from the pivot to the record holder has length at most \(h_{i+1}(v)+2i-1\).
On success, the message uses the retrieved label to reach \(v\).
The distance from the record holder to \(v\) is at most
\(h_{i+1}(v)-1+d(p_i(u),v)\), so the total route length is at most
\[
    h_i(u)+2h_{i+1}(v)+2i-2+d(p_i(u),v)
    \le d(u,v)+2h_i(u)+2h_{i+1}(v)+2i-2.
\]
On failure, the message returns to \(u\) and takes the
fallback route in \(T(C(p_\tau(v)))\).
The total route length is then at most
\[
    d(u,v)+2h_i(u)+2h_{i+1}(v)+2i-2+2h_\tau(v).
\]

A failure implies \(d(u,v)\ge\delta_i\).
Indeed, below this threshold, \(v\) belongs to the cluster
by \Cref{L-Pivot-Cluster}, and its distance from the pivot
is less than \(h_{i+1}(v)\), so \(v\in V(T_j)\).
Combining the two cases gives
\[
    \hat d(u,v)\le d(u,v)+2h_i(u)+2h_{i+1}(v)+2i-2
    +2H\,
      \one[d(u,v)\ge\delta_i].
\]

\paragraph{Choosing the visit and the radius.}
The search radius need not be \(h_{i+1}(v)\).
For each \(1\le i<\tau\), the source considers every integer
distance guess \(t\) with \(M<t\le\delta_i\).
For this choice, the pivot searches the first \(T_j\)
containing all cluster vertices at distance less than
\(h_i(u)+t\). The guess \(t=\delta_i\) gives the visit above.

The search takes place in \(T_{j-1}\), whose radius is
at most \(h_i(u)+t-1\), since distances are integers.
Thus the extra cost is at most
\(4h_i(u)+2t+2i-2\) on success, and at most
\(4h_i(u)+2t+2i-2+2H\) on failure.
Failure implies \(d(u,v)\ge t\).
The source therefore minimizes
\[
\beta^*=
\min\left\{
    \frac{2H}{M},\
    \min_{\substack{1\le i<\tau\\
                    t\in\mathbb Z,\ M<t\le\delta_i}}
    \max\left\{
        \frac{4h_i(u)+2t+2i-2}{M},
        \frac{4h_i(u)+2t+2i-2+2H}{t}
    \right\}
\right\}.
\]
The first choice is the fallback route.
The resulting stretch is at most \(1+\beta^*\).

\begin{restatable}{lemma}{lemWorstDetOpt}\label{L-worst-det-opt}
Let
\[
    c_M=
    \frac{8}
    {1+(4+1/M)\displaystyle\sum_{r=2M+1}^{3M}\frac1r}.
\]
Then
\[
    \beta^*
    \le c_M\tau+2+\frac{c_M(M-1)}{4M+2},
    \qquad
    c_M\le\frac{8}{1+4\ln(3/2)}.
\]
\end{restatable}

\begin{proof}
The proof is deferred to~\Cref{S-worst-opt}.
\end{proof}

\begin{proof}[Proof of~\Cref{T-worst-det}]
The tables have \(\Ot(n^{1/k})\) words at every vertex by \Cref{S-worst-prepr}, and the labels and headers are as in \Cref{S-avg}.
Let \(c=8/(1+4\ln(3/2))\).
By \Cref{L-worst-det-opt} and \(\tau\le k-1\), the stretch is at most
\[
    1+\beta^*
    \le
    c_M\left(\tau+\frac{M-1}{4M+2}\right)+3
    \le c(k-3/4)+3
    <ck+1,
\]
using \(c_M\le c\) and \(\frac{M-1}{4M+2}<\frac14\) for the second step, and \(c>8/3\) for the last step.
\end{proof}

\subsection{Randomized routing}\label{S-worst-rand}
\thmWCrandom*

We use the same visits as in the deterministic scheme.
The source chooses either the fallback route or a pair
\((i,t)\), where \(1\le i<\tau\) and
\(M<t\le\delta_i\) is an integer.
The pair \((i,t)\) specifies a visit to \(p_i(u)\)
with search radius \(h_i(u)+t\).
As proved above, its extra cost is at most
\[
    4h_i(u)+2t+2i-2
    +2H\,\one[d\ge t],
\]
where \(d=d(u,v)\).
The fallback route has extra cost at most \(2H\).

Let
\[
    \gamma(d)=\frac{8d+4}{3d+1}\,d.
\]
We bound the term \(2t-2\le6M-2\) separately and measure \(d\) by \(\gamma(d)\approx\frac83d\). Since \(\gamma(d)=\frac83d+\frac{4d}{3(3d+1)}\le\frac83d+\frac49\), this gives the bound and the additive \(\frac49k\) of \Cref{L-worst-rand-opt}.
The source minimizes \(z\ge0\) over distributions
\(\pi_0,\pi_{i,t}\) on the available choices, subject to
\begin{equation}\label{e-worst-rand-LP}
\begin{aligned}
    2H\pi_0
    +\sum_{i,t}\pi_{i,t}
       \bigl(
          4h_i(u)+2t+2i-2
          +2H\,\one[d\ge t]
       \bigr)
    \le z \gamma(d)+6M-2
\end{aligned}
\end{equation}
for every integer \(M\le d\le3M\).
Here \(\pi_0\) is the probability of taking the fallback route.
Since every available guess satisfies \(t\le3M\) (indeed, \(t\le\delta_i=h_{i+1}(v)-h_i(u)\le(h_{i+1}(u)+M)-h_i(u)\le(h_i(v)+2M)-h_i(u)\le3M\)),
the left-hand side is constant for \(d\ge3M\).
As \(\gamma\) is increasing, the constraints therefore hold
for every integer \(d\ge M\).

\begin{restatable}{lemma}{lemWorstRandOpt}\label{L-worst-rand-opt}
The optimal value \(z^*\) of~\eqref{e-worst-rand-LP} satisfies
\[
    z^*\le\tau+\frac12.
\]
Consequently, for \(d=d(u,v)\),
\[
    \mathbb{E}[\hat d(u,v)]
    \le
    \left(\frac{8d+4}{3d+1}\,k+6\right)d
    \le
    \left(\frac83 k+6\right)d+\frac49 k.
\]
The expectation is over the source's choice,
with the preprocessing fixed.
\end{restatable}

\begin{proof}
The proof is deferred to~\Cref{S-worst-opt}.
\end{proof}
\begin{proof}[Proof of~\Cref{T-worst-random}]
The storage, label, and header bounds are as in the proof of~\Cref{T-worst-det}.
By the second inequality of \Cref{L-worst-rand-opt}, \(\EE[\hat d(u,v)]\le(\frac83k+6)d(u,v)+\frac49k\).
\end{proof}

\section{Name-independent routing with average storage}\label{S-Name-Independent}
In this section, we present our name-independent schemes with average storage: a randomized scheme with expected stretch $4k-5$, and a deterministic scheme with stretch $\approx5.77k$.
Both improve the $96k$ stretch of Abraham, Gavoille, and Malkhi~\cite{AbrahamGM04} by a factor of more than $16$.

\subsection{Expected \texorpdfstring{$4k-5$}{4k-5} stretch}\label{S-NI-avg}
Fix a source $u$ and a destination $v\ne u$ for the rest of this section.
In a name-independent scheme, the source $u$ receives only the name $\id(v)$ of $v$.
Thus it knows neither the pivots of $v$ nor their heights, which the labeled schemes of~\Cref{S-avg} read from $\L(v)$.
In particular, it can compute neither the fallback level $\tau$ of~\Cref{S-avg-framework} nor the lower bound $M$ of~\Cref{S-avg-selection}.
Instead, the source visits several of its own pivots, in increasing order of level, until one of them has $v$ in its cluster.
It chooses these levels by guessing $d(u,v)$ at randomly shifted geometric scales.

\thmNIavg*

\paragraph{Preprocessing.}
We use the routing tables of~\Cref{S-avg-prepr}, and no labels.
In the first part of $\RT(u)$, we also store $\L(w,T(C(w)))$ for every $w\in B(u)$. This does not change the storage bound.
The second part of $\RT(w)$ stores the whole cluster of $w$, keyed by identifiers.
Hence, given only $\id(v)$, a pivot $p_i(u)$ can test whether $v\in C(p_i(u))$, and if so, find $\L(v,T(C(p_i(u))))$.
This is where we use average storage. With worst-case storage, a cluster cannot be stored at its root, and the same test needs a name-independent search in the cluster tree (\Cref{D-S-NI-Tree}).

\paragraph{The naive case.}
If $v\in B(u)$, then $u\in C(v)$, and $u$ routes to $v$ in $T(C(v))$, using $\L(v,T(C(v)))$ from the first part of $\RT(u)$.
If $v\in C(u)$, then $u$ routes to $v$ in $T(C(u))$, as in~\Cref{S-avg-framework}.
In both cases, the message follows a shortest path.
For the rest of the section, we assume that $v\notin B(u)\cup C(u)$.
Since $v\notin B_0(u)$, we have
\begin{equation}\label{e-NI-h1}
h_1(u)\le d(u,v).
\end{equation}

\paragraph{Visiting a pivot.}
In a \emph{visit} at level $i\in[k]$, as in~\Cref{A-avg-route}, the source routes to $p_i(u)$ in $T(C(p_i(u)))$, and $p_i(u)$ checks whether $v\in C(p_i(u))$.
If so, the visit succeeds, and $p_i(u)$ routes to $v$ in $T(C(p_i(u)))$.
Otherwise, the visit fails, and the message returns to $u$ in $T(C(p_i(u)))$.
Since $T(C(p_i(u)))$ is a shortest-path tree rooted at $p_i(u)$, a failed visit costs $2h_i(u)$, and a successful visit costs $h_i(u)+d(p_i(u),v)\le2h_i(u)+d(u,v)$.

In the labeled schemes, a visit at level $i$ fails only if $d(u,v)$ is at least the threshold $h_{i+1}(v)-h_i(u)$ (\Cref{L-Pivot-Cluster}).
Here, the source does not know $h_{i+1}(v)$. Instead, it uses the \emph{gap} $g_i=h_{i+1}(u)-h_i(u)$ of level $i$, which it knows.
\begin{lemma}\label{L-NI-success}
If $g_i>2d(u,v)$, then $v\in C(p_i(u))$.
\end{lemma}
\begin{proof}
By the triangle inequality, $h_{i+1}(v)\ge h_{i+1}(u)-d(u,v)>h_i(u)+d(u,v)$, and~\Cref{L-Pivot-Cluster} gives $v\in C(p_i(u))$.
\end{proof}

\paragraph{If the distance were known.}
For $D>0$, let
\[
i(D)=\min\{i\in[k]\mid g_i>2D\}
\]
be the first level whose gap exceeds $2D$.
It is nondecreasing in $D$, and by its minimality, every gap below $i(D)$ is at most $2D$. Hence
\begin{equation}\label{e-NI-gaps}
h_{i(D)}(u)=\sum_{r<i(D)}g_r\le\sum_{r<k-1}g_r\cdot\one[g_r\le2D].
\end{equation}
A source that knows $d(u,v)$ could visit level $i(d(u,v))$.
By~\Cref{L-NI-success}, this visit succeeds.
By~\eqref{e-NI-h1}, the first gap is $g_0=h_1(u)\le d(u,v)$, and every other gap below $i(d(u,v))$ is at most $2d(u,v)$. So $h_{i(d(u,v))}(u)\le(2k-3)d(u,v)$, and the visit costs at most $2h_{i(d(u,v))}(u)+d(u,v)\le(4k-5)d(u,v)$.
Here a gap below the chosen level can be up to $2d(u,v)$, since the proof of \Cref{L-NI-success} uses only $h_{i+1}(v)\ge h_{i+1}(u)-d(u,v)$, which loses $d(u,v)$.
No choice of levels, even with the knowledge of $d(u,v)$, gets a better bound from~\Cref{L-NI-success}.
Indeed, if $g_0=d(u,v)$ and all the other finite gaps equal $2d(u,v)$, the lemma guarantees success only at level $k-1$, whose visit can cost $2h_{k-1}(u)+d(u,v)=(4k-5)d(u,v)$.
This matches the stretch $4k-5$ of the labeled scheme of Thorup and Zwick~\cite{DBLP:conf/spaa/ThorupZ01}.

\begin{algorithm2e}[t]
\caption{Name-independent $\Route(u,\id(v))$}\label{A-NI-route}
\SetKw{KwRoute}{Route}
\If{$v\in B(u)$}{\KwRoute $u\leadsto v$ in $T(C(v))$\;\KwRet\;}
\If{$v\in C(u)$}{\KwRoute $u\leadsto v$ in $T(C(u))$\;\KwRet\;}
Sample $X$ uniformly from $[0,1)$\;
Let $i_1<\cdots<i_m$ be the distinct values of $i(e^{j+X})$, for $j\in\mathbb Z$ with $e^{j+X}\ge h_1(u)$\;
\For{$s=1,\ldots,m$}{
$i\gets i_s$\;
\KwRoute $u\leadsto p_i(u)$ in $T(C(p_i(u)))$\;
\If(\tcp*[h]{tested at $p_i(u)$}){$v\in C(p_i(u))$}{\KwRoute $p_i(u)\leadsto v$ in $T(C(p_i(u)))$\;\KwRet\;}
\KwRoute $p_i(u)\leadsto u$ in $T(C(p_i(u)))$\;
}
\end{algorithm2e}

\paragraph{Random scales.}
The source does not know $d(u,v)$, so it guesses it at the geometric scales $D_j=e^{j+X}$, for $j\in\mathbb Z$, where $X$ is uniform in $[0,1)$.\footnote{The levels depend only on how $X$ compares with the fractional parts of $\ln g_0$ and of $\ln(g_r/2)$, for $1\le r<k-1$, which the source determines exactly by drawing the bits of $X$ one at a time.}
By~\eqref{e-NI-h1}, the source skips the scales below $h_1(u)$.
Let $i_1<\cdots<i_m$ be the distinct values of $i(D_j)$, for $j\in\mathbb Z$ with $D_j\ge h_1(u)$.
The source visits these levels in this order until a visit succeeds (\Cref{A-NI-route}).
As in the scheme of Kadria and Roditty~\cite{kadria2025compact}, the source visits a sequence of its pivots, and every failure raises a lower bound on $d(u,v)$.
Indeed, by~\Cref{L-NI-success}, a failed visit at level $i(D_j)$ shows that $d(u,v)>D_j$.

The shift $X$ is the only random choice, and it plays the role of the random index in~\Cref{S-avg-rand}.
The graph, and hence $d(u,v)$, is fixed before the source samples $X$.
With fixed scales, $d(u,v)$ could lie just above a scale. Then the visit at this scale may fail, and the next scale is almost $e$ times larger than $d(u,v)$.
With the random shift, the fractional part of $\ln d(u,v)-X$ is uniform in $[0,1)$, and we bound the cost averaged over it.

\paragraph{Stretch analysis.}
\begin{lemma}\label{L-NI-walk}
$\displaystyle\hat d(u,v)\le d(u,v)+2\sum_{j\in\mathbb Z:\ h_1(u)\le D_j<e\cdot d(u,v)}h_{i(D_j)}(u)$.
\end{lemma}
\begin{proof}
Let $j^*$ be the smallest $j$ with $D_j\ge d(u,v)$, so $D_j<e\cdot d(u,v)$ exactly for $j\le j^*$. By~\eqref{e-NI-h1}, $D_{j^*}\ge h_1(u)$.
The gap of level $i(D_{j^*})$ exceeds $2D_{j^*}\ge2d(u,v)$, so the visit at this level succeeds by~\Cref{L-NI-success}.
Hence every visited level is at most $i(D_{j^*})$. Since $i(D)$ is nondecreasing, every visited level is $i(D_j)$ for some $j\le j^*$ with $D_j\ge h_1(u)$.
The bound follows from the costs of the visits: $2h_i(u)$ for every visited level $i$, plus $d(u,v)$ for the successful visit.
\end{proof}

\begin{lemma}\label{L-NI-expected}
$\EE[\hat d(u,v)]\le(4k-5)d(u,v)$.
\end{lemma}
\begin{proof}
Write $d=d(u,v)$.
For every nonnegative function $\varphi$ on $(0,\infty)$, since $X$ is uniform in $[0,1)$,
\[
\EE\Big[\sum_{j\in\mathbb Z}\varphi(e^{j+X})\Big]=\sum_{j\in\mathbb Z}\int_0^1\varphi(e^{j+x})\,dx=\int_{-\infty}^\infty\varphi(e^t)\,dt=\int_0^\infty\frac{\varphi(D)}D\,dD.
\]
By~\Cref{L-NI-walk} and this identity for $\varphi(D)=h_{i(D)}(u)\cdot\one[h_1(u)\le D<e\cdot d]$,
\[
\EE[\hat d(u,v)]\le d+2\int_{h_1(u)}^{e\cdot d}\frac{h_{i(D)}(u)}D\,dD.
\]
By~\eqref{e-NI-gaps}, the first gap $g_0=h_1(u)$ contributes
\[
\int_{g_0}^{e\cdot d}\frac{g_0}D\,dD=g_0\ln\frac{e\cdot d}{g_0}\le d,
\]
since $x\ln(e\cdot d/x)$ is increasing for $0<x\le d$, and $g_0\le d$ by~\eqref{e-NI-h1}.
For $r\ge1$, a gap $g_r\ge2e\cdot d$ contributes nothing to the integral, and a gap $g_r<2e\cdot d$ contributes at most
\[
\int_{g_r/2}^{e\cdot d}\frac{g_r}D\,dD=g_r\ln\frac{2e\cdot d}{g_r}\le2d,
\]
since $x\ln(c/x)\le c/e$ for every $x>0$.
There are $k-2$ such finite gaps, so the integral is at most $d+2(k-2)d=(2k-3)d$, and $\EE[\hat d(u,v)]\le d+2(2k-3)d=(4k-5)d$.
\end{proof}

\begin{proof}[Proof of~\Cref{T-NI-avg}]
The storage bound follows from~\Cref{L-avg-size}, since the root labels add $\Ot(1)$ words to every entry of the first part.
Every vertex on the walk computes the next port as in the general case of~\Cref{S-avg-framework}.
By~\Cref{L-NI-success}, the visit at level $i(D_j)$ succeeds once $D_j\ge d(u,v)$, so the message is always delivered.
Besides $\id(v)$, the header contains the levels $i_1,\ldots,i_m$, the index $s$ of the current visit, and, as in~\Cref{S-avg-framework}, the current tree identifier, target label, and return label, that is, $\Ot(1)$ words.
The expected stretch follows from~\Cref{L-NI-expected}.
\end{proof}

\subsection{Deterministic \texorpdfstring{$\approx5.77k$}{5.77k} stretch}\label{S-NI-det}
We now choose the sequence of visited levels deterministically.
\thmNIdet*
We use the preprocessing, the visits, and the notation of~\Cref{S-NI-avg}.

\paragraph{The cost of a sequence.}
Let $\sigma=(i_1<\cdots<i_m)$ be a sequence of levels with $i_1=0$ and $i_m=k-1$, which the source visits in this order until a visit succeeds.
For $d>0$, let
\[
s_\sigma(d)=\min\{s\mid g_{i_s}>2d\}
\]
be the position of the first level of $\sigma$ whose gap exceeds $2d$.
By~\Cref{L-NI-success}, the visit at level $i_{s_\sigma(d(u,v))}$ succeeds, so the costs of the visits give
\begin{equation}\label{e-NI-seq}
\hat d(u,v)\le d(u,v)+2\sum_{s\le s_\sigma(d(u,v))}h_{i_s}(u).
\end{equation}
As in~\Cref{S-avg-selection}, the source knows every term of this bound except $d(u,v)$.

\paragraph{The deterministic choice.}
Let
\[
\beta(\sigma)=\sup_{d>0}\frac2d\sum_{s\le s_\sigma(d)}h_{i_s}(u)
\]
be the largest extra stretch that~\eqref{e-NI-seq} allows for $\sigma$, and let $\beta^*=\min_\sigma\beta(\sigma)$.
The source chooses a sequence attaining $\beta^*$ and visits its levels as in~\Cref{A-NI-route}, without sampling $X$.
The sum in $\beta(\sigma)$ changes only when $2d$ crosses the gap of a level of $\sigma$, and between these values the bound decreases in $d$.
So, as with $\Theta$ in~\Cref{S-avg-selection}, the source computes $\beta(\sigma)$ from finitely many values of $d$.

\paragraph{Stretch analysis.}
By~\eqref{e-NI-seq}, the stretch is at most $1+\beta^*$, and the following bound suffices.
\begin{restatable}{lemma}{lemNIDetBound}\label{L-NI-det-bound}
$\beta^*\le\frac4{2^{1/(k-1)}-1}$.
\end{restatable}
Let $\lambda=1/(2^{1/(k-1)}-1)$.
If every prefix $(i_1,\ldots,i_t)$ of $\sigma$ with $t\ge2$ has $\sum_{s\le t}h_{i_s}(u)\le\lambda g_{i_{t-1}}$, then $\beta(\sigma)\le4\lambda$, since the sum in~\eqref{e-NI-seq} includes $h_{i_t}(u)$ only if $g_{i_{t-1}}\le2d(u,v)$.
For the geometric gaps $g_a=2^{a/(k-1)}$, every sequence has $\beta(\sigma)\ge4\lambda$, so the bound is tight.
\begin{proof}
The proof is deferred to~\Cref{S-NI-opt}.
\end{proof}

\begin{proof}[Proof of~\Cref{T-NI-det}]
By~\eqref{e-NI-seq} and~\Cref{L-NI-det-bound}, the stretch is at most $1+\beta^*\le1+4/(2^{1/(k-1)}-1)$, and $2^{1/(k-1)}-1\ge\ln2/(k-1)$.
The storage and header bounds follow as in the proof of~\Cref{T-NI-avg}.
\end{proof}

\section{Name-independent routing with worst-case storage}\label{S-NI-worst}

In this section, we present our name-independent schemes with
$\Ot(n^{1/k})$ local storage at every vertex in unweighted graphs, and prove \Cref{T-NI-worst-det,T-NI-worst-random}.
We combine the search structures of
\Cref{S-worst-prepr} with the sequence of visits
of \Cref{S-Name-Independent}.
The source receives only $\id(v)$, so it knows neither
$p_i(v)$ nor $h_i(v)$.
It therefore uses its guesses of $d(u,v)$ both to choose a pivot
and to choose the radius of the search.

\paragraph{Preprocessing.}
We use the preprocessing of \Cref{S-worst-prepr}, without labels.
Every record in a search structure of a cluster tree $T(C(w))$ contains $\id(v)$ and $\L(v,T(C(w)))$.

\paragraph{Routing.}
If $v\in C_0(u)$, route to $v$ along a shortest path.
Otherwise, the source starts with the lower bound
\[
    M=\max\left\{1,\left\lceil\frac{h_1(u)}2\right\rceil\right\}
    \le d(u,v).
\]
Indeed, $v\notin C_0(u)$ implies $d(u,v)\ge h_1(v)$,
and hence
\[
    h_1(u)\le d(u,v)+h_1(v)\le2d(u,v).
\]

The source makes a sequence of visits, using increasing
guesses $D\ge M$.
In a visit at level $i$, it first routes to $p_i(u)$
and requests a search with radius
\[
    R=h_i(u)+D.
\]
The pivot selects the smallest prefix tree $T_j$
containing every cluster vertex at distance at most $R$.
It checks the record locally if $j=0$, and otherwise
uses the search structure hosted on $T_{j-1}$.
By the minimality of $j$, the host tree has radius at most $R$.

If the search succeeds, the message uses the retrieved
label to reach $v$.
Otherwise, it returns to $u$, which increases its guess
and makes another visit.
A failed search concerns only the searched prefix:
it does not necessarily show that $v$ is outside the
whole cluster.

\paragraph{The gain from visiting level $i$.}
Write
\[
    g_i=h_{i+1}(u)-h_i(u),
    \qquad 0\le i<k,
\]
where $g_{k-1}=\infty$.
These gaps are known to the source.

Suppose that $d=d(u,v)\le D$ and $g_i>2D$.
Then
\[
    h_{i+1}(v)
    \ge h_{i+1}(u)-d
    >h_i(u)+d
    \ge d(p_i(u),v).
\]
Thus $v\in C(p_i(u))$.
Moreover,
$d(p_i(u),v)\le h_i(u)+D=R$,
so the searched prefix contains its record.
The visit therefore succeeds.
Hence, if $g_i>2D$, a failure proves $d>D$.

For each guess, we choose
\[
    i(D)=
    \min\{1\le i<k:g_i>2D\}.
\]
This level always exists because $g_{k-1}=\infty$.
Since $g_i>0$, we have $p_i(u)\in A_i\setminus A_{i+1}$, so its cluster has at most $S^{i+1}$ vertices. By \Cref{T-S-NI-Tree} with radius at most $R$ and $\lceil\log_SN\rceil\le i+1$, the walk to the record holder has length at most $R+2i$.

A failed visit costs at most
\[
    2h_i(u)+2R+2i
    =4h_i(u)+2D+2i.
\]
On success, the search and subsequent delivery cost at most
\[
    h_i(u)+2R+2i+d(p_i(u),v)
    \le d+4h_i(u)+2D+2i.
\]
We therefore write
\[
    c(D)=
    4h_{i(D)}(u)+2D+2i(D).
\]
A failed visit costs at most $c(D)$,
and a successful visit costs at most $d+c(D)$.
Among the levels with $g_i>2D$, the first one
minimizes this bound, since both $h_i(u)$ and $i$
are nondecreasing.

\subsection{Deterministic routing}
\thmNIworstDet*

As in the average-storage name-independent scheme,
the source chooses a sequence that minimizes its largest
extra-stretch bound.
Let
\[
    \sigma=(D_1<\cdots<D_m)
\]
be a sequence of integer guesses with
$D_1\ge M$ and $D_m=n-1$.
The source visits $i(D_1),\ldots,i(D_m)$
until a search succeeds.
For a possible distance $d$, let
\[
    s_\sigma(d)=\min\{s:D_s\ge d\}.
\]
The visit at position $s_\sigma(d)$ is guaranteed
to succeed. Hence
\[
    \hat d(u,v)
    \le d(u,v)+
       \sum_{r\le s_\sigma(d(u,v))}c(D_r).
\]

The deterministic optimization is therefore
\[
    \beta_{\mathrm{det}}^*
    =
    \min_\sigma
    \max_{d\in\{M,\ldots,n-1\}}
    \frac{1}{d}
       \sum_{r\le s_\sigma(d)}c(D_r).
\]
Every term is determined by the source's own heights.
A minimizing sequence gives stretch at most
$1+\beta_{\mathrm{det}}^*$.

\begin{restatable}{lemma}{lemNIWorstDet}\label{L-NI-worst-det}
The deterministic optimization satisfies
\[
    \beta_{\mathrm{det}}^*
    \le \frac{8(k-1)}{\ln2}+8.
\]
Consequently, the deterministic routing scheme has
stretch at most
\[
    \frac{8(k-1)}{\ln2}+9.
\]
\end{restatable}

\begin{proof}
The proof is deferred to~\Cref{S-NI-worst-opt}.
\end{proof}

\subsection{Randomized routing}
\thmNIworstRandom*

We use randomly shifted geometric guesses, as in the
average-storage name-independent scheme.
The source samples $X$ uniformly from $[0,1)$ and uses the guesses $D_j$ of \Cref{L-NI-worst-rand}.
At each guess, it visits $i(D_j)$ and performs
the corresponding radius-limited search.
It stops when a search succeeds.

\begin{restatable}{lemma}{lemNIWorstRand}\label{L-NI-worst-rand}
Using the guesses
\[
    D_j=\left(M+\frac14\right)e^{j+X}-\frac14,
    \qquad j=0,1,\ldots,
\]
where \(X\) is uniform on \([0,1)\), the randomized
routing scheme satisfies
\[
    \mathbb{E}[\hat d(u,v)]
    \le (8k+2e-7)d(u,v)+2k-2.
\]
The expectation is over the source's random shift,
with the preprocessing fixed.
\end{restatable}

\begin{proof}
The proof is deferred to~\Cref{S-NI-worst-opt}.
\end{proof}

\begin{proof}[Proof of~\Cref{T-NI-worst-det,T-NI-worst-random}]
The storage bound is as in \Cref{S-worst-prepr}.
Besides $\id(v)$, the header contains the current guess, the selected level and search structure, and the current tree identifier, target label, and return label, that is, $\Ot(1)$ words.
The visit with the first guess $D\ge d(u,v)$ succeeds, so the message is always delivered.
The stretch bounds are \Cref{L-NI-worst-det,L-NI-worst-rand}.
\end{proof}

\section{Discussion}\label{S-Discussion}
With average storage, randomized routing achieves the optimal expected stretch $2k-1$, while our deterministic stretch is $\approx2.47k$.
The coefficient $1/\ln(3/2)$ of our deterministic scheme cannot be improved using only the inequalities of our analysis (\Cref{S-avg-opt-tight}).
\begin{problem}
Is there a labeled routing scheme with deterministic routing, stretch $2k-1$, and $\Ot(n^{1/k})$ average local storage?
\end{problem}
With worst-case storage, our bound is further from the lower bound. Our scheme has $(\frac83k+6,\frac49k)\approx(2.67k+6,0.44k)$-expected stretch, and the lower bound is $2k-1$.
\begin{problem}
Is there a labeled routing scheme with expected stretch $2k-1$ and $\Ot(n^{1/k})$ worst-case local storage?
\end{problem}
For name-independent routing, our expected stretch $4k-5$ with average storage is the best bound that the distances of the source to its own pivots give, and the lower bound is $2k-1$~\cite{DBLP:conf/spaa/AbrahamGM06}. Closing this gap would require an analysis that uses more than the distances of the source to its pivots.

\paragraph{Use of AI.}
ChatGPT and Claude were used to assist with drafting and revising the entire paper, checking mathematical arguments, verifying references and novelty, and solving the optimization problems in the appendices. 
The authors reviewed the suggestions and take full responsibility for the final content, including all results, proofs, and citations.

\bibliographystyle{alpha}
\bibliography{bibliography}

@inproceedings{DBLP:conf/spaa/ThorupZ01,
  author       = {Mikkel Thorup and
                  Uri Zwick},
  editor       = {Arnold L. Rosenberg},
  title        = {Compact routing schemes},
  booktitle    = {Proceedings of the Thirteenth Annual {ACM} Symposium on Parallel Algorithms
                  and Architectures, {SPAA} 2001, Heraklion, Crete Island, Greece, July
                  4-6, 2001},
  pages        = {1--10},
  publisher    = {{ACM}},
  year         = {2001},
  url          = {https://doi.org/10.1145/378580.378581},
  doi          = {10.1145/378580.378581},
  bibsource    = {dblp computer science bibliography, https://dblp.org}
}

@article{DBLP:journals/jacm/ThorupZ05,
  author       = {Mikkel Thorup and
                  Uri Zwick},
  title        = {Approximate distance oracles},
  journal      = {J. {ACM}},
  volume       = {52},
  number       = {1},
  pages        = {1--24},
  year         = {2005},
  url          = {https://doi.org/10.1145/1044731.1044732},
  doi          = {10.1145/1044731.1044732},
  bibsource    = {dblp computer science bibliography, https://dblp.org}
}

@inproceedings{DBLP:conf/icalp/RodittyTZ05,
  author       = {Liam Roditty and
                  Mikkel Thorup and
                  Uri Zwick},
  editor       = {Lu{\'{\i}}s Caires and
                  Giuseppe F. Italiano and
                  Lu{\'{\i}}s Monteiro and
                  Catuscia Palamidessi and
                  Moti Yung},
  title        = {Deterministic Constructions of Approximate Distance Oracles and Spanners},
  booktitle    = {Automata, Languages and Programming, 32nd International Colloquium,
                  {ICALP} 2005, Lisbon, Portugal, July 11-15, 2005, Proceedings},
  series       = {Lecture Notes in Computer Science},
  volume       = {3580},
  pages        = {261--272},
  publisher    = {Springer},
  year         = {2005},
  url          = {https://doi.org/10.1007/11523468\_22},
  doi          = {10.1007/11523468\_22},
  bibsource    = {dblp computer science bibliography, https://dblp.org}
}

@inproceedings{DBLP:conf/podc/Chechik13,
  author       = {Shiri Chechik},
  editor       = {Panagiota Fatourou and
                  Gadi Taubenfeld},
  title        = {Compact routing schemes with improved stretch},
  booktitle    = {{ACM} Symposium on Principles of Distributed Computing, {PODC} '13,
                  Montreal, QC, Canada, July 22-24, 2013},
  pages        = {33--41},
  publisher    = {{ACM}},
  year         = {2013},
  url          = {https://doi.org/10.1145/2484239.2484268},
  doi          = {10.1145/2484239.2484268},
  bibsource    = {dblp computer science bibliography, https://dblp.org}
}

@inproceedings{DBLP:conf/podc/RodittyT15,
  author       = {Liam Roditty and
                  Roei Tov},
  editor       = {Chryssis Georgiou and
                  Paul G. Spirakis},
  title        = {New Routing Techniques and their Applications},
  booktitle    = {Proceedings of the 2015 {ACM} Symposium on Principles of Distributed
                  Computing, {PODC} 2015, Donostia-San Sebasti{\'{a}}n, Spain,
                  July 21 - 23, 2015},
  pages        = {23--32},
  publisher    = {{ACM}},
  year         = {2015},
  url          = {https://doi.org/10.1145/2767386.2767409},
  doi          = {10.1145/2767386.2767409},
  bibsource    = {dblp computer science bibliography, https://dblp.org}
}

@article{DBLP:journals/jal/Cowen01,
  author       = {Lenore Cowen},
  title        = {Compact Routing with Minimum Stretch},
  journal      = {J. Algorithms},
  volume       = {38},
  number       = {1},
  pages        = {170--183},
  year         = {2001},
  url          = {https://doi.org/10.1006/jagm.2000.1134},
  doi          = {10.1006/JAGM.2000.1134},
  bibsource    = {dblp computer science bibliography, https://dblp.org}
}

@article{DBLP:journals/jal/CowenW04,
  author       = {Lenore Cowen and
                  Christopher G. Wagner},
  title        = {Compact roundtrip routing in directed networks},
  journal      = {J. Algorithms},
  volume       = {50},
  number       = {1},
  pages        = {79--95},
  year         = {2004},
  url          = {https://doi.org/10.1016/j.jalgor.2003.08.001},
  doi          = {10.1016/J.JALGOR.2003.08.001},
  bibsource    = {dblp computer science bibliography, https://dblp.org}
}

@article{DBLP:journals/talg/RodittyTZ08,
  author       = {Liam Roditty and
                  Mikkel Thorup and
                  Uri Zwick},
  title        = {Roundtrip spanners and roundtrip routing in directed graphs},
  journal      = {{ACM} Trans. Algorithms},
  volume       = {4},
  number       = {3},
  pages        = {29:1--29:17},
  year         = {2008},
  url          = {https://doi.org/10.1145/1367064.1367069},
  doi          = {10.1145/1367064.1367069},
  bibsource    = {dblp computer science bibliography, https://dblp.org}
}

@article{DBLP:journals/jacm/PelegU89,
  author       = {David Peleg and
                  Eli Upfal},
  title        = {A trade-off between space and efficiency for routing tables},
  journal      = {J. {ACM}},
  volume       = {36},
  number       = {3},
  pages        = {510--530},
  year         = {1989},
  url          = {https://doi.org/10.1145/65950.65953},
  doi          = {10.1145/65950.65953},
  bibsource    = {dblp computer science bibliography, https://dblp.org}
}

@article{DBLP:journals/jal/AwerbuchBLP90,
  author       = {Baruch Awerbuch and
                  Amotz Bar{-}Noy and
                  Nathan Linial and
                  David Peleg},
  title        = {Improved Routing Strategies with Succinct Tables},
  journal      = {J. Algorithms},
  volume       = {11},
  number       = {3},
  pages        = {307--341},
  year         = {1990},
  url          = {https://doi.org/10.1016/0196-6774(90)90017-9},
  doi          = {10.1016/0196-6774(90)90017-9},
  bibsource    = {dblp computer science bibliography, https://dblp.org}
}

@inproceedings{DBLP:conf/stacs/FraigniaudG02,
  author       = {Pierre Fraigniaud and
                  Cyril Gavoille},
  editor       = {Helmut Alt and
                  Afonso Ferreira},
  title        = {A Space Lower Bound for Routing in Trees},
  booktitle    = {{STACS} 2002, 19th Annual Symposium on Theoretical Aspects of Computer
                  Science, Antibes - Juan les Pins, France, March 14-16, 2002, Proceedings},
  series       = {Lecture Notes in Computer Science},
  volume       = {2285},
  pages        = {65--75},
  publisher    = {Springer},
  year         = {2002},
  url          = {https://doi.org/10.1007/3-540-45841-7\_4},
  doi          = {10.1007/3-540-45841-7\_4},
  bibsource    = {dblp computer science bibliography, https://dblp.org}
}

@article{DBLP:journals/siamdm/AwerbuchP92,
  author       = {Baruch Awerbuch and
                  David Peleg},
  title        = {Routing with Polynomial Communication-Space Trade-Off},
  journal      = {{SIAM} J. Discret. Math.},
  volume       = {5},
  number       = {2},
  pages        = {151--162},
  year         = {1992},
  url          = {https://doi.org/10.1137/0405013},
  doi          = {10.1137/0405013},
  bibsource    = {dblp computer science bibliography, https://dblp.org}
}

@article{DBLP:journals/jal/EilamGP03,
  author       = {Tamar Eilam and
                  Cyril Gavoille and
                  David Peleg},
  title        = {Compact routing schemes with low stretch factor},
  journal      = {J. Algorithms},
  volume       = {46},
  number       = {2},
  pages        = {97--114},
  year         = {2003},
  url          = {https://doi.org/10.1016/S0196-6774(03)00002-6},
  doi          = {10.1016/S0196-6774(03)00002-6},
  bibsource    = {dblp computer science bibliography, https://dblp.org}
}

@article{DBLP:journals/ipl/Laing07,
  author       = {Kofi A. Laing},
  title        = {Name-independent compact routing in trees},
  journal      = {Inf. Process. Lett.},
  volume       = {103},
  number       = {2},
  pages        = {57--60},
  year         = {2007},
  url          = {https://doi.org/10.1016/j.ipl.2007.02.015},
  doi          = {10.1016/J.IPL.2007.02.015},
  bibsource    = {dblp computer science bibliography, https://dblp.org}
}

@article{DBLP:journals/talg/AbrahamGMNT08,
  author       = {Ittai Abraham and
                  Cyril Gavoille and
                  Dahlia Malkhi and
                  Noam Nisan and
                  Mikkel Thorup},
  title        = {Compact name-independent routing with minimum stretch},
  journal      = {{ACM} Trans. Algorithms},
  volume       = {4},
  number       = {3},
  pages        = {37:1--37:12},
  year         = {2008},
  url          = {https://doi.org/10.1145/1367064.1367077},
  doi          = {10.1145/1367064.1367077},
  bibsource    = {dblp computer science bibliography, https://dblp.org}
}

@inproceedings{kadria2025compact,
  author    = {Avi Kadria and Liam Roditty},
  title     = {Compact Routing Schemes in Undirected and Directed Graphs},
  booktitle = {39th International Symposium on Distributed Computing, {DISC} 2025},
  series    = {LIPIcs},
  volume    = {356},
  pages     = {38:1--38:19},
  publisher = {Schloss Dagstuhl -- Leibniz-Zentrum f{\"{u}}r Informatik},
  year      = {2025},
  doi       = {10.4230/LIPIcs.DISC.2025.38}
}

@inproceedings{DBLP:conf/icalp/FraigniaudG01,
  author       = {Pierre Fraigniaud and
                  Cyril Gavoille},
  editor       = {Fernando Orejas and
                  Paul G. Spirakis and
                  Jan van Leeuwen},
  title        = {Routing in Trees},
  booktitle    = {Automata, Languages and Programming, 28th International Colloquium,
                  {ICALP} 2001, Crete, Greece, July 8-12, 2001, Proceedings},
  series       = {Lecture Notes in Computer Science},
  volume       = {2076},
  pages        = {757--772},
  publisher    = {Springer},
  year         = {2001},
  url          = {https://doi.org/10.1007/3-540-48224-5\_62},
  doi          = {10.1007/3-540-48224-5\_62},
  bibsource    = {dblp computer science bibliography, https://dblp.org}
}

@inproceedings{DBLP:conf/podc/KonjevodRX06,
  author       = {Goran Konjevod and
                  Andr{\'{e}}a W. Richa and
                  Donglin Xia},
  editor       = {Eric Ruppert and
                  Dahlia Malkhi},
  title        = {Optimal-stretch name-independent compact routing in doubling metrics},
  booktitle    = {Proceedings of the Twenty-Fifth Annual {ACM} Symposium on Principles
                  of Distributed Computing, {PODC} 2006, Denver, CO, USA, July 23-26,
                  2006},
  pages        = {198--207},
  publisher    = {{ACM}},
  year         = {2006},
  url          = {https://doi.org/10.1145/1146381.1146412},
  doi          = {10.1145/1146381.1146412},
  bibsource    = {dblp computer science bibliography, https://dblp.org}
}

@inproceedings{DBLP:conf/wdag/GavoilleGHI13,
  author       = {Cyril Gavoille and
                  Christian Glacet and
                  Nicolas Hanusse and
                  David Ilcinkas},
  editor       = {Yehuda Afek},
  title        = {On the Communication Complexity of Distributed Name-Independent Routing
                  Schemes},
  booktitle    = {Distributed Computing - 27th International Symposium, {DISC} 2013,
                  Jerusalem, Israel, October 14-18, 2013. Proceedings},
  series       = {Lecture Notes in Computer Science},
  volume       = {8205},
  pages        = {418--432},
  publisher    = {Springer},
  year         = {2013},
  url          = {https://doi.org/10.1007/978-3-642-41527-2\_29},
  doi          = {10.1007/978-3-642-41527-2\_29},
  bibsource    = {dblp computer science bibliography, https://dblp.org}
}

@incollection{Erdos64,
  author    = {Paul Erd{\H{o}}s},
  title     = {Extremal problems in graph theory},
  booktitle = {Theory of Graphs and its Applications (Proc. Sympos. Smolenice, 1963)},
  pages     = {29--36},
  publisher = {Publ. House Czechoslovak Acad. Sci., Prague},
  year      = {1964}
}

@inproceedings{DBLP:conf/spaa/AbrahamGM06,
  author    = {Ittai Abraham and Cyril Gavoille and Dahlia Malkhi},
  title     = {On space-stretch trade-offs: lower bounds},
  booktitle = {Proceedings of the 18th Annual {ACM} Symposium on Parallelism in Algorithms and Architectures, {SPAA} 2006},
  pages     = {207--216},
  publisher = {{ACM}},
  year      = {2006},
  doi       = {10.1145/1148109.1148143}
}

@inproceedings{DBLP:conf/wdag/Zinovyev22,
  author    = {Anatoliy Zinovyev},
  title     = {Space-Stretch Tradeoff in Routing Revisited},
  booktitle = {36th International Symposium on Distributed Computing, {DISC} 2022},
  series    = {LIPIcs},
  volume    = {246},
  pages     = {37:1--37:16},
  publisher = {Schloss Dagstuhl -- Leibniz-Zentrum f{\"{u}}r Informatik},
  year      = {2022},
  doi       = {10.4230/LIPIcs.DISC.2022.37}
}

@article{KleinrockKamoun77,
  author  = {Leonard Kleinrock and Farouk Kamoun},
  title   = {Hierarchical Routing for Large Networks:
             Performance Evaluation and Optimization},
  journal = {Computer Networks},
  volume  = {1},
  number  = {3},
  pages   = {155--174},
  year    = {1977},
  doi     = {10.1016/0376-5075(77)90002-2}
}

@article{DBLP:journals/jpdc/GavoilleG01,
  author    = {Cyril Gavoille and Marc Gengler},
  title     = {Space-Efficiency for Routing Schemes of Stretch Factor Three},
  journal   = {J. Parallel Distributed Comput.},
  volume    = {61},
  number    = {5},
  pages     = {679--687},
  year      = {2001}
}

@inproceedings{DBLP:conf/stoc/AwerbuchBLP89,
  author       = {Baruch Awerbuch and
                  Amotz Bar{-}Noy and
                  Nathan Linial and
                  David Peleg},
  editor       = {David S. Johnson},
  title        = {Compact Distributed Data Structures for Adaptive Routing (Extended
                  Abstract)},
  booktitle    = {Proceedings of the 21st Annual {ACM} Symposium on Theory of Computing,
                  May 14-17, 1989, Seattle, Washington, {USA}},
  pages        = {479--489},
  publisher    = {{ACM}},
  year         = {1989},
  url          = {https://doi.org/10.1145/73007.73053},
  doi          = {10.1145/73007.73053},
  bibsource    = {dblp computer science bibliography, https://dblp.org}
}

@inproceedings{DBLP:conf/podc/IwamaO02,
  author       = {Kazuo Iwama and
                  Masaki Okita},
  editor       = {Aleta Ricciardi},
  title        = {Compact routing for average-case networks},
  booktitle    = {Proceedings of the Twenty-First Annual {ACM} Symposium on Principles
                  of Distributed Computing, {PODC} 2002, Monterey, California, USA,
                  July 21-24, 2002},
  pages        = {255},
  publisher    = {{ACM}},
  year         = {2002},
  url          = {https://doi.org/10.1145/571825.571868},
  doi          = {10.1145/571825.571868},
  bibsource    = {dblp computer science bibliography, https://dblp.org}
}

@inproceedings{DBLP:conf/wdag/IwamaO03,
  author       = {Kazuo Iwama and
                  Masaki Okita},
  editor       = {Faith Ellen Fich},
  title        = {Compact Routing for Flat Networks},
  booktitle    = {Distributed Computing, 17th International Conference, {DISC} 2003,
                  Sorrento, Italy, October 1-3, 2003, Proceedings},
  series       = {Lecture Notes in Computer Science},
  volume       = {2848},
  pages        = {196--210},
  publisher    = {Springer},
  year         = {2003},
  url          = {https://doi.org/10.1007/978-3-540-39989-6\_14},
  doi          = {10.1007/978-3-540-39989-6\_14},
  bibsource    = {dblp computer science bibliography, https://dblp.org}
}

@inproceedings{DBLP:conf/wdag/AbrahamGM05,
  author       = {Ittai Abraham and
                  Cyril Gavoille and
                  Dahlia Malkhi},
  editor       = {Pierre Fraigniaud},
  title        = {Compact Routing for Graphs Excluding a Fixed Minor},
  booktitle    = {Distributed Computing, 19th International Conference, {DISC} 2005,
                  Cracow, Poland, September 26-29, 2005, Proceedings},
  series       = {Lecture Notes in Computer Science},
  volume       = {3724},
  pages        = {442--456},
  publisher    = {Springer},
  year         = {2005},
  url          = {https://doi.org/10.1007/11561927\_32},
  doi          = {10.1007/11561927\_32},
  bibsource    = {dblp computer science bibliography, https://dblp.org}
}

@inproceedings{DBLP:conf/isaac/MulzerW20,
  author       = {Wolfgang Mulzer and
                  Max Willert},
  editor       = {Yixin Cao and
                  Siu{-}Wing Cheng and
                  Minming Li},
  title        = {Compact Routing in Unit Disk Graphs},
  booktitle    = {31st International Symposium on Algorithms and Computation, {ISAAC}
                  2020, December 14-18, 2020, Hong Kong, China (Virtual Conference)},
  series       = {LIPIcs},
  volume       = {181},
  pages        = {16:1--16:14},
  publisher    = {Schloss Dagstuhl - Leibniz-Zentrum f{\"{u}}r Informatik},
  year         = {2020},
  url          = {https://doi.org/10.4230/LIPIcs.ISAAC.2020.16},
  doi          = {10.4230/LIPICS.ISAAC.2020.16},
  bibsource    = {dblp computer science bibliography, https://dblp.org}
}

@article{DBLP:journals/algorithmica/NarayananO99,
  author       = {Lata Narayanan and
                  Jaroslav Opatrny},
  title        = {Compact Routing on Chordal Rings of Degree 4},
  journal      = {Algorithmica},
  volume       = {23},
  number       = {1},
  pages        = {72--96},
  year         = {1999},
  url          = {https://doi.org/10.1007/PL00009251},
  doi          = {10.1007/PL00009251},
  bibsource    = {dblp computer science bibliography, https://dblp.org}
}

@inproceedings{DBLP:conf/podc/AbrahamM04,
  author       = {Ittai Abraham and
                  Dahlia Malkhi},
  editor       = {Soma Chaudhuri and
                  Shay Kutten},
  title        = {Compact routing on euclidian metrics},
  booktitle    = {Proceedings of the Twenty-Third Annual {ACM} Symposium on Principles
                  of Distributed Computing, {PODC} 2004, St. John's, Newfoundland, Canada,
                  July 25-28, 2004},
  pages        = {141--149},
  publisher    = {{ACM}},
  year         = {2004},
  url          = {https://doi.org/10.1145/1011767.1011789},
  doi          = {10.1145/1011767.1011789},
  bibsource    = {dblp computer science bibliography, https://dblp.org}
}

@inproceedings{DBLP:conf/infocom/KrioukovFY04,
  author       = {Dmitri V. Krioukov and
                  Kevin R. Fall and
                  Xiaowei Yang},
  title        = {Compact Routing on Internet-like Graphs},
  booktitle    = {Proceedings {IEEE} {INFOCOM} 2004, The 23rd Annual Joint Conference
                  of the {IEEE} Computer and Communications Societies, Hong Kong, China,
                  March 7-11, 2004},
  publisher    = {{IEEE}},
  year         = {2004},
  url          = {https://doi.org/10.1109/INFCOM.2004.1354495},
  doi          = {10.1109/INFCOM.2004.1354495},
  bibsource    = {dblp computer science bibliography, https://dblp.org}
}

@inproceedings{DBLP:conf/dasc/TangYZ09,
  author       = {Mingdong Tang and
                  Jing Yang and
                  Guoqing Zhang},
  title        = {Compact Routing on Random Power Law Graphs},
  booktitle    = {Eighth {IEEE} International Conference on Dependable, Autonomic and
                  Secure Computing, {DASC} 2009, Chengdu, China, 12-14 December, 2009},
  pages        = {575--578},
  publisher    = {{IEEE} Computer Society},
  year         = {2009},
  url          = {https://doi.org/10.1109/DASC.2009.133},
  doi          = {10.1109/DASC.2009.133},
  bibsource    = {dblp computer science bibliography, https://dblp.org}
}

@inproceedings{DBLP:conf/icalp/GavoilleH99,
  author       = {Cyril Gavoille and
                  Nicolas Hanusse},
  editor       = {Jir{\'{\i}} Wiedermann and
                  Peter van Emde Boas and
                  Mogens Nielsen},
  title        = {Compact Routing Tables for Graphs of Bounded Genus},
  booktitle    = {Automata, Languages and Programming, 26th International Colloquium,
                  ICALP'99, Prague, Czech Republic, July 11-15, 1999, Proceedings},
  series       = {Lecture Notes in Computer Science},
  volume       = {1644},
  pages        = {351--360},
  publisher    = {Springer},
  year         = {1999},
  url          = {https://doi.org/10.1007/3-540-48523-6\_32},
  doi          = {10.1007/3-540-48523-6\_32},
  bibsource    = {dblp computer science bibliography, https://dblp.org}
}

@inproceedings{DBLP:conf/wdag/DourisboureG02,
  author       = {Yon Dourisboure and
                  Cyril Gavoille},
  editor       = {Dahlia Malkhi},
  title        = {Improved Compact Routing Scheme for Chordal Graphs},
  booktitle    = {Distributed Computing, 16th International Conference, {DISC} 2002,
                  Toulouse, France, October 28-30, 2002 Proceedings},
  series       = {Lecture Notes in Computer Science},
  volume       = {2508},
  pages        = {252--264},
  publisher    = {Springer},
  year         = {2002},
  url          = {https://doi.org/10.1007/3-540-36108-1\_17},
  doi          = {10.1007/3-540-36108-1\_17},
  bibsource    = {dblp computer science bibliography, https://dblp.org}
}

@article{DBLP:conf/cocoon/Lu02,
  author    = {Hsueh{-}I Lu},
  title     = {Improved Compact Routing Tables for Planar Networks via Orderly Spanning Trees},
  journal   = {{SIAM} J. Discret. Math.},
  volume    = {23},
  number    = {4},
  pages     = {2079--2092},
  year      = {2010},
  doi       = {10.1137/070703041}
}

@article{DBLP:conf/isaac/DraganL04,
  author    = {Feodor F. Dragan and Irina Lomonosov},
  title     = {On compact and efficient routing in certain graph classes},
  journal   = {Discret. Appl. Math.},
  volume    = {155},
  number    = {11},
  pages     = {1458--1470},
  year      = {2007},
  doi       = {10.1016/j.dam.2007.03.011}
}

@article{DBLP:journals/jpdc/CalamoneriI03,
  author       = {Tiziana Calamoneri and
                  Miriam Di Ianni},
  title        = {Interval routing {\&} layered cross product: compact routing schemes
                  for butterflies, meshes of trees, fat trees and Benes networks},
  journal      = {J. Parallel Distributed Comput.},
  volume       = {63},
  number       = {11},
  pages        = {1017--1025},
  year         = {2003},
  url          = {https://doi.org/10.1016/S0743-7315(03)00110-2},
  doi          = {10.1016/S0743-7315(03)00110-2},
  bibsource    = {dblp computer science bibliography, https://dblp.org}
}

@inproceedings{DBLP:conf/wdag/AbrahamG11,
  author       = {Ittai Abraham and
                  Cyril Gavoille},
  editor       = {David Peleg},
  title        = {On Approximate Distance Labels and Routing Schemes with Affine Stretch},
  booktitle    = {Distributed Computing - 25th International Symposium, {DISC} 2011,
                  Rome, Italy, September 20-22, 2011. Proceedings},
  series       = {Lecture Notes in Computer Science},
  volume       = {6950},
  pages        = {404--415},
  publisher    = {Springer},
  year         = {2011},
  url          = {https://doi.org/10.1007/978-3-642-24100-0\_39},
  doi          = {10.1007/978-3-642-24100-0\_39},
  bibsource    = {dblp computer science bibliography, https://dblp.org}
}

@inproceedings{DBLP:conf/infocom/AgarwalGH11,
  author       = {Rachit Agarwal and
                  Philip Brighten Godfrey and
                  Sariel Har{-}Peled},
  title        = {Approximate distance queries and compact routing in sparse graphs},
  booktitle    = {{INFOCOM} 2011. 30th {IEEE} International Conference on Computer Communications,
                  Joint Conference of the {IEEE} Computer and Communications Societies,
                  10-15 April 2011, Shanghai, China},
  pages        = {1754--1762},
  publisher    = {{IEEE}},
  year         = {2011},
  url          = {https://doi.org/10.1109/INFCOM.2011.5934973},
  doi          = {10.1109/INFCOM.2011.5934973},
  bibsource    = {dblp computer science bibliography, https://dblp.org}
}

@article{DBLP:journals/corr/abs-1201-2703,
  author       = {Rachit Agarwal and
                  Brighten Godfrey and
                  Sariel Har{-}Peled},
  title        = {Faster Approximate Distance Queries and Compact Routing in Sparse
                  Graphs},
  journal      = {CoRR},
  volume       = {abs/1201.2703},
  year         = {2012},
  url          = {http://arxiv.org/abs/1201.2703},
  eprinttype    = {arXiv},
  eprint       = {1201.2703},
  bibsource    = {dblp computer science bibliography, https://dblp.org}
}

@inproceedings{DBLP:conf/podc/GavoilleP96,
  author       = {Cyril Gavoille and
                  Stephane Perennes},
  editor       = {James E. Burns and
                  Yoram Moses},
  title        = {Memory Requirements for Routing in Distributed Networks (Extended
                  Abstract)},
  booktitle    = {Proceedings of the Fifteenth Annual {ACM} Symposium on Principles
                  of Distributed Computing, Philadelphia, Pennsylvania, USA, May 23-26,
                  1996},
  pages        = {125--133},
  publisher    = {{ACM}},
  year         = {1996},
  url          = {https://doi.org/10.1145/248052.248075},
  doi          = {10.1145/248052.248075},
  bibsource    = {dblp computer science bibliography, https://dblp.org}
}

@article{DBLP:journals/algorithmica/ChechikLPR12,
  author       = {Shiri Chechik and
                  Michael Langberg and
                  David Peleg and
                  Liam Roditty},
  title        = {f-Sensitivity Distance Oracles and Routing Schemes},
  journal      = {Algorithmica},
  volume       = {63},
  number       = {4},
  pages        = {861--882},
  year         = {2012},
  url          = {https://doi.org/10.1007/s00453-011-9543-0},
  doi          = {10.1007/S00453-011-9543-0},
  bibsource    = {dblp computer science bibliography, https://dblp.org}
}

@article{DBLP:journals/dc/HassinP01,
  author       = {Yehuda Hassin and
                  David Peleg},
  title        = {Sparse communication networks and efficient routing in the plane},
  journal      = {Distributed Comput.},
  volume       = {14},
  number       = {4},
  pages        = {205--215},
  year         = {2001},
  url          = {https://doi.org/10.1007/s004460100054},
  doi          = {10.1007/S004460100054},
  bibsource    = {dblp computer science bibliography, https://dblp.org}
}

@article{thorup2004compact,
  title={Compact oracles for reachability and approximate distances in planar digraphs},
  author={Thorup, Mikkel},
  journal={Journal of the ACM (JACM)},
  volume={51},
  number={6},
  pages={993--1024},
  year={2004},
  publisher={ACM New York, NY, USA}
}

@article{ABCP93,
  author    = {Baruch Awerbuch and Bonnie Berger and Lenore Cowen and David Peleg},
  title     = {Near-Linear Time Construction of Sparse Neighborhood Covers},
  journal   = {{SIAM} J. Comput.},
  volume    = {28},
  number    = {1},
  pages     = {263--277},
  year      = {1998},
  doi       = {10.1137/S0097539794271898}
}

@article{SarmaDP15,
  author       = {Das Sarma, Atish and Dinitz, Michael and Pandurangan, Gopal},
  title        = {Efficient Distributed Computation of Distance Sketches in Networks},
  journal      = {Distributed Computing},
  volume       = {28},
  number       = {5},
  pages        = {309--320},
  year         = {2015},
  doi          = {10.1007/s00446-015-0246-7},
  url          = {https://doi.org/10.1007/s00446-015-0246-7}
}

@inproceedings{CenDG20,
  author       = {Ruoxu Cen and
                  Ran Duan and
                  Yong Gu},
  editor       = {Artur Czumaj and
                  Anuj Dawar and
                  Emanuela Merelli},
  title        = {Roundtrip Spanners with (2k-1) Stretch},
  booktitle    = {47th International Colloquium on Automata, Languages, and Programming,
                  {ICALP} 2020, July 8-11, 2020, Saarbr{\"{u}}cken, Germany (Virtual
                  Conference)},
  series       = {LIPIcs},
  volume       = {168},
  pages        = {24:1--24:11},
  publisher    = {Schloss Dagstuhl - Leibniz-Zentrum f{\"{u}}r Informatik},
  year         = {2020},
  url          = {https://doi.org/10.4230/LIPIcs.ICALP.2020.24},
  doi          = {10.4230/LIPICS.ICALP.2020.24},
  bibsource    = {dblp computer science bibliography, https://dblp.org}
}

@inproceedings{DBLP:conf/soda/PachockiRSTW18,
  author       = {Jakub Pachocki and
                  Liam Roditty and
                  Aaron Sidford and
                  Roei Tov and
                  Virginia {Vassilevska Williams}},
  editor       = {Artur Czumaj},
  title        = {Approximating Cycles in Directed Graphs: Fast Algorithms for Girth
                  and Roundtrip Spanners},
  booktitle    = {Proceedings of the Twenty-Ninth Annual {ACM-SIAM} Symposium on Discrete
                  Algorithms, {SODA} 2018, New Orleans, LA, USA, January 7-10, 2018},
  pages        = {1374--1392},
  publisher    = {{SIAM}},
  year         = {2018},
  url          = {https://doi.org/10.1137/1.9781611975031.91},
  doi          = {10.1137/1.9781611975031.91},
  bibsource    = {dblp computer science bibliography, https://dblp.org}
}

@article{ZhuLam18,
  author = {Zhu, Chun Jiang and Lam, Kam-Yiu},
  title = {Deterministic Improved Round-Trip Spanners},
  journal = {Information Processing Letters},
  year = {2018},
  volume = {129},
  pages = {57--60},
  doi = {10.1016/j.ipl.2017.09.008}
}

@inproceedings{DBLP:conf/cocoon/StaffordZ23,
  author       = {Eli Stafford and
                  Chunjiang Zhu},
  editor       = {Weili Wu and
                  Guangmo Tong},
  title        = {Improved Sourcewise Roundtrip Spanners with Constant Stretch},
  booktitle    = {Computing and Combinatorics - 29th International Conference, {COCOON}
                  2023, Hawaii, HI, USA, December 15-17, 2023, Proceedings, Part {I}},
  series       = {Lecture Notes in Computer Science},
  volume       = {14422},
  pages        = {297--309},
  publisher    = {Springer},
  year         = {2023},
  url          = {https://doi.org/10.1007/978-3-031-49190-0\_21},
  doi          = {10.1007/978-3-031-49190-0\_21},
  bibsource    = {dblp computer science bibliography, https://dblp.org}
}

@inproceedings{DBLP:conf/soda/HarbuzovaJWX24,
  author       = {Alina Harbuzova and
                  Ce Jin and
                  Virginia Vassilevska Williams and
                  Zixuan Xu},
  editor       = {David P. Woodruff},
  title        = {Improved Roundtrip Spanners, Emulators, and Directed Girth Approximation},
  booktitle    = {Proceedings of the 2024 {ACM-SIAM} Symposium on Discrete Algorithms,
                  {SODA} 2024, Alexandria, VA, USA, January 7-10, 2024},
  pages        = {4641--4669},
  publisher    = {{SIAM}},
  year         = {2024},
  url          = {https://doi.org/10.1137/1.9781611977912.166},
  doi          = {10.1137/1.9781611977912.166},
  bibsource    = {dblp computer science bibliography, https://dblp.org}
}

@article{ZhuLam17,
  author = {Zhu, Chun Jiang and Lam, Kam-Yiu},
  title = {Source-Wise Round-Trip Spanners},
  journal = {Information Processing Letters},
  year = {2017},
  volume = {124},
  pages = {42--45},
  doi = {10.1016/j.ipl.2017.04.009}
}

@article{ZhuHanLam21,
  author = {Zhu, Chun Jiang and Han, Song and Lam, Kam-Yiu},
  title = {A Fast Algorithm for Source-Wise Round-Trip Spanners},
  journal = {Theoretical Computer Science},
  year = {2021},
  volume = {876},
  pages = {34--44},
  doi = {10.1016/j.tcs.2021.05.019}
}

@inproceedings{DBLP:conf/soda/Wulff-Nilsen12,
  author       = {Christian Wulff{-}Nilsen},
  editor       = {Yuval Rabani},
  title        = {Approximate distance oracles with improved preprocessing time},
  booktitle    = {Proceedings of the Twenty-Third Annual {ACM-SIAM} Symposium on Discrete
                  Algorithms, {SODA} 2012, Kyoto, Japan, January 17-19, 2012},
  pages        = {202--208},
  publisher    = {{SIAM}},
  year         = {2012},
  url          = {https://doi.org/10.1137/1.9781611973099.18},
  doi          = {10.1137/1.9781611973099.18},
  bibsource    = {dblp computer science bibliography, https://dblp.org}
}

@inproceedings{DBLP:conf/soda/Wulff-Nilsen13,
  author       = {Christian Wulff{-}Nilsen},
  editor       = {Sanjeev Khanna},
  title        = {Approximate Distance Oracles with Improved Query Time},
  booktitle    = {Proceedings of the Twenty-Fourth Annual {ACM-SIAM} Symposium on Discrete
                  Algorithms, {SODA} 2013, New Orleans, Louisiana, USA, January 6-8,
                  2013},
  pages        = {539--549},
  publisher    = {{SIAM}},
  year         = {2013},
  url          = {https://doi.org/10.1137/1.9781611973105.39},
  doi          = {10.1137/1.9781611973105.39},
  bibsource    = {dblp computer science bibliography, https://dblp.org}
}

@inproceedings{DBLP:conf/stoc/Chechik14,
  author       = {Shiri Chechik},
  editor       = {David B. Shmoys},
  title        = {Approximate distance oracles with constant query time},
  booktitle    = {Symposium on Theory of Computing, {STOC} 2014, New York, NY, USA,
                  May 31 - June 03, 2014},
  pages        = {654--663},
  publisher    = {{ACM}},
  year         = {2014},
  url          = {https://doi.org/10.1145/2591796.2591801},
  doi          = {10.1145/2591796.2591801},
  bibsource    = {dblp computer science bibliography, https://dblp.org}
}

@inproceedings{DBLP:conf/stoc/Chechik15,
  author       = {Shiri Chechik},
  editor       = {Rocco A. Servedio and
                  Ronitt Rubinfeld},
  title        = {Approximate Distance Oracles with Improved Bounds},
  booktitle    = {Proceedings of the Forty-Seventh Annual {ACM} on Symposium on Theory
                  of Computing, {STOC} 2015, Portland, OR, USA, June 14-17, 2015},
  pages        = {1--10},
  publisher    = {{ACM}},
  year         = {2015},
  url          = {https://doi.org/10.1145/2746539.2746562},
  doi          = {10.1145/2746539.2746562},
  bibsource    = {dblp computer science bibliography, https://dblp.org}
}

@article{DBLP:conf/focs/MendelN06,
  author    = {Manor Mendel and Assaf Naor},
  title     = {Ramsey partitions and proximity data structures},
  journal   = {J. Eur. Math. Soc.},
  volume    = {9},
  number    = {2},
  pages     = {253--275},
  year      = {2007},
  doi       = {10.4171/JEMS/79}
}

@article{PatrascuR14,
  author       = {Mihai P\v{a}tra\c{s}cu and
                  Liam Roditty},
  title        = {Distance Oracles beyond the Thorup-Zwick Bound},
  journal      = {{SIAM} J. Comput.},
  volume       = {43},
  number       = {1},
  pages        = {300--311},
  year         = {2014},
  url          = {https://doi.org/10.1137/11084128X},
  doi          = {10.1137/11084128X},
  bibsource    = {dblp computer science bibliography, https://dblp.org}
}

@inproceedings{PatrascuRT12,
  author       = {Mihai P\v{a}tra\c{s}cu and
                  Liam Roditty and
                  Mikkel Thorup},
  title        = {A New Infinity of Distance Oracles for Sparse Graphs},
  booktitle    = {53rd Annual {IEEE} Symposium on Foundations of Computer Science, {FOCS}
                  2012, New Brunswick, NJ, USA, October 20-23, 2012},
  pages        = {738--747},
  publisher    = {{IEEE} Computer Society},
  year         = {2012},
  url          = {https://doi.org/10.1109/FOCS.2012.44},
  doi          = {10.1109/FOCS.2012.44},
  bibsource    = {dblp computer science bibliography, https://dblp.org}
}

@inproceedings{DBLP:conf/focs/SommerVY09,
  author       = {Christian Sommer and
                  Elad Verbin and
                  Wei Yu},
  title        = {Distance Oracles for Sparse Graphs},
  booktitle    = {50th Annual {IEEE} Symposium on Foundations of Computer Science, {FOCS}
                  2009, October 25-27, 2009, Atlanta, Georgia, {USA}},
  pages        = {703--712},
  publisher    = {{IEEE} Computer Society},
  year         = {2009},
  url          = {https://doi.org/10.1109/FOCS.2009.27},
  doi          = {10.1109/FOCS.2009.27},
  bibsource    = {dblp computer science bibliography, https://dblp.org}
}

@article{RodittyT23,
  author  = {Liam Roditty and Roei Tov},
  title   = {Approximate distance oracles with improved stretch for sparse graphs},
  journal = {Theoretical Computer Science},
  volume  = {943},
  pages   = {89--101},
  year    = {2023},
  doi     = {10.1016/j.tcs.2022.11.016}
}

@inproceedings{KadriaR25new,
  author    = {Avi Kadria and Liam Roditty},
  title     = {New Approximate Distance Oracles and Their Applications},
  booktitle = {Proceedings of the 36th International Symposium on Algorithms and Computation ({ISAAC})},
  series    = {LIPIcs},
  volume    = {359},
  pages     = {43:1--43:17},
  year      = {2025},
  doi       = {10.4230/LIPIcs.ISAAC.2025.43}
}

@inproceedings{KadriaR25faster,
  author       = {Avi Kadria and
                  Liam Roditty},
  title        = {Faster Algorithms for $(2k-1)$-Stretch Distance Oracles},
  booktitle    = {53rd International Colloquium on Automata, Languages, and Programming,
                  {ICALP} 2026},
  series       = {LIPIcs},
  volume       = {374},
  pages        = {121:1--121:22},
  publisher    = {Schloss Dagstuhl - Leibniz-Zentrum f{\"{u}}r Informatik},
  year         = {2026},
  doi          = {10.4230/LIPIcs.ICALP.2026.121}
}

@misc{KadriaR26pivot,
  author        = {Avi Kadria and Liam Roditty},
  title         = {Tight Space Bounds for Pivot-Based Distance Oracles in Sparse Graphs},
  howpublished  = {arXiv preprint arXiv:2610.29624},
  year          = {2026},
  eprint        = {2610.29624},
  archivePrefix = {arXiv}
}

@incollection{DBLP:reference/algo/Roditty16a,
  author       = {Liam Roditty},
  title        = {Distance Oracles for Sparse Graphs},
  booktitle    = {Encyclopedia of Algorithms},
  pages        = {563--564},
  publisher    = {Springer},
  year         = {2016},
  url          = {https://doi.org/10.1007/978-1-4939-2864-4\_571},
  doi          = {10.1007/978-1-4939-2864-4\_571},
  bibsource    = {dblp computer science bibliography, https://dblp.org}
}

@article{Sommer14,
  author       = {Christian Sommer},
  title        = {Shortest-Path Queries in Static Networks},
  journal      = {{ACM} Computing Surveys},
  volume       = {46},
  number       = {4},
  pages        = {45:1--45:31},
  year         = {2014},
  doi          = {10.1145/2530531}
}

@article{ElkinP16,
  author       = {Michael Elkin and
                  Seth Pettie},
  title        = {A Linear-Size Logarithmic Stretch Path-Reporting Distance Oracle for
                  General Graphs},
  journal      = {{ACM} Trans. Algorithms},
  volume       = {12},
  number       = {4},
  pages        = {50:1--50:31},
  year         = {2016},
  url          = {https://doi.org/10.1145/2888397},
  doi          = {10.1145/2888397},
  bibsource    = {dblp computer science bibliography, https://dblp.org}
}

@article{ElkinNW16,
  author       = {Michael Elkin and
                  Ofer Neiman and
                  Christian Wulff{-}Nilsen},
  title        = {Space-efficient path-reporting approximate distance oracles},
  journal      = {Theor. Comput. Sci.},
  volume       = {651},
  pages        = {1--10},
  year         = {2016},
  url          = {https://doi.org/10.1016/j.tcs.2016.07.038},
  doi          = {10.1016/J.TCS.2016.07.038},
  bibsource    = {dblp computer science bibliography, https://dblp.org}
}

@article{DBLP:conf/focs/ElkinS23,
  author    = {Michael Elkin and Idan Shabat},
  title     = {Path-Reporting Distance Oracles with Logarithmic Stretch and Size {$O(n \log\log n)$}},
  journal   = {{SIAM} J. Comput.},
  volume    = {55},
  number    = {6},
  pages     = {1229--1292},
  year      = {2026},
  doi       = {10.1137/24M1708954}
}

@inproceedings{DBLP:conf/icalp/Chechik024a,
  author       = {Shiri Chechik and
                  Tianyi Zhang},
  editor       = {Karl Bringmann and
                  Martin Grohe and
                  Gabriele Puppis and
                  Ola Svensson},
  title        = {Path-Reporting Distance Oracles with Logarithmic Stretch and Linear
                  Size},
  booktitle    = {51st International Colloquium on Automata, Languages, and Programming,
                  {ICALP} 2024, July 8-12, 2024, Tallinn, Estonia},
  series       = {LIPIcs},
  volume       = {297},
  pages        = {42:1--42:18},
  publisher    = {Schloss Dagstuhl - Leibniz-Zentrum f{\"{u}}r Informatik},
  year         = {2024},
  url          = {https://doi.org/10.4230/LIPIcs.ICALP.2024.42},
  doi          = {10.4230/LIPICS.ICALP.2024.42},
  bibsource    = {dblp computer science bibliography, https://dblp.org}
}

@inproceedings{NeimanShabat24,
  author = {Neiman, Ofer and Shabat, Idan},
  title = {Path-Reporting Distance Oracles with Linear Size},
  booktitle = {19th Scandinavian Symposium and Workshops on Algorithm Theory (SWAT 2024)},
  year = {2024},
  series = {Leibniz International Proceedings in Informatics},
  volume = {294},
  pages = {36:1--36:16},
  publisher = {Schloss Dagstuhl -- Leibniz-Zentrum f{\"u}r Informatik},
  doi = {10.4230/LIPIcs.SWAT.2024.36}
}

@article{DBLP:journals/jal/GavoillePPR04,
  author  = {Cyril Gavoille and David Peleg and St{\'{e}}phane P{\'{e}}rennes and Ran Raz},
  title   = {Distance Labeling in Graphs},
  journal = {Journal of Algorithms},
  volume  = {53},
  number  = {1},
  pages   = {85--112},
  year    = {2004},
  doi     = {10.1016/j.jalgor.2004.05.002}
}

@inproceedings{GavoilleKKPP01,
  author    = {Cyril Gavoille and Michal Katz and Nir A. Katz and Christophe Paul and David Peleg},
  title     = {Approximate Distance Labeling Schemes},
  booktitle = {Proceedings of the 9th Annual European Symposium on Algorithms ({ESA})},
  series    = {Lecture Notes in Computer Science},
  volume    = {2161},
  pages     = {476--487},
  year      = {2001},
  doi       = {10.1007/3-540-44676-1_40}
}

@article{Althofer93,
  author = {Alth{\"o}fer, Ingo and Das, Gautam and Dobkin, David and Joseph, Deborah and Soares, Jos{\'e}},
  title = {On Sparse Spanners of Weighted Graphs},
  journal = {Discrete \& Computational Geometry},
  year = {1993},
  volume = {9},
  pages = {81--100},
  doi = {10.1007/BF02189308}
}

@article{DBLP:journals/rsa/BaswanaS07,
  author       = {Surender Baswana and
                  Sandeep Sen},
  title        = {A simple and linear time randomized algorithm for computing sparse
                  spanners in weighted graphs},
  journal      = {Random Struct. Algorithms},
  volume       = {30},
  number       = {4},
  pages        = {532--563},
  year         = {2007},
  url          = {https://doi.org/10.1002/rsa.20130},
  doi          = {10.1002/RSA.20130},
  bibsource    = {dblp computer science bibliography, https://dblp.org}
}

@incollection{DBLP:reference/algo/Elkin08a,
  author       = {Michael Elkin},
  editor       = {Ming{-}Yang Kao},
  title        = {Sparse Graph Spanners},
  booktitle    = {Encyclopedia of Algorithms - 2008 Edition},
  publisher    = {Springer},
  year         = {2008},
  url          = {https://doi.org/10.1007/978-0-387-30162-4\_387},
  doi          = {10.1007/978-0-387-30162-4\_387},
  bibsource    = {dblp computer science bibliography, https://dblp.org}
}

@article{ElkinFN18,
  author       = {Michael Elkin and Arnold Filtser and Ofer Neiman},
  title        = {Prioritized Metric Structures and Embedding},
  journal      = {{SIAM} Journal on Computing},
  volume       = {47},
  number       = {3},
  pages        = {829--858},
  year         = {2018},
  doi          = {10.1137/17M1118749},
  url          = {https://doi.org/10.1137/17M1118749}
}

@article{AriasCLRT06,
  author    = {Marta Arias and Lenore J. Cowen and Kofi A. Laing and Rajmohan Rajaraman and Orjeta Taka},
  title     = {Compact Routing with Name Independence},
  journal   = {{SIAM} J. Discret. Math.},
  volume    = {20},
  number    = {3},
  pages     = {705--726},
  year      = {2006},
  doi       = {10.1137/04062053}
}

@inproceedings{AbrahamGM06upper,
  author    = {Ittai Abraham and Cyril Gavoille and Dahlia Malkhi},
  title     = {On space-stretch trade-offs: upper bounds},
  booktitle = {Proceedings of the 18th Annual {ACM} Symposium on Parallelism in Algorithms and Architectures, {SPAA} 2006},
  pages     = {217--224},
  publisher = {{ACM}},
  year      = {2006},
  doi       = {10.1145/1148109.1148144}
}

@inproceedings{AbrahamGM04,
  author    = {Ittai Abraham and Cyril Gavoille and Dahlia Malkhi},
  title     = {Routing with Improved Communication-Space Trade-Off},
  booktitle = {Distributed Computing, 18th International Conference, {DISC} 2004},
  series    = {Lecture Notes in Computer Science},
  volume    = {3274},
  pages     = {305--319},
  publisher = {Springer},
  year      = {2004},
  doi       = {10.1007/978-3-540-30186-8_22}
}

@inproceedings{AbrahamGGM06,
  author    = {Ittai Abraham and Cyril Gavoille and Andrew V. Goldberg and Dahlia Malkhi},
  title     = {Routing in Networks with Low Doubling Dimension},
  booktitle = {26th {IEEE} International Conference on Distributed Computing Systems, {ICDCS} 2006},
  pages     = {75},
  publisher = {{IEEE} Computer Society},
  year      = {2006},
  doi       = {10.1109/ICDCS.2006.72}
}

@inproceedings{Talwar04,
  author    = {Kunal Talwar},
  title     = {Bypassing the embedding: algorithms for low dimensional metrics},
  booktitle = {Proceedings of the 36th Annual {ACM} Symposium on Theory of Computing, {STOC} 2004},
  pages     = {281--290},
  publisher = {{ACM}},
  year      = {2004},
  doi       = {10.1145/1007352.1007399}
}

@article{Slivkins07,
  author    = {Aleksandrs Slivkins},
  title     = {Distance estimation and object location via rings of neighbors},
  journal   = {Distributed Comput.},
  volume    = {19},
  number    = {4},
  pages     = {313--333},
  year      = {2007},
  doi       = {10.1007/s00446-006-0015-8}
}

@article{ChanGMZ16,
  author    = {T.{-}H. Hubert Chan and Anupam Gupta and Bruce M. Maggs and Shuheng Zhou},
  title     = {On Hierarchical Routing in Doubling Metrics},
  journal   = {{ACM} Trans. Algorithms},
  volume    = {12},
  number    = {4},
  year      = {2016},
  doi       = {10.1145/2915183}
}

@article{KonjevodRX16,
  author    = {Goran Konjevod and Andr{\'{e}}a W. Richa and Donglin Xia},
  title     = {Scale-Free Compact Routing Schemes in Networks of Low Doubling Dimension},
  journal   = {{ACM} Trans. Algorithms},
  volume    = {12},
  number    = {3},
  year      = {2016},
  doi       = {10.1145/2876055}
}

@inproceedings{KonjevodRXY07,
  author    = {Goran Konjevod and Andr{\'{e}}a W. Richa and Donglin Xia and Hai Yu},
  title     = {Compact routing with slack in low doubling dimension},
  booktitle = {Proceedings of the 26th Annual {ACM} Symposium on Principles of Distributed Computing, {PODC} 2007},
  pages     = {71--80},
  publisher = {{ACM}},
  year      = {2007},
  doi       = {10.1145/1281100.1281113}
}

@inproceedings{AbrahamM05,
  author    = {Ittai Abraham and Dahlia Malkhi},
  title     = {Name independent routing for growth bounded networks},
  booktitle = {Proceedings of the 17th Annual {ACM} Symposium on Parallelism in Algorithms and Architectures, {SPAA} 2005},
  pages     = {49--55},
  publisher = {{ACM}},
  year      = {2005},
  doi       = {10.1145/1073970.1073978}
}

@article{LaingR07,
  author    = {Kofi A. Laing and Rajmohan Rajaraman},
  title     = {A Space Lower Bound for Name-Independent Compact Routing in Trees},
  journal   = {J. Interconnect. Networks},
  volume    = {8},
  number    = {3},
  pages     = {229--251},
  year      = {2007},
  doi       = {10.1142/S0219265907002004}
}

@article{SantoroK85,
  author    = {Nicola Santoro and Ramez Khatib},
  title     = {Labelling and Implicit Routing in Networks},
  journal   = {Comput. J.},
  volume    = {28},
  number    = {1},
  pages     = {5--8},
  year      = {1985},
  doi       = {10.1093/comjnl/28.1.5}
}

@article{vanLeeuwenT87,
  author    = {Jan van Leeuwen and Richard B. Tan},
  title     = {Interval Routing},
  journal   = {Comput. J.},
  volume    = {30},
  number    = {4},
  pages     = {298--307},
  year      = {1987},
  doi       = {10.1093/comjnl/30.4.298}
}

@article{Gavoille00survey,
  author    = {Cyril Gavoille},
  title     = {A survey on interval routing},
  journal   = {Theor. Comput. Sci.},
  volume    = {245},
  number    = {2},
  pages     = {217--253},
  year      = {2000},
  doi       = {10.1016/S0304-3975(99)00283-2}
}

@article{Gavoille01,
  author    = {Cyril Gavoille},
  title     = {Routing in distributed networks: overview and open problems},
  journal   = {{SIGACT} News},
  volume    = {32},
  number    = {1},
  pages     = {36--52},
  year      = {2001},
  doi       = {10.1145/568438.568451}
}

@article{GavoilleP03,
  author    = {Cyril Gavoille and David Peleg},
  title     = {Compact and localized distributed data structures},
  journal   = {Distributed Comput.},
  volume    = {16},
  number    = {2-3},
  pages     = {111--120},
  year      = {2003},
  doi       = {10.1007/s00446-002-0073-5}
}

@book{Peleg00book,
  author    = {David Peleg},
  title     = {Distributed Computing: A Locality-Sensitive Approach},
  series    = {{SIAM} Monographs on Discrete Mathematics and Applications},
  publisher = {{SIAM}},
  year      = {2000},
  doi       = {10.1137/1.9780898719772}
}

@article{Peleg00labeling,
  author    = {David Peleg},
  title     = {Proximity-preserving labeling schemes},
  journal   = {J. Graph Theory},
  volume    = {33},
  number    = {3},
  pages     = {167--176},
  year      = {2000},
  doi       = {10.1002/(SICI)1097-0118(200003)33:3<167::AID-JGT7>3.0.CO;2-5}
}

@article{EilamGP07,
  author    = {Tamar Eilam and Cyril Gavoille and David Peleg},
  title     = {Average stretch analysis of compact routing schemes},
  journal   = {Discret. Appl. Math.},
  volume    = {155},
  number    = {5},
  pages     = {598--610},
  year      = {2007},
  doi       = {10.1016/j.dam.2006.09.010}
}

@article{Chechik13ft,
  author    = {Shiri Chechik},
  title     = {Fault-tolerant compact routing schemes for general graphs},
  journal   = {Inf. Comput.},
  volume    = {222},
  pages     = {36--44},
  year      = {2013},
  doi       = {10.1016/j.ic.2012.10.009}
}

@article{DoryP21,
  author    = {Michal Dory and Merav Parter},
  title     = {Fault-Tolerant Labeling and Compact Routing Schemes},
  journal   = {{SIAM} J. Comput.},
  volume    = {55},
  number    = {4},
  pages     = {671--707},
  year      = {2026},
  doi       = {10.1137/25M1722494}
}

@article{RodittyT16,
  author    = {Liam Roditty and Roei Tov},
  title     = {Close to linear space routing schemes},
  journal   = {Distributed Comput.},
  volume    = {29},
  number    = {1},
  pages     = {65--74},
  year      = {2016},
  doi       = {10.1007/s00446-015-0256-5}
}

@article{ElkinN18,
  author    = {Michael Elkin and Ofer Neiman},
  title     = {On efficient distributed construction of near optimal routing schemes},
  journal   = {Distributed Comput.},
  volume    = {31},
  number    = {2},
  pages     = {119--137},
  year      = {2018},
  doi       = {10.1007/s00446-017-0304-4}
}

@inproceedings{ElkinN18podc,
  author    = {Michael Elkin and Ofer Neiman},
  title     = {Near-Optimal Distributed Routing with Low Memory},
  booktitle = {Proceedings of the 2018 {ACM} Symposium on Principles of Distributed Computing, {PODC} 2018},
  pages     = {207--216},
  publisher = {{ACM}},
  year      = {2018},
  doi       = {10.1145/3212734.3212761}
}

@inproceedings{LenzenP13,
  author    = {Christoph Lenzen and Boaz Patt{-}Shamir},
  title     = {Fast routing table construction using small messages},
  booktitle = {Symposium on Theory of Computing Conference, {STOC} 2013},
  pages     = {381--390},
  publisher = {{ACM}},
  year      = {2013},
  doi       = {10.1145/2488608.2488656}
}

@article{LenzenPP19,
  author    = {Christoph Lenzen and Boaz Patt{-}Shamir and David Peleg},
  title     = {Distributed distance computation and routing with small messages},
  journal   = {Distributed Comput.},
  volume    = {32},
  number    = {2},
  pages     = {133--157},
  year      = {2019},
  doi       = {10.1007/s00446-018-0326-6}
}

@inproceedings{Bartal96,
  author    = {Yair Bartal},
  title     = {Probabilistic Approximations of Metric Spaces and Its Algorithmic Applications},
  booktitle = {37th Annual Symposium on Foundations of Computer Science, {FOCS} 1996},
  pages     = {184--193},
  publisher = {{IEEE} Computer Society},
  year      = {1996},
  doi       = {10.1109/SFCS.1996.548477}
}

@article{FakcharoenpholRT04,
  author    = {Jittat Fakcharoenphol and Satish Rao and Kunal Talwar},
  title     = {A tight bound on approximating arbitrary metrics by tree metrics},
  journal   = {J. Comput. Syst. Sci.},
  volume    = {69},
  number    = {3},
  pages     = {485--497},
  year      = {2004},
  doi       = {10.1016/j.jcss.2004.04.011}
}

@article{AlonKPW95,
  author    = {Noga Alon and Richard M. Karp and David Peleg and Douglas B. West},
  title     = {A Graph-Theoretic Game and Its Application to the $k$-Server Problem},
  journal   = {{SIAM} J. Comput.},
  volume    = {24},
  number    = {1},
  pages     = {78--100},
  year      = {1995},
  doi       = {10.1137/S0097539792224474}
}

@article{KaoRT96,
  author    = {Ming{-}Yang Kao and John H. Reif and Stephen R. Tate},
  title     = {Searching in an Unknown Environment: An Optimal Randomized Algorithm for the Cow-Path Problem},
  journal   = {Inf. Comput.},
  volume    = {131},
  number    = {1},
  pages     = {63--79},
  year      = {1996},
  doi       = {10.1006/inco.1996.0092}
}

@article{ChrobakKNY08,
  author    = {Marek Chrobak and Claire Kenyon and John Noga and Neal E. Young},
  title     = {Incremental Medians via Online Bidding},
  journal   = {Algorithmica},
  volume    = {50},
  number    = {4},
  pages     = {455--478},
  year      = {2008},
  doi       = {10.1007/s00453-007-9005-x}
}

@inproceedings{AwerbuchP90,
  author    = {Baruch Awerbuch and David Peleg},
  title     = {Sparse Partitions},
  booktitle = {31st Annual Symposium on Foundations of Computer Science, {FOCS} 1990},
  pages     = {503--513},
  publisher = {{IEEE} Computer Society},
  year      = {1990},
  doi       = {10.1109/FSCS.1990.89571}
}

@inproceedings{ValiantB81,
  author    = {Leslie G. Valiant and Gordon J. Brebner},
  title     = {Universal Schemes for Parallel Communication},
  booktitle = {Proceedings of the 13th Annual {ACM} Symposium on Theory of Computing, {STOC} 1981},
  pages     = {263--277},
  publisher = {{ACM}},
  year      = {1981},
  doi       = {10.1145/800076.802479}
}

@inproceedings{Racke02,
  author    = {Harald R{\"{a}}cke},
  title     = {Minimizing Congestion in General Networks},
  booktitle = {43rd Symposium on Foundations of Computer Science, {FOCS} 2002},
  pages     = {43--52},
  publisher = {{IEEE} Computer Society},
  year      = {2002},
  doi       = {10.1109/SFCS.2002.1181881}
}
\appendix

\section{Summary of our results}\label{S-summary}
\Cref{tab-our} lists the stretch of our schemes in all four models, for unweighted and weighted graphs.
\begin{table}[h]
\centering
\small
\begin{tabular}{@{}lllll@{}}
\hline
Stretch & Model & Storage & Graphs & Reference \\
\hline
$2k-1$ & labeled & $\Ot(n^{1/k})$ average & weighted & \Cref{T-avg-random}$^\star$ \\
$\approx2.47k-1.47$ & labeled & $\Ot(n^{1/k})$ average & weighted & \Cref{T-avg-det} \\
$4k-5$ & name-independent & $\Ot(n^{1/k})$ average & weighted & \Cref{T-NI-avg}$^\star$ \\
$\approx5.77k-4.77$ & name-independent & $\Ot(n^{1/k})$ average & weighted & \Cref{T-NI-det} \\
\hline
$(\approx2.67k+6,\approx0.44k)$ & labeled & $\Ot(n^{1/k})$ worst-case & unweighted & \Cref{T-worst-random}$^\star$ \\
$\approx3.05k+1$ & labeled & $\Ot(n^{1/k})$ worst-case & unweighted & \Cref{T-worst-det} \\
$(\approx8k-1.56,2k-2)$ & name-independent & $\Ot(n^{1/k})$ worst-case & unweighted & \Cref{T-NI-worst-random}$^\star$ \\
$\approx11.54k-2.54$ & name-independent & $\Ot(n^{1/k})$ worst-case & unweighted & \Cref{T-NI-worst-det} \\
\hline
$\approx(1+\eps)3.07k+O(1)$ & labeled & $\Ot(\eps^{-1}n^{1/k}\log\Delta)$ worst-case & weighted & \Cref{T-weighted-rand}$^\star$ \\
$\approx(1+\eps)3.42k+O(1)$ & labeled & $\Ot(\eps^{-1}n^{1/k}\log\Delta)$ worst-case & weighted & \Cref{T-weighted-det} \\
$\approx14.62(1+\eps)^2k+O(1)$ & name-independent & $\Ot(\eps^{-1}n^{1/k}\log\Delta)$ worst-case & weighted & \Cref{T-weighted-ni-rand}$^\star$ \\
$25.6(1+\eps)^2k+O(1)$ & name-independent & $\Ot(\eps^{-1}n^{1/k}\log\Delta)$ worst-case & weighted & \Cref{T-weighted-ni-det} \\
\hline
\end{tabular}
\caption{Our routing schemes in all four models, where $\Delta$ is the aspect ratio of the graph and $0<\eps<1$. A pair $(\alpha,\beta)$ denotes $(\alpha,\beta)$-stretch, that is, $\hat d(u,v)\le\alpha\,d(u,v)+\beta$, or $\EE[\hat d(u,v)]\le\alpha\,d(u,v)+\beta$ for rows marked $^\star$. $^\star$Expected stretch.}
\label{tab-our}
\end{table}

\section{Proofs of the optimization bounds}\label{S-avg-opt}
This appendix proves~\Cref{L-avg-LP-value} in~\Cref{S-avg-opt-rand} and~\Cref{L-avg-det-bound} in~\Cref{S-avg-opt-det}.
\Cref{S-avg-opt-compare} compares the deterministic bound with the previous bound of Kadria and Roditty~\cite{kadria2025compact}.
\Cref{S-avg-opt-tight} shows that the coefficient $1/\ln(3/2)$ in~\Cref{L-avg-det-bound} cannot be improved using only the inequalities of~\Cref{L-avg-normalized}.

Fix $u$, $v$, $\tau$, and $M$ as in~\Cref{S-avg-selection}. In this appendix, we divide all distances by $M$. Write
\[
x_i=\frac{h_i(u)}M,\qquad y_i=\frac{h_i(v)}M\quad(0\le i\le\tau),\qquad H=y_\tau,
\]
and $\delta_i=y_{i+1}-x_i$ for $i\in[\tau]$.
So $\delta_i$ is the threshold of level $i$, divided by $M$.

\begin{lemma}\label{L-avg-normalized}
The normalized heights are nondecreasing, $x_0=y_0=0$, $0<H\le2\tau-1$, and
\begin{align}
&x_1\le1\text{ and }y_1\le1,\tag{N1}\label{e-avg-N1}\\
&x_{i+1}\le y_i+1\qquad(0\le i<\tau),\tag{N2}\label{e-avg-N2}\\
&|x_i-y_i|\le1\qquad(0\le i\le\tau),\tag{N3}\label{e-avg-N3}\\
&0\le y_{i+1}-y_i\le2\qquad(0\le i<\tau).\tag{N4}\label{e-avg-N4}
\end{align}
\end{lemma}
\begin{proof}
The bound on $H$ follows from~\eqref{e-avg-M}.
The definition of $M$ gives~\eqref{e-avg-N1}--\eqref{e-avg-N3}, and monotonicity together with $y_{i+1}\le x_{i+1}+1\le y_i+2$ gives~\eqref{e-avg-N4}.
\end{proof}

\paragraph{Penalties.}
Let $d=\theta M$ be a possible value of $d(u,v)$, so $\theta\ge1$, and let $i\in[\tau]$.
Dividing~\eqref{e-avg-penalties} by $d$ shows that the extra stretch $(\hat d(u,v)-d)/d$ of level $i$ is at most its \emph{penalty}
\[
\chi_i(\theta)=\frac{2x_i+2H\cdot\one[\theta\ge\delta_i]}\theta\qquad(\theta\ge1).
\]
Since $\chi_i$ is decreasing on $[1,\delta_i)$ and on $[\max\{1,\delta_i\},\infty)$, the quantity $\beta_i$ of~\Cref{S-avg-det} is the largest penalty of level $i$:
\[
\beta_i=\max\left\{2x_i,\frac{2(x_i+H)}{\max\{1,\delta_i\}}\right\}=\sup_{\theta\ge1}\chi_i(\theta).
\]
Since $x_0=0$ and $\delta_0=y_1\le1$ by~\eqref{e-avg-N1}, $\beta_0=2H$.
After substituting $d=\theta M$ and dividing the constraints by $d$, we can write~\eqref{e-avg-LP} as
\begin{equation}\label{e-avg-LP-norm}
z^*=\min_{\pi\in\mathcal P([\tau])}\ \max_{\theta\in\Theta/M}\ \sum_{i\in[\tau]}\pi_i\chi_i(\theta),
\end{equation}
where $\Theta/M=\{d/M:d\in\Theta\}=\{1\}\cup\{\delta_i:i\in[\tau],\ \delta_i>1\}$.

\subsection{The randomized bound}\label{S-avg-opt-rand}
By LP duality, it suffices to show that for every distribution of the normalized distance $\theta$, some level has expected penalty at most $2\tau$.
Reweighting the distribution by $1/\theta$ turns the expected penalty of a level into its expected extra cost divided by the mean $m$ of the reweighted distribution.
Every level whose visit, on success, costs less than the smallest expected extra cost must fail with large probability.
So its threshold is at most a quantile of the distribution.
By~\eqref{e-avg-N2}, $x_{i+2}-x_i$ is then at most $1$ plus this quantile.
As $x_i$ grows, the required failure probability decreases, so different levels use different parts of the distribution.
Measured in units of these quantiles, the height grows by at most $1$ per level.
The Cauchy--Schwarz inequality turns this into a bound in terms of $m$.

\paragraph{Quantiles.}
For a distribution $\nu$ on a finite subset of $[1,\infty)$, let $m=\EE_{\theta\sim\nu}[\theta]$ and $\bar F(\zeta)=\Pr_{\theta\sim\nu}[\theta\ge\zeta]$.
Its quantile function is $f(\alpha)=\min\{\zeta\in\operatorname{supp}(\nu)\mid\Pr_{\theta\sim\nu}[\theta\le\zeta]>\alpha\}$ for $0\le\alpha<1$.
It is nondecreasing, $f\ge1$, and $m=\int_0^1f(\alpha)\,d\alpha$.
Moreover,
\begin{equation}\label{e-avg-quantile}
\bar F(\zeta)\ge1-\alpha\ \Longrightarrow\ \zeta\le f(\alpha)\qquad(0\le\alpha<1).
\end{equation}
Indeed, if $f(\alpha)<\zeta$, then $\Pr_{\theta\sim\nu}[\theta\le f(\alpha)]\le1-\bar F(\zeta)\le\alpha$, contradicting the definition of $f(\alpha)$.

For such $\nu$, let
\[
C=\min_{i\in[\tau]}\big(2x_i+2H\,\bar F(\delta_i)\big).
\]

\lemLPvalue*
\begin{proof}
LP duality, applied to~\eqref{e-avg-LP-norm}, gives
\begin{equation}\label{e-avg-minimax}
z^*=\max_{\lambda\in\mathcal P(\Theta/M)}\ \min_{i\in[\tau]}\sum_{\theta\in\Theta/M}\lambda(\theta)\chi_i(\theta).
\end{equation}
Fix $\lambda$, let $Z=\sum_{\theta\in\Theta/M}\lambda(\theta)/\theta>0$, and set $\nu(\theta)=\lambda(\theta)/(\theta Z)$.
Then $\nu$ is a distribution with mean $m=1/Z$.
Since $\lambda(\theta)=\theta\nu(\theta)/m$, every level $i$ has
\[
\sum_{\theta\in\Theta/M}\lambda(\theta)\chi_i(\theta)=\frac{\EE_{\theta\sim\nu}[\theta \chi_i(\theta)]}m=\frac{2x_i+2H\,\bar F(\delta_i)}m.
\]
Hence the minimum over $i$ is $C/m$, which is at most $2\tau$ by~\Cref{L-avg-T} below.
Since this holds for every $\lambda$, \eqref{e-avg-minimax} proves $z^*\le2\tau$.
\end{proof}

\begin{lemma}\label{L-avg-T}
For every distribution $\nu$ on a finite subset of $[1,\infty)$ with mean $m$, $\min_{i\in[\tau]}\big(2x_i+2H\Pr_{\theta\sim\nu}[\theta\ge\delta_i]\big)\le2\tau m$.
\end{lemma}
\begin{proof}
If $C\le2\tau$, the lemma holds since $m\ge1$.
Otherwise, $\bar F(\delta_0)=1$ since $\delta_0\le1$, so $2\tau<C\le2H\le2(2\tau-1)$ by~\Cref{L-avg-normalized}.
Let $r=C/(2H)\in(1/2,1]$ and $\alpha_i=1-r+x_i/H$.
For every $i\in[\tau]$ with $x_i<rH$, the definition of $C$ gives $\bar F(\delta_i)\ge r-x_i/H=1-\alpha_i$, so~\eqref{e-avg-quantile} gives
\begin{equation}\label{e-avg-delta-quantile}
\delta_i\le f(\alpha_i).
\end{equation}
Let
\[
\varphi(w)=1-r+rf(1-r+w/H)\quad(0\le w<rH),\qquad
\Psi(z)=\int_0^{\min\{z,rH\}}\frac{dw}{\varphi(w)}\quad(z\ge0).
\]
Since $f$ is nondecreasing and $f\ge1$, $\varphi$ is nondecreasing and $\varphi\ge1$.

\begin{claim}\label{C-avg-pairs}
$\Psi(x_j)\le j\qquad(0\le j<\tau)$.
\end{claim}
\begin{proof}
Since $\varphi\ge1$, we have $\Psi(z)\le z$, so $\Psi(x_0)=0$ and $\Psi(x_1)\le1$ by~\eqref{e-avg-N1}.
Let $0\le i\le\tau-3$. Since $\Psi$ is constant on $[rH,\infty)$, we may assume $x_i<rH$.
By~\eqref{e-avg-N2} and~\eqref{e-avg-delta-quantile},
\[
x_{i+2}-x_i\le1+\delta_i\le1+f(\alpha_i)\le2\varphi(x_i),
\]
where the last inequality uses $r\ge1/2$ and $f\ge1$.
Since $\varphi\ge\varphi(x_i)$ on $[x_i,rH)$, we get $\Psi(x_{i+2})-\Psi(x_i)\le2$, and induction proves the claim.
\end{proof}

\begin{claim}\label{C-avg-last}
$\Psi(rH)\le\tau$.
\end{claim}
\begin{proof}
By~\Cref{C-avg-pairs}, it suffices to show $\Psi(rH)-\Psi(x_{\tau-1})\le1$, and we may assume $x_{\tau-1}<rH$.
Since $\delta_{\tau-1}=H-x_{\tau-1}$, \eqref{e-avg-delta-quantile} gives
\[
rH-x_{\tau-1}\le r\delta_{\tau-1}\le rf(\alpha_{\tau-1})\le\varphi(x_{\tau-1}).
\]
Hence $\Psi(rH)-\Psi(x_{\tau-1})\le(rH-x_{\tau-1})/\varphi(x_{\tau-1})\le1$.
\end{proof}

Substituting $\alpha=1-r+w/H$ and using $f\ge1$ on $[0,1-r)$, we get
\[
\int_0^{rH}\varphi(w)\,dw=rH\Big(1-r+\int_{1-r}^1f(\alpha)\,d\alpha\Big)\le rHm.
\]
By the Cauchy--Schwarz inequality and~\Cref{C-avg-last},
\[
(rH)^2\le\Psi(rH)\int_0^{rH}\varphi(w)\,dw\le\tau\,rHm.
\]
Hence $C=2rH\le2\tau m$.
\end{proof}

\subsection{The deterministic bound}\label{S-avg-opt-det}
If $\tau=1$, then $\beta^*\le\beta_0=2H\le2<1/\ln(3/2)$.
Otherwise, write $\tau=2q+s$, with $q\ge1$ and $s\in\{0,1\}$.
Let $t_\tau>0$ solve
\begin{equation}\label{e-avg-tL}
\Big(1+\frac1{t_\tau}\Big)^q=\frac{3t_\tau+1}{2t_\tau+s+1}.
\end{equation}
The left side decreases from $\infty$ to $1$, and the right side increases to $3/2$, so increasing $q$ or $s$ increases the solution.
For $\tau=2$, the equation reads $1+1/t=(3t+1)/(2t+1)$, whose solution is $1+\sqrt2$. Hence $t_\tau\ge1+\sqrt2$.
Going from $(q,1)$ to $(q+1,0)$ multiplies the left side of $(2t+s+1)\rho^q=3t+1$ by $(2t+1)(t+1)/\big(t(2t+2)\big)=(2t+1)/(2t)\ge1$, so $t_\tau$ is nondecreasing in $\tau$.
Let $\rho=1+1/t_\tau$.
We show $\beta^*\le2t_\tau$ and then $2t_\tau<\tau/\ln(3/2)$.

\begin{lemma}\label{L-avg-det-exact}
$\beta^*\le2t_\tau$.
\end{lemma}
\begin{proof}
Let $t=t_\tau$, and suppose that $\beta^*>2t$.
Then $H>t$, since $\beta_0=2H$.
A level $i+1$ with $x_{i+1}\le t$ must have $\delta_{i+1}<(x_{i+1}+H)/t$, since otherwise $\beta_{i+1}\le2t$.
Since $x_{i+1}\le y_i+1$ by~\eqref{e-avg-N2}, this bounds $y_{i+2}$ in terms of $y_i$ while $y_i\le t-1$.
After that, $y$ grows by at most $2$ per level by~\eqref{e-avg-N4}.
Together, these bounds contradict $H>t$.

Equation~\eqref{e-avg-tL} gives $(2t+s+1)\rho^q=3t+1$ and $\rho^q<3/2$.
Subtracting $t\rho^q+t+1$ from both sides of the first identity gives
\begin{equation}\label{e-avg-tL-alt}
(2-\rho^q)t=s+(t+s+1)(\rho^q-1).
\end{equation}

\paragraph{Growth over pairs of levels.}
If $0\le i\le\tau-2$ and $y_i\le t-1$, then $x_{i+1}\le y_i+1\le t$ by~\eqref{e-avg-N2}.
Since $2x_{i+1}\le2t<\beta_{i+1}$, the second term of $\beta_{i+1}$ exceeds $2t$.
Hence $\delta_{i+1}\le\max\{1,\delta_{i+1}\}<(x_{i+1}+H)/t$, and
\[
y_{i+2}=x_{i+1}+\delta_{i+1}<\rho(y_i+1)+H/t.
\]
Equivalently,
\begin{equation}\label{e-avg-det-recurrence}
y_{i+2}+H+t+1<\rho(y_i+H+t+1).
\end{equation}

\paragraph{Following the recurrence.}
Since $y_s\le s\le1<t-1$ by~\eqref{e-avg-N1} and $y_{s+2q}=H>t$, there is a first $j^*\in\{1,\ldots,q\}$ with $y_{s+2j^*}>t-1$.
Let $y^*=y_{s+2j^*}$.
Applying~\eqref{e-avg-det-recurrence} for $i=s,s+2,\ldots,s+2j^*-2$, starting from $y_s\le s$, gives
\begin{equation}\label{e-avg-det-envelope}
y^*<s+(H+t+s+1)(\rho^{j^*}-1).
\end{equation}
Let $a=q-j^*$ be the number of remaining pairs.
By~\eqref{e-avg-N4}, $H\le y^*+4a$.

If $a=0$, then $y^*=H$, so~\eqref{e-avg-det-envelope} and~\eqref{e-avg-tL-alt} give
\[
(2-\rho^q)H<s+(t+s+1)(\rho^q-1)=(2-\rho^q)t,
\]
contradicting $H>t$.

If $a\ge1$, we substitute $H\le y^*+4a$ into~\eqref{e-avg-det-envelope} and use $y^*>t-1$.
Since $2-\rho^{j^*}>0$, this yields
\[
(2-\rho^{j^*})(t-1)<s+(4a+t+s+1)(\rho^{j^*}-1).
\]
Rearranging gives $3t-1+4a<(2t+s+4a)\rho^{j^*}$, and multiplying by $\rho^a$, with $j^*+a=q$, gives
\begin{equation}\label{e-avg-det-crossing}
(3t-1+4a)\rho^a<(2t+s+4a)\rho^q.
\end{equation}
But Bernoulli's inequality bounds the left side from below by
\[
(3t-1+4a)(1+a/t)=3t-1+7a+\frac{(4a-1)a}{t}\ge3t-1+7a,
\]
whereas $(2t+s+1)\rho^q=3t+1$ and $\rho^q<3/2$ bound the right side by
\[
(2t+s+4a)\rho^q=3t+1+(4a-1)\rho^q<3t+6a-\tfrac12.
\]
These bounds contradict~\eqref{e-avg-det-crossing}, since $a\ge1$.
\end{proof}

It remains to compare $2t_\tau$ with $\tau/\ln(3/2)$.
\lemDetBound*
\begin{proof}
By~\Cref{L-avg-det-exact} and $2t_\tau=2/(\rho-1)$, it suffices to show that $\tau(\rho-1)/2>\ln(3/2)$.
Here $1<\rho\le\sqrt2$, since $t_\tau\ge1+\sqrt2$.
We use the standard bounds
\begin{equation}\label{e-avg-log}
\frac{2(\omega-1)}{\omega+1}\le\ln\omega\le\frac12\Big(\omega-\frac1\omega\Big)\qquad(\omega\ge1).
\end{equation}
Substituting $t_\tau=1/(\rho-1)$ into~\eqref{e-avg-tL} gives $\rho^{\tau/2}=(\rho+2)/(\rho+1)$ when $s=0$, and $\rho^{\tau/2}=(\rho+2)/(2\sqrt\rho)$ when $s=1$.
Since $2\sqrt\rho\le\rho+1$, both imply $\rho^{\tau/2}\ge(\rho+2)/(\rho+1)$.
The upper bound in~\eqref{e-avg-log}, applied to $\sqrt\rho$, now gives
\[
\frac{\tau(\rho-1)}2\ge\sqrt\rho\,\frac\tau2\ln\rho
\ge\sqrt\rho\ln\frac{\rho+2}{\rho+1}.
\]
Let $\eta(\xi)=\sqrt\xi\ln((\xi+2)/(\xi+1))$ for $1\le\xi\le\sqrt2$.
The lower bound in~\eqref{e-avg-log} gives
\[
\eta'(\xi)=\frac1{2\sqrt\xi}\left(\ln\frac{\xi+2}{\xi+1}-\frac{2\xi}{(\xi+1)(\xi+2)}\right)
\ge\frac{2-\xi^2}{\sqrt\xi(2\xi+3)(\xi+1)(\xi+2)}.
\]
Thus $\eta$ is strictly increasing on $[1,\sqrt2]$, so $\eta(\rho)>\eta(1)=\ln(3/2)$.
\end{proof}

\subsection{Comparison with the previous bound}\label{S-avg-opt-compare}
For $k\ge3$, we compare~\Cref{T-avg-det} with Corollary~4.1 of Kadria and Roditty~\cite{kadria2025compact}.
Asymptotically, their bound is at least $\frac52k-O(1)$, while ours is $\frac1{\ln(3/2)}k+O(1)\approx2.47k$.
We show that ours is smaller for every $k\ge3$.
Write $k-1=2\ell+s'$, with $\ell\ge1$ and $s'\in\{0,1\}$.
Their scheme visits $p_{2i}(u)$ only if the threshold $h_{2i+1}(v)-h_{2i}(u)$ exceeds $c_i$ times its current lower bound on $d(u,v)$.
Their Corollary~4.1 is the case $\tau=k-1$, where the constants are
\[
c_0=1,\qquad c_i=2-\frac{\ell-i}{\ell+\sum_{j<i}c_j}\quad(1\le i\le\ell).
\]
It bounds the stretch by $2\sum_{i=0}^\ell c_i+2\ell+1$ for $k=2\ell+2$, and by $2\sum_{i=0}^\ell c_i+2\ell-1$ for $k=2\ell+1$.
In both cases, the bound is $2\ell-1+2s'+2\sum_{i=0}^\ell c_i$, and since $c_\ell=2$, this is $2\ell+3+2s'+2\sum_{i=0}^{\ell-1}c_i$.
By induction, $c_i\ge1$ for every $i$, so $\sum_{j<i}c_j\ge i$.
Therefore
\[
c_i\ge2-\frac{\ell-i}{\ell+i}=1+\frac{2i}{\ell+i}\ge1+\frac i\ell\qquad(0\le i<\ell).
\]
Summing gives $\sum_{i<\ell}c_i\ge(3\ell-1)/2$, so their bound is at least
\[
5\ell+2+2s'>1+\tfrac52(k-1),
\]
since $\tfrac52(k-1)=5\ell+\tfrac52s'$ and $s'\le1$.
Our bound is at most $1+(k-1)/\ln(3/2)<1+\tfrac52(k-1)$, since $\ln(3/2)>2/5$.
Thus our stretch bound is strictly smaller for every $k\ge3$.

\subsection{A tight profile for the deterministic bound}\label{S-avg-opt-tight}
We give normalized heights for which every level has $\beta_i=2t_\tau$.
These heights satisfy~\eqref{e-avg-N1}--\eqref{e-avg-N4} but need not come from a graph.
Hence \Cref{L-avg-det-exact} cannot be improved using only these inequalities.

Write $t=t_\tau$.
Set $x_0=y_0=0$, $y_1=1$, and
\[
x_i=y_{i-1}+1\quad(1\le i\le\tau),\qquad
y_{i+1}=x_i+1+\frac{x_i}t\quad(1\le i<\tau).
\]

\paragraph{The heights.}
The two rules give $y_{i+1}=\rho(y_{i-1}+1)+1$, that is, $y_{i+1}+2t+1=\rho(y_{i-1}+2t+1)$.
Starting from $y_0=0$ and $y_1=1$, we get
\[
y_{2j}=(2t+1)(\rho^j-1),\qquad
y_{2j+1}=(2t+2)\rho^j-(2t+1).
\]
The equation $y_\tau=t$ reads $(2t+s+1)\rho^q=3t+1$, which is~\eqref{e-avg-tL}.
Hence $H=y_\tau=t$.
The increments of $y$ are $1,\rho,\rho,\rho^2,\rho^2,\ldots$, ending with $\rho^q$.
Since $\rho^q<3/2$ by~\eqref{e-avg-tL}, they lie in $[1,3/2)$.

Since $x_i=y_{i-1}+1$, we have $y_i-x_i=(y_i-y_{i-1})-1\in[0,1/2)$ for $i\ge1$.
Hence both sequences are nondecreasing, and~\eqref{e-avg-N1}--\eqref{e-avg-N4} hold, with equality in~\eqref{e-avg-N2}.
Also $H<1+\tfrac32(\tau-1)\le2\tau-1$.

\paragraph{The penalties.}
For $1\le i<\tau$, the second rule gives $\delta_i=y_{i+1}-x_i=1+x_i/t>1$, and $x_i\le y_i\le t$.
Hence
\[
\beta_i=\max\left\{2x_i,\frac{2(x_i+t)}{1+x_i/t}\right\}=2t=\beta_0,
\]
so $\beta^*=2t_\tau$.

\paragraph{Asymptotics.}
Since every increment is at least $1$, we have $t=H\ge\tau$, so $q=O(t)$.
Taking logarithms in~\eqref{e-avg-tL} gives
\[
\frac qt+O\Big(\frac1t\Big)
=q\ln\Big(1+\frac1t\Big)
=\ln\frac{3t+1}{2t+s+1}
=\ln\frac32+O\Big(\frac1t\Big).
\]
Thus $t_\tau=q/\ln(3/2)+O(1)$, so $2t_\tau=\tau/\ln(3/2)+O(1)$, and the coefficient $1/\ln(3/2)$ in~\Cref{L-avg-det-bound} cannot be improved using only the inequalities of~\Cref{L-avg-normalized}.

\section{Proofs of the worst-case optimization bounds}\label{S-worst-opt}
This appendix proves \Cref{L-worst-det-opt} in \Cref{S-worst-opt-det} and \Cref{L-worst-rand-opt} in \Cref{S-worst-opt-rand}. We use the notation of \Cref{S-worst-det,S-worst-rand}.

\subsection{The deterministic bound}\label{S-worst-opt-det}
\lemWorstDetOpt*
\begin{proof}
Recall that the heights are nondecreasing integers and satisfy
\[
    h_1(u),h_1(v)\le M,\qquad
    h_{i+1}(u)\le h_i(v)+M,\qquad
    |h_i(u)-h_i(v)|\le M.
\]
Consequently,
\[
    H\le(2\tau-1)M,\qquad
    \delta_i\le3M,\qquad
    h_{i+2}(u)-h_i(u)\le M+\delta_i
    \quad(0\le i\le\tau-2).
\]

Suppose that the first claimed bound is false, and choose
\[
    c_M\tau+2+\frac{c_M(M-1)}{4M+2}
    <\beta<\beta^*.
\]
Since $\beta<\beta^*\le2H/M$, we have $2H>\beta M$.
Also $c_M>2$, since
$(4+1/M)\sum_{r=2M+1}^{3M}1/r<5/2<3$, so $\beta>2\tau+2$.

Set
\[
    z_i=h_i(u)+i/2\quad(0\le i\le\tau),
    \qquad
    z_{\tau+1}=H+M+(\tau+1)/2.
\]
The height inequalities give
\[
    z_{i+2}-z_i\le M+1+\delta_i
    \qquad(0\le i<\tau).
\]
For $i=\tau-1$, this follows directly from the definition
of $z_{\tau+1}$.

We first bound $\delta_i$ while $z_i$ is small.
Let
\[
    b=\frac{M(\beta-2)-H+1}{2}>0.
\]
If $1\le i<\tau$ and $z_i<b$, define
\[
    t=
    \left\lfloor
        \frac{4z_i+2H-2}{\beta-2}
    \right\rfloor+1.
\]
The inequalities $2H>\beta M$ and $z_i<b$ imply
$M<t\le2M$.
Indeed, $4z_i+2H-2\ge M(\beta-2)$ since $2H-2>\beta M-2\ge M(\beta-2)$, so $t>M$.
And $z_i<b$ means $4z_i+2H-2<2M(\beta-2)$, so $t\le2M$.
If $\delta_i\ge t$, this is an available distance guess.
Moreover,
\[
    4z_i+2t-2+2H<\beta t
\]
and, since $t\le2M$ and $2H>\beta M$,
\[
    4z_i+2t-2<\beta t-2H<\beta M.
\]
Hence both terms of the maximum in the definition of $\beta^*$ for the pair $(i,t)$ are less than $\beta$ (the first term by $4z_i+2t-2<\beta M$, the second by $4z_i+2t-2+2H<\beta t$), so $\beta^*<\beta$, contradicting $\beta<\beta^*$.
Hence
\[
    \delta_i\le
    \left\lfloor
        \frac{4z_i+2H-2}{\beta-2}
    \right\rfloor
    \qquad\text{whenever }z_i<b.
\]
This inequality also holds for $i=0$, because
$\delta_0=h_1(v)\le M$.

Define
\[
    \phi(z)=
    \begin{cases}
      \displaystyle
      \frac{M+1+
        \left\lfloor(4z+2H-2)/(\beta-2)\right\rfloor}{2},
          &0\le z<b,\\[2mm]
      \displaystyle\frac{4M+1}{2},
          &z\ge b,
    \end{cases}
    \qquad
    \Phi(z)=\int_0^z\frac{ds}{\phi(s)}.
\]
The function $\phi$ is nondecreasing and
$\phi(z)\ge M+1/2$.
The preceding bounds on $\delta_i$ give
$z_{i+2}-z_i\le2\phi(z_i)$.
Therefore
\[
    \Phi(z_{i+2})-\Phi(z_i)
    \le\frac{z_{i+2}-z_i}{\phi(z_i)}
    \le2.
\]
Since $\Phi(z_0)=0$ and
$\Phi(z_1)\le1$, induction on the even and odd indices yields
\[
    \Phi(z_{\tau+1})\le\tau+1.
\]

We now evaluate this integral.
Since $z_{\tau+1}>b$, substitution in its first part gives
\[
\Phi(z_{\tau+1})
=
\frac{\beta-2}{2}
\int_{(2H-2)/(\beta-2)}^{2M}
    \frac{ds}{M+1+\lfloor s\rfloor}
+
\frac{3H+4M+\tau-M\beta}{4M+1}.
\]
For fixed $\beta$, this expression is increasing in $H$.
Indeed, away from its finitely many breakpoints, its derivative is
\[
    \frac{3}{4M+1}
    -
    \frac{1}
    {M+1+\left\lfloor(2H-2)/(\beta-2)\right\rfloor}
    \ge
    \frac{3}{4M+1}-\frac{1}{2M+1}>0.
\]
Here we used $H\ge\beta M/2$ and $M\ge1$.
We may therefore lower-bound the expression by substituting
$H=\beta M/2$.

At this value, the lower endpoint of the integral is
$M+(2M-2)/(\beta-2)$.
Also,
\[
    \int_M^{2M}\frac{ds}{M+1+\lfloor s\rfloor}
    =\sum_{r=2M+1}^{3M}\frac1r.
\]
The omitted initial interval contributes at most
$(2M-2)/((\beta-2)(2M+1))$.
Consequently,
\[
\Phi(z_{\tau+1})
\ge
\frac{\beta-2}{2}\sum_{r=2M+1}^{3M}\frac1r
-\frac{M-1}{2M+1}
+\frac{\beta M/2+4M+\tau}{4M+1}.
\]
Combining this with $\Phi(z_{\tau+1})\le\tau+1$
and rearranging gives
\[
    \beta\le
    c_M\tau+2+\frac{c_M(M-1)}{4M+2},
\]
contradicting our choice of $\beta$.

It remains to bound $c_M$.
For $M=1$, we have $c_M=3\le8/(1+4\ln(3/2))\approx3.05$.
For $M\ge2$, convexity of $1/x$ and the trapezoidal rule give
\[
    \sum_{r=2M+1}^{3M}\frac1r
    \ge \ln(3/2)-\frac1{12M}.
\]
Hence
\[
\left(4+\frac1M\right)
\sum_{r=2M+1}^{3M}\frac1r
\ge
4\ln(3/2)
+\frac{\ln(3/2)-1/3}{M}
-\frac1{12M^2}
\ge4\ln(3/2).
\]
The last inequality uses $(\ln(3/2)-\tfrac13)/M\ge1/(12M^2)$, which holds for $M\ge2$ and fails for $M=1$.
This proves the second bound.
\end{proof}

\subsection{The randomized bound}\label{S-worst-opt-rand}
\lemWorstRandOpt*
\begin{proof}
By the finite minimax theorem, it suffices to show that
for every distribution of an integer $D\in[M,3M]$,
some available choice has expected extra cost at most
\[
    \left(\tau+\frac12\right)\EE[\gamma(D)]+6M-2.
\]
Fix such a distribution, and put
\[
    C=\min_{0\le i<\tau}
       \left\{
           4h_i(u)+2i
           +2H\Pr[D\ge\delta_i]
       \right\}.
\]
Since $h_0(u)=0$ and $\delta_0=h_1(v)\le M$,
index $0$ gives $C\le2H$ and represents the fallback route.
Every other index with $\delta_i\le M$ has value greater
than $2H$, so it cannot minimize this expression.
For a minimizing index $i\ge1$, the choice $t=\delta_i$
is available and has expected extra cost
\[
    C+2\delta_i-2\le C+6M-2.
\]
Thus it remains to prove
\[
    C\le
    \left(\tau+\frac12\right)\EE[\gamma(D)].
\]

Let $f$ be the quantile function of $D$:
\[
    f(\alpha)=
    \min\{d\in\operatorname{supp}(D):
                  \Pr[D\le d]>\alpha\},
    \qquad 0\le\alpha<1.
\]
It is nondecreasing and satisfies $M\le f\le3M$.
Moreover,
\[
    \Pr[D\ge a]\ge1-\alpha
    \quad\Longrightarrow\quad
    a\le f(\alpha).
\]
Indeed, if $f(\alpha)<a$, then
$\Pr[D\le f(\alpha)]\le\Pr[D<a]\le\alpha$,
contrary to its definition.

Let $r=C/(2H)\in(0,1]$.
As in the deterministic proof, set
\[
    z_i=h_i(u)+\frac{i}{2}
       \quad(0\le i\le\tau),
    \qquad
    z_{\tau+1}=H+M+\frac{\tau+1}{2}.
\]
The height inequalities give
\[
    z_0=0,\qquad z_1\le M+\frac12,
    \qquad
    z_{i+2}-z_i\le M+1+\delta_i
       \quad(0\le i<\tau).
\]
For every $i$ with $z_i<rH/2$, the definition of $C$
implies
\[
    \Pr[D\ge\delta_i]\ge r-\frac{2z_i}{H}.
\]
The quantile property therefore gives
\[
    \delta_i\le
    f\left(1-r+\frac{2z_i}{H}\right).
\]

Define
\[
    \phi(z)=
    \begin{cases}
        \displaystyle
        \frac{M+1+f(1-r+2z/H)}{2},
            &0\le z<rH/2,\\[2mm]
        \displaystyle\frac{4M+1}{2},
            &z\ge rH/2,
    \end{cases}
    \qquad
    \Phi(z)=\int_0^z\frac{ds}{\phi(s)}.
\]
The function $\phi$ is nondecreasing and
$\phi\ge M+\tfrac12$.
The preceding inequalities, together with $\delta_i\le3M$,
give
\[
    z_{i+2}-z_i\le2\phi(z_i).
\]
Hence
$\Phi(z_{i+2})-\Phi(z_i)\le2$.
Since $\Phi(z_0)=0$ and $\Phi(z_1)\le1$,
induction on the two parities yields
\[
    \Phi(z_{\tau+1})\le\tau+1.
\]

Evaluating the integral gives
\[
    H\int_{1-r}^1
        \frac{d\alpha}{M+1+f(\alpha)}
    +\frac{(2-r)H+2M+\tau+1}{4M+1}
    \le\tau+1.
\]
Equivalently,
\[
    H\left(
        (4M+1)\int_{1-r}^1
            \frac{d\alpha}{M+1+f(\alpha)}
        +2-r
    \right)
    \le4M\tau+2M.
\]
Put
\[
    I=\EE\left[\frac{1}{M+1+D}\right]
     =\int_0^1\frac{d\alpha}{M+1+f(\alpha)}.
\]
Since $f\ge M$, the part $\int_0^{1-r}d\alpha/(M+1+f(\alpha))$ of $I$
is at most $(1-r)/(2M+1)$, and $I\ge1/(4M+1)$ since $f\le3M$. Consequently,
\[
    (4M+1)\int_{1-r}^1
        \frac{d\alpha}{M+1+f(\alpha)}
        +2-r
    \ge r\bigl(1+(4M+1)I\bigr).
\]
Indeed, the left side minus the right side is at least $(1-r)\big[(4M+1)\big(I-\tfrac1{2M+1}\big)+2\big]\ge(1-r)\tfrac{2M+2}{2M+1}\ge0$.
Using $C=2rH$, we obtain
\[
    C\le
    \frac{8M\tau+4M}{1+(4M+1)I}.
\]

For every $d\ge M$,
\[
    \gamma(d)\left(1+\frac{4M+1}{M+1+d}\right)\ge8M.
\]
Indeed, subtracting $8M$ from the left-hand side gives
\[
    \frac{
        4(d-M)\bigl(2d^2+(6M+5)d+2M+2\bigr)
    }{
        (3d+1)(M+1+d)
    }\ge0.
\]
By the Cauchy--Schwarz inequality,
\[
\begin{aligned}
    \EE[\gamma(D)]\bigl(1+(4M+1)I\bigr)
    &\ge
    \left(
        \EE\left[
            \sqrt{
                \gamma(D)\left(1+\frac{4M+1}{M+1+D}\right)
            }
        \right]
    \right)^2\\
    &\ge8M.
\end{aligned}
\]
Combining the last two bounds proves
\[
    C\le
    \left(\tau+\frac12\right)\EE[\gamma(D)].
\]
Minimax now gives $z^*\le\tau+\tfrac12$.

Finally, the program's constraints imply
\[
    \EE[\hat d(u,v)]
    \le d+\left(\tau+\frac12\right)\gamma(d)+6M-2.
\]
Using $\tau\le k-1$, $M\le d$, and
$\gamma(d)\ge\tfrac83d$, this is at most $k\gamma(d)+6d$.
The second bound in the statement follows from
\[
    \gamma(d)=\frac83d+\frac{4d}{3(3d+1)}
         \le\frac83d+\frac49.
\]
\end{proof}

\section{The deterministic name-independent bound}\label{S-NI-opt}
This appendix proves~\Cref{L-NI-det-bound} and shows that it is tight.
Fix $u$ as in~\Cref{S-NI-avg}, with the gaps $g_a$ as there, and $s_\sigma(d)$ and $\beta(\sigma)$ as in~\Cref{S-NI-det}. We write $h_a=h_a(u)$.
Let $\rho=2^{1/(k-1)}$ and $\lambda=1/(\rho-1)$, so $\rho=1+1/\lambda$ and $\rho^{k-1}=2$.

\paragraph{Valid sequences.}
For a sequence $\sigma=(i_1<\cdots<i_m)$ of levels with $i_1=0$, let $S_t=\sum_{s\le t}h_{i_s}$.
We call $\sigma$ \emph{valid} if $S_t\le\lambda g_{i_{t-1}}$ for every $2\le t\le m$.
If a valid $\sigma$ ends at $i_m=k-1$, then $\beta(\sigma)\le4\lambda$.
Indeed, fix $d>0$ and let $t=s_\sigma(d)$. If $t=1$, the sum in $\beta(\sigma)$ is $h_0=0$.
Otherwise, $g_{i_{t-1}}\le2d$ by the minimality of $t$, so $\frac2dS_t\le\frac2d\lambda g_{i_{t-1}}\le4\lambda$.
Since $4\lambda=4/(2^{1/(k-1)}-1)$, it suffices to find a valid sequence that ends at $k-1$.

\paragraph{Cheapest valid sequences.}
A level $a$ is \emph{reachable} if a valid sequence ends at $a$. Its \emph{cost} $C_a$ is the smallest sum $S_m$ of such a sequence, and its \emph{budget} is $R_a=\lambda g_a-C_a$.
Level $0$ is reachable with $C_0=0$.
Appending a level $b>a$ to a valid sequence that ends at $a$ keeps it valid if and only if its sum plus $h_b$ is at most $\lambda g_a$.
Hence $b$ is reachable if and only if $h_b\le R_a$ for some reachable $a<b$. A cheapest valid sequence ending at $a$ has the largest budget among all valid sequences ending at $a$, so
\[
C_b=h_b+e_b,\qquad\text{where }e_b=\min\{C_a\mid a<b\text{ is reachable and }R_a\ge h_b\}.
\]
Since level $0$ has cost $0$, $e_b>0$ implies $h_b>R_0$.

When the levels $0,\ldots,L$ are reachable, let $X_a=\max_{a'\le a}R_{a'}$ for $a\le L$, and for $X_0<y\le X_L$, let $u(y)=\min\{a\mid X_a\ge y\}$.
The level $u(y)$ is a new maximum, so $R_{u(y)}=X_{u(y)}\ge y$.
Since $\lambda g_a=h_a+R_a+e_a$, we have $h_{a+1}=h_a+g_a=\rho h_a+(R_a+e_a)/\lambda$, and hence
\begin{equation}\label{e-NI-height}
h_{L+1}=\frac1\lambda\sum_{a\le L}\rho^{L-a}(R_a+e_a).
\end{equation}

\begin{lemma}\label{L-NI-reach}
For every $0\le L\le k-2$, the levels $0,\ldots,L+1$ are reachable and $h_{L+1}\le(\rho^{L+1}-1)X_L$.
\end{lemma}
For $L=k-2$, the lemma shows that $k-1$ is reachable, which proves~\Cref{L-NI-det-bound}.
\begin{proof}
We use induction on $L$.
For $L=0$, $h_1=g_0=R_0/\lambda=(\rho-1)X_0\le X_0$, so level $1$ is reachable.
Let $L\ge1$.
By the induction hypothesis, the levels $0,\ldots,L$ are reachable, and $h_a\le(\rho^a-1)X_{a-1}$ for $1\le a\le L$.
It suffices to show $h_{L+1}\le(\rho^{L+1}-1)X_L$. Then $h_{L+1}\le X_L$ since $\rho^{L+1}\le2$, so level $L+1$ is reachable.

\paragraph{The slack.}
Let $n_a=e_a-(X_a-R_a)$ be the part of the cost $e_a$ that exceeds the deficit $X_a-R_a$ of the budget of $a$.
By~\eqref{e-NI-height} and $\sum_{a\le L}\rho^{L-a}=\lambda(\rho^{L+1}-1)$,
\begin{equation}\label{e-NI-slack}
\lambda\big((\rho^{L+1}-1)X_L-h_{L+1}\big)=\sum_{a\le L}\rho^{L-a}(X_L-R_a-e_a)=G-\sum_{a\le L}\rho^{L-a}n_a,
\end{equation}
where
\[
G=\sum_{a\le L}\rho^{L-a}(X_L-X_a)=\sum_{1\le a\le L}\lambda\big(\rho^{L+1}-\rho^{L+1-a}\big)(X_a-X_{a-1})=\int_{X_0}^{X_L}\gamma(y)\,dy
\]
for $\gamma(y)=\lambda(\rho^{L+1}-\rho^{L+1-u(y)})$.
We also set $\gamma(y)=\lambda(\rho^{L+1}-1)$ for $y>X_L$.
Since $u(y)$ is nondecreasing and at most $L$, $\gamma$ is nondecreasing.
So $G$ grows whenever the largest budget grows, and it remains to show that $G$ pays for the costs $n_a$.

\paragraph{The cost of a predecessor.}
Let $1\le a\le L$ with $e_a>0$, let $q=u(h_a)$, and let $\eps=\rho^q-1$.
We claim that $1\le q<a$ and $e_a<\eps h_a/(1-\eps)$.
Since $e_a>0$, $h_a>R_0=X_0$, so $q\ge1$.
Since $a$ is reachable, some $a'<a$ has $R_{a'}\ge h_a$, so $q<a$, and $\eps<\rho^{k-1}-1=1$.
Level $q$ has $R_q\ge h_a$, so $e_a\le C_q=h_q+e_q$.
By the induction hypothesis and $X_{q-1}<h_a$, we have $h_q\le(\rho^q-1)X_{q-1}<\eps h_a$.
If $e_q>0$, the same argument for $q$ gives a level $q'<q$ with $e_q\le h_{q'}+e_{q'}$ and $h_{q'}<(\rho^{q'}-1)h_q\le\eps h_q<\eps^2h_a$.
We repeat until the cost is $0$, and get $e_a<h_a(\eps+\eps^2+\cdots)=\eps h_a/(1-\eps)$.

\paragraph{One level.}
Let $1\le a\le L$ with $n_a>0$. Then $e_a\ge n_a>0$, and we let $q$ and $\eps$ be as above.
We claim that
\begin{equation}\label{e-NI-window}
\rho^{L-a}n_a<g_a\,\gamma(h_a)=\lambda g_a\,\rho^{L+1-q}(\rho^q-1).
\end{equation}
Let $\psi=\rho^{a+1-q}(\rho^q-1)$. Since $\lambda g_a=h_a+R_a+e_a=h_a+X_a+n_a$, the claim reads $n_a<(h_a+X_a+n_a)\psi$.
It holds if $\psi\ge1$. Otherwise, it reads $n_a(1-\psi)<(h_a+X_a)\psi$.
Let $x=\rho^a$ and $v=\rho^q$, so $\eps=v-1$ and $\psi=\rho x(v-1)/v$.
By the induction hypothesis, $h_a\le(x-1)X_{a-1}\le(x-1)X_a$, so $h_a+X_a\ge h_ax/(x-1)$.
Since $n_a\le e_a<h_a(v-1)/(2-v)$, it suffices that
\[
\frac{v-1}{2-v}\le\frac{x}{x-1}\cdot\frac{\psi}{1-\psi}.
\]
Multiplying out, this is $(x-1)\big(v-\rho x(v-1)\big)\le\rho x^2(2-v)$, that is,
\[
f(v)=\rho x(x+1)-v(\rho x+x-1)\ge0.
\]
Since $q<a\le k-2$, we have $v\le x/\rho$ and $x\le\rho^{k-2}=2/\rho$.
The function $f$ is decreasing, so it suffices that $f(x/\rho)=\frac x\rho\big(\rho^2+1-x(1+\rho-\rho^2)\big)\ge0$.
This is clear if $1+\rho-\rho^2\le0$. Otherwise, $x\le2/\rho$ gives
\[
f(x/\rho)\ge\frac x\rho\Big(\rho^2+1-\frac2\rho(1+\rho-\rho^2)\Big)=\frac{x(\rho-1)(\rho+1)(\rho+2)}{\rho^2}\ge0.
\]

\paragraph{All levels.}
The intervals $(h_a,h_{a+1}]$, for $a\le L$, are disjoint.
If $n_a>0$, then $h_a>X_0$, and $\gamma(y)\ge\gamma(h_a)$ for $y>h_a$ since $\gamma$ is nondecreasing.
Hence~\eqref{e-NI-window} gives $\rho^{L-a}n_a<\int_{h_a}^{h_{a+1}}\gamma(y)\,dy$.
Summing over these levels gives
\[
\sum_{a\le L}\rho^{L-a}n_a\le\int_{X_0}^{\max\{X_L,h_{L+1}\}}\gamma(y)\,dy=G+\lambda(\rho^{L+1}-1)\max\{0,h_{L+1}-X_L\},
\]
with strict inequality if some $n_a>0$.
Suppose that $h_{L+1}>X_L$.
If no $n_a$ is positive, the right side of~\eqref{e-NI-slack} is at least $G\ge0$, so $h_{L+1}\le(\rho^{L+1}-1)X_L\le X_L$, a contradiction.
Otherwise,~\eqref{e-NI-slack} and the strict inequality give $(2-\rho^{L+1})h_{L+1}<0$, contradicting $\rho^{L+1}\le2$.
Hence $h_{L+1}\le X_L$, and~\eqref{e-NI-slack} gives $(\rho^{L+1}-1)X_L-h_{L+1}\ge0$.
\end{proof}

\paragraph{A tight example.}
The bound of~\Cref{L-NI-det-bound} is tight.
Let $g_a=\rho^a$ for $a<k-1$, so $h_a=\lambda(\rho^a-1)$ and $h_{k-1}=\lambda$.
Let $\sigma$ be a sequence, and let $a$ be its last level before $k-1$.
Then $h_a+h_{k-1}=\lambda\rho^a=\lambda g_a$, and every level of $\sigma$ before $k-1$ has gap at most $g_a$.
So for $d>g_a/2$, we have $s_\sigma(d)=m$, and
\[
\frac2d\sum_{s\le m}h_{i_s}\ge\frac{2\lambda g_a}d,
\]
which tends to $4\lambda$ as $d$ tends to $g_a/2$.
Hence $\beta(\sigma)\ge4\lambda$ for every $\sigma$, and $\beta^*=4\lambda$ for these gaps.

\section{Proofs of the name-independent worst-case bounds}\label{S-NI-worst-opt}
In this section, we prove the bounds of \Cref{S-NI-worst}, using its notation.

\subsection{The deterministic bound}
\lemNIWorstDet*
\begin{proof}
We exhibit a suitable sequence of rounded geometric
guesses. For a shift \(x\in(0,1)\), let
\[
    t_j=M2^{j+x},
    \qquad
    D_j=\lceil t_j\rceil-1,
    \qquad j=0,1,\ldots.
\]
Stop at the first guess at least \(n-1\), replacing
that guess by \(n-1\).
Since \(c\) is nondecreasing, this replacement can
only decrease the cost bound.
We analyze the sequence before the replacement
and then choose its shift.

Fix an integer \(d\ge M\), and let \(J\) be the first
index with \(D_J\ge d\).
If \(J\ge1\), the preceding guess failed to cover \(d\),
so
\[
    d\ge\lceil t_{J-1}\rceil=\lceil t_J/2\rceil.
\]
This also holds when \(J=0\), since \(t_0/2<M\le d\).
In particular, \(t_J\le2d\), and the radius terms satisfy
\[
    2\sum_{j=0}^{J}D_j
    \le2\sum_{j=0}^{J}t_j
    <4t_J
    \le8d.
\]

We bound the height and index costs together:
\[
    4h_i(u)+2i=\sum_{r<i}(4g_r+2).
\]
For each finite gap, define
\[
    a_r=\max\left\{M,\left\lceil\frac{g_r}{2}\right\rceil\right\},
\]
and let \(t_r^*(x)\) be the first value \(t_j>a_r\).
A gap can contribute at guess \(D_j\) only if
\(g_r\le2D_j\), which, by integrality, is equivalent
to \(t_j>\lceil g_r/2\rceil\).
Thus it contributes only from \(t_r^*(x)\) onward.

Suppose that there are \(\ell\ge1\) guesses from
\(t_r^*(x)\) through \(t_J\).
The gap's total contribution is at most
\(\ell(4g_r+2)\), and
\[
    d\ge
    \left\lceil2^{\ell-2}t_r^*(x)\right\rceil.
\]
We use the elementary inequality
\[
    \frac{\ell}{\lceil2^{\ell-2}t\rceil}
    \le\frac{2}{\lceil t\rceil},
    \qquad t>1,\quad \ell\ge1.
\]
For \(\ell=1\), it follows from
\(\lceil t\rceil\le2\lceil t/2\rceil\).
For \(\ell\ge2\), writing \(m=\lceil t\rceil\ge2\) gives
\[
    \lceil2^{\ell-2}t\rceil
    \ge2^{\ell-2}(m-1)+1
    \ge\frac{\ell m}{2}.
\]
Consequently, the contribution of gap \(r\) is at most
\[
    \frac{8g_r+4}{\lceil t_r^*(x)\rceil}\,d.
\]

Define
\[
    F(x)=
    \sum_{r=0}^{k-2}
        \frac{8g_r+4}{\lceil t_r^*(x)\rceil}.
\]
The preceding bounds hold simultaneously for every
integer \(d\ge M\), and give
\[
    \sum_{j=0}^{J}c(D_j)\le(F(x)+8)d.
\]
It remains to choose a shift with
\(F(x)\le8(k-1)/\ln2\).

For this purpose only, let \(X\) be uniform on \((0,1)\).
For each \(r\), the ratio \(t_r^*(X)/a_r\) has the same
distribution as \(2^X\). Hence
\[
    \mathbb{E}\left[
        \frac{8g_r+4}{\lceil t_r^*(X)\rceil}
    \right]
    =
    \frac{8g_r+4}{\ln2}
    \int_{a_r}^{2a_r}\frac{dt}{t\lceil t\rceil}.
\]
For every positive integer \(a\),
\[
\begin{aligned}
    \int_a^{2a}\frac{dt}{t\lceil t\rceil}
    &=\sum_{m=a+1}^{2a}
        \frac1m\ln\frac{m}{m-1}\\
    &\le
      \sum_{m=a+1}^{2a}
        \frac{1-1/(2m)}{m(m-1)}\\
    &\le
      \left(1-\frac1{4a}\right)
      \sum_{m=a+1}^{2a}\frac1{m(m-1)}\\
    &=\frac{4a-1}{8a^2}
    \le\frac2{4a+1}.
\end{aligned}
\]
The first inequality follows by integrating the convex
function \(1/t\) over each interval \([m-1,m]\)
and bounding its integral by the trapezoid area.

Since \(g_r\le2a_r\), we conclude that
\[
    \mathbb{E}\left[
        \frac{8g_r+4}{\lceil t_r^*(X)\rceil}
    \right]
    \le
    \frac{16a_r+4}{\ln2}\cdot\frac2{4a_r+1}
    =\frac8{\ln2}.
\]
Summing over the \(k-1\) finite gaps yields
\[
    \mathbb{E}[F(X)]\le\frac{8(k-1)}{\ln2}.
\]
Therefore some shift \(x\) satisfies the same bound.
The source can choose such a shift deterministically,
since \(F\) depends only on its own gaps.

The resulting sequence has extra stretch at most
\(8(k-1)/\ln2+8\).
Hence $\beta^*_{\mathrm{det}}\le8(k-1)/\ln2+8$, and the stretch is at most $1+\beta^*_{\mathrm{det}}$.
\end{proof}

\subsection{The randomized bound}
\lemNIWorstRand*
\begin{proof}
Fix \(d=d(u,v)\ge M\).
Let \(j^*\) be the first index with \(D_{j^*}\ge d\).
The visit at this index succeeds, so
\(\hat d(u,v)\le d+\sum_{D_j<e(d+1/4)-1/4}c(D_j)\).
Substituting \(t=(M+\tfrac14)e^{j+X}\) and averaging over the random shift gives
\[
    \mathbb{E}[\hat d(u,v)]
    \le d+
    \int_{M+1/4}^{e(d+1/4)}
        \frac{c(t-1/4)}{t}\,dt.
\]

We now bound the height and index costs together:
\[
    4h_i(u)+2i
    =4\sum_{r<i}\left(g_r+\frac12\right).
\]
For every \(r<i(t-1/4)\), the choice of the level gives
\[
    g_r\le2(t-1/4),
    \qquad\text{and hence}\qquad
    g_r+\frac12\le2t.
\]
For \(r=0\), the same conclusion follows from
\(g_0\le2M\) and \(t\ge M+1/4\).

Fix a finite gap \(g_r\), and write \(a=g_r+\tfrac12\).
Its contribution to the integral is zero if
\(a\ge2e(d+\tfrac14)\).
Otherwise, its contribution is at most
\[
\begin{aligned}
    4a\ln
       \frac{e(d+1/4)}{\max\{M+1/4,a/2\}}
    &\le
    4a\ln\frac{2e(d+1/4)}{a}\\
    &\le8\left(d+\frac14\right).
\end{aligned}
\]
The last inequality follows by maximizing
\(a\ln(2e(d+\tfrac14)/a)\) over \(a>0\).
Summing over the \(k-1\) finite gaps bounds the combined
height and index costs by
\[
    8(k-1)d+2(k-1).
\]

The remaining term in \(c(t-1/4)\) is \(2t-\tfrac12\).
Since \(M\ge1\),
\[
\begin{aligned}
    \int_{M+1/4}^{e(d+1/4)}
        \frac{2t-1/2}{t}\,dt
    &\le
    2\left(e(d+1/4)-(M+1/4)\right)\\
    &\le2ed.
\end{aligned}
\]
Adding the shortest-path distance and the two integral
bounds yields
\[
    \mathbb{E}[\hat d(u,v)]
    \le
    \bigl(8(k-1)+2e+1\bigr)d+2(k-1),
\]
as claimed.
\end{proof}

\section{Weighted graphs}\label{S-Weighted}

Our average-storage schemes work in positively weighted graphs with the same stretch bounds.
In this section, we extend our schemes with worst-case storage to weighted graphs.
For every $0<\eps<1$, our labeled schemes have stretch $\approx(1+\eps)3.42k+O(1)$ and expected stretch $\approx(1+\eps)3.07k+O(1)$ (\Cref{T-weighted-det,T-weighted-rand}), and our name-independent schemes have stretch $\frac{128}{5}(1+\eps)^2k+O(1)\approx25.6(1+\eps)^2k$ and expected stretch $\frac{4e+20\sqrt e}{3}(1+\eps)^2k+O(1)\approx14.62(1+\eps)^2k$ (\Cref{T-weighted-ni-det,T-weighted-ni-rand}).
All of them use $\Ot(\eps^{-1}n^{1/k}\log\Delta)$ local storage at every vertex, where $\Delta$ is the aspect ratio of the graph.
For small enough $\eps$, the labeled schemes improve the $(4-\alpha)k-\beta$ of Chechik~\cite{DBLP:conf/podc/Chechik13}, for small constants $\alpha,\beta>0$. The name-independent schemes improve the $96k$ of Abraham, Gavoille, and Malkhi~\cite{AbrahamGM04}.

\subsection{Labeled routing}
The additional traversals made by the distributed tree
search may be expensive in a weighted graph.
To control their weight, we preprocess graphs obtained
by deleting heavy edges.
The source uses the destination label to select one
of these graphs, and performs both the visit and the
fallback route in the selected graph.

\paragraph{Preprocessing.}
Normalize the minimum edge weight to $1$, and write
$\Delta=\operatorname{diam}(G)$.
Fix $0<\eps<1$, and define
\[
    W_j=(1+\eps)^j,
    \qquad
    G_j=(V,\{e\in E:\wt(e)\le W_j\}),
\]
for
\[
    0\le j\le J
    =\left\lceil\log_{1+\eps}(2\Delta)\right\rceil.
\]
We call each $G_j$ a layer.
Its edges retain their original weights and ports.

Perform the preprocessing of the preceding sections
separately in every connected component of every layer.
Use the global parameter $S=\lceil n^{1/k}\rceil$.
As in \Cref{S-worst-prepr}, every vertex belongs to $\Ot(S)$ cluster
trees in each layer, and every level-$i$ cluster
has at most $4S^{i+1}$ vertices.
Write $p_i^j(x)$, $h_i^j(x)$, and $B_j(x)$
for the pivots, heights, and bunches in $G_j$.

For each cluster tree, construct the prefix search
structures of \Cref{S-worst-prepr}.
A prefix contains the first
$\min\{2^\ell,|C_j(w)|\}$ vertices in nondecreasing
weighted distance from the root $w$.
Since edge weights are positive, each prefix contains
the ancestors of its vertices and forms a subtree.
The records store names and routing labels in the
full cluster tree.

Each vertex stores the information of \Cref{S-worst-prepr}
for every layer.
In addition, for every root in its bunch, it stores
its exact distance to that root.
The explicit level-$0$ dictionaries also store exact
root distances.
These distances allow the source to bound the fallback
route before sending the message.

The label of $v$ contains its component identifier
and its pivot and tree-label information for every layer.
Thus the source can identify layers in which $u$
and $v$ are connected and compute the usual routing
quantities within each such layer.

\paragraph{Weighted tree searches.}
Use the branching parameter $8S$ in the distributed indices.
The same counting argument as in
\Cref{T-S-NI-Tree} bounds the number of additional
two-edge traversals in a level-$i$ search by $i$.
Indeed, each additional index step reduces the number
of possible record holders by a factor of at least $8S$.
One further traversal may be needed to read the root
table stored off the initial root path, which gives $i+1$ in total.

If the host tree has radius at most $R$ and lies in
$G_j$, a lookup therefore costs at most
\[
    R+2(i+1)W_j.
\]
An unsuccessful lookup followed by a return to the root
costs at most
\[
    2R+2(i+1)W_j.
\]
After a successful lookup, the retrieved full-tree label
allows delivery from the record holder to $v$ at
additional cost at most
\[
    R+d_{G_j}(w,v).
\]

For a requested radius $R$, the root selects the smallest
prefix containing all cluster vertices at distance
at most $R$.
The preceding prefix, which hosts the search structure,
has radius at most $R$.
Thus the prefix construction supports every requested
radius, without rounding $R$.

There are $O(\log n)$ prefixes per cluster.
Choose their preorder blocks independently and fix a
placement satisfying the simultaneous storage bound,
as in \Cref{S-worst-prepr}.
The resulting tables use
$\Ot((J+1)n^{1/k})$ words at every vertex.
The labels use $\Ot(k(J+1))$ words.
The message carries the same routing state as in \Cref{S-worst-det},
together with the selected layer and search-structure
identifiers.

\paragraph{Choosing a layer.}
Fix a parameter $1<c\le2$, to be specified separately
for the deterministic and randomized choices.
The source skips layers in which its component identifier
differs from the destination's.

Consider a remaining layer $G_j$.
If its bunch or explicit level-$0$ dictionary supplies
a shortest route to $v$, set
\[
    M_j=d_{G_j}(u,v),
\]
using the stored distance.
Otherwise, let $\tau_j$ be the fallback level in this
layer and define
\[
\begin{aligned}
    M_j=\max\biggl\{&
        h_1^j(v),\
        \max_{0\le i<\tau_j}
            \bigl(h_{i+1}^j(u)-h_i^j(v)\bigr),\
        \max_{1\le i\le\tau_j}
            |h_i^j(u)-h_i^j(v)|,\\
        &\left|
            d_{G_j}(u,p_{\tau_j}^j(v))
            -h_{\tau_j}^j(v)
        \right|
    \biggr\}.
\end{aligned}
\]
The first three terms are the lower bounds used in
\Cref{S-avg-selection}.
The last term is also a lower bound, by the reverse
triangle inequality.
Hence
\[
    M_j\le d_{G_j}(u,v).
\]
Every term is available at the source, including the
distance to the fallback pivot, which belongs to its bunch.

Select the first connected layer satisfying
\[
    M_j\le W_j/c.
\]
For the rest of the section, write
\[
    W=W_j,\qquad M=M_j,\qquad \mu=W/c,
\]
and suppress the layer superscripts on pivots and heights.

Let $d=d_G(u,v)$.
The selected layer satisfies
\begin{equation}\label{eq:weighted-layer-selection}
    M\le d,
    \qquad
    \mu\le(1+\eps)d.
\end{equation}
For the first inequality, if $d\le W$, the layer retains
a shortest path and $M\le d_{G_j}(u,v)=d$.
If $d>W$, the selection condition gives $M\le W/c<d$.

For the second inequality, every layer with $W_j\ge cd$
retains a shortest path and satisfies
$M_j\le d\le W_j/c$.
The weight of the selected layer is therefore at most
$c(1+\eps)d$.
Such a layer exists because $c\le2$ and the
preprocessing reaches $2\Delta$.

If the selected layer supplies a direct route, its length
is $M\le W/c<W$.
Any path using a deleted edge has weight greater than $W$,
so this route is globally shortest.
We henceforth consider the remaining case.

\paragraph{The fallback route.}
Let $\tau$ be the fallback level in the selected layer,
and write
\[
    H=h_\tau(v).
\]
The additional term in $M$ gives
\[
    d_{G_j}(u,p_\tau(v))\le M+H.
\]
Consequently, the fallback route in this layer has length
at most
\[
    d_{G_j}(u,p_\tau(v))+H
    \le M+2H
    \le d+2H.
\]
This bound holds even when the selected layer does not
preserve $d_G(u,v)$.

The definition of $M$, together with $M\le\mu$, also gives
\[
\begin{gathered}
    h_1(u),h_1(v)\le\mu,\\
    h_{i+1}(u)\le h_i(v)+\mu
        \qquad(0\le i<\tau),\\
    |h_i(u)-h_i(v)|\le\mu
        \qquad(1\le i\le\tau).
\end{gathered}
\]
In particular,
\[
    H\le(2\tau-1)\mu,
    \qquad
    h_{i+1}(v)-h_i(u)\le3\mu
        \qquad(0\le i<\tau).
\]

\paragraph{The cost of a visit.}
Call a level $1\le i<\tau$ \emph{considered} if $h_{i+1}(v)-h_i(u)>\mu$.
Let $i$ be a considered level.
Then
\[
    h_{i+1}(u)
    \ge h_{i+1}(v)-M
    >h_i(u).
\]
Thus $p_i(u)$ is a level-$i$ center and
$u\in C_j(p_i(u))$.

The message routes to $p_i(u)$ and searches with radius
\[
    R=h_{i+1}(v).
\]
If the search succeeds, it uses the retrieved label
to reach $v$.
Otherwise, it returns to $u$ and takes the fallback
route in the selected layer.

Failure implies
\[
    d\ge
    \min\{W,h_{i+1}(v)-h_i(u)\}.
\]
Indeed, below this threshold the layer preserves
the original distance, and
\[
    d_{G_j}(p_i(u),v)
    \le h_i(u)+d
    <h_{i+1}(v)=R.
\]
The destination then belongs to the pivot's cluster
and lies within the searched radius.

The complete route satisfies
\begin{equation}\label{eq:weighted-visit}
\begin{aligned}
    \hat d(u,v)\le d
        &+2h_i(u)+2h_{i+1}(v)+2(i+1)W\\
        &+2H\,\one
          [d\ge\min\{W,h_{i+1}(v)-h_i(u)\}].
\end{aligned}
\end{equation}
On failure, this follows from the return cost and the
fallback bound.

On success with $d<W$, the layer preserves the distance,
so delivery from the record holder gives the same bound
without the indicator term.
On success with $d\ge W$, cluster membership gives
\[
    d_{G_j}(p_i(u),v)<h_{i+1}(v)=R.
\]
The successful route therefore costs at most
$h_i(u)+3R+2(i+1)W$.
Since $R\le H$, this is bounded by
\eqref{eq:weighted-visit}, whose indicator is then $1$.

\paragraph{The quantities used in the optimization.}
For each considered level, define
\[
    a_i=4h_i(u)+2iW,
    \qquad
    t_i=\min\{W,h_{i+1}(v)-h_i(u)\}.
\]
Here $a_i$ is the part of the visit cost retained
in the optimization, and $t_i$ is its failure threshold.
The remaining cost satisfies
\[
\begin{aligned}
    2h_i(u)+2h_{i+1}(v)+2(i+1)W-a_i
    &=2\bigl(h_{i+1}(v)-h_i(u)\bigr)+2W\\
    &\le(6+2c)\mu.
\end{aligned}
\]
Thus
\begin{equation}\label{eq:weighted-core}
    \hat d(u,v)
    \le d+a_i+2H\,\one[d\ge t_i]
          +(6+2c)\mu.
\end{equation}

We optimize over $s\ge\mu$, where $s$ stands for $\max\{d,\mu\}$.
Every considered threshold satisfies
\[
    \mu<t_i\le W=c\mu.
\]
A level with $h_{i+1}(v)-h_i(u)\le\mu$ would have
its indicator equal to $1$ throughout this range,
and would therefore be dominated by the fallback route.

Although the actual distance may be smaller than $\mu$,
\eqref{eq:weighted-layer-selection} gives
\[
    \max\{d,\mu\}\le(1+\eps)d.
\]
Replacing $d$ by $\max\{d,\mu\}$ in the indicator
can only increase the bound.
We can therefore use an optimization over $s\ge\mu$
and account for this replacement by a factor
of $1+\eps$.

\paragraph{Deterministic routing.}
Take $c=2$.
For a considered level $i$, the largest ratio of
the retained extra cost to $s\ge\mu$ is
\[
    \max\left\{
        \frac{a_i}{\mu},\
        \frac{a_i+2H}{t_i}
    \right\}.
\]
The first term is attained at the beginning of the range.
The second is attained at the failure threshold.
Between these points and beyond the threshold,
the ratio is nonincreasing.

The source compares these values with the fallback bound
and computes
\[
    \beta_{\mathrm{det}}
    =
    \min\left\{
        \frac{2H}{\mu},\
        \min_i
        \max\left\{
            \frac{a_i}{\mu},\
            \frac{a_i+2H}{t_i}
        \right\}
    \right\}.
\]
It then follows a minimizing route.
By \eqref{eq:weighted-core},
\[
\begin{aligned}
    \hat d(u,v)
    &\le
    d+\beta_{\mathrm{det}}\max\{d,\mu\}
      +(6+2c)\mu\\
    &\le
    \left[
        1+(1+\eps)
          \bigl(\beta_{\mathrm{det}}+6+2c\bigr)
    \right]d.
\end{aligned}
\]
It remains to bound $\beta_{\mathrm{det}}$ using
the height inequalities above.

\begin{lemma}\label{lem:weighted-det-opt}
Let $W=2\mu$ and
\[
    \beta_{\mathrm{det}}=\min\left\{\frac{2H}{\mu},\ \min_i\max\left\{\frac{a_i}{\mu},\frac{a_i+2H}{t_i}\right\}\right\}
\]
over the considered levels $i$, where $a_i=4h_i(u)+2iW$ and $t_i=\min\{W,h_{i+1}(v)-h_i(u)\}$.
Assume $h_1(u),h_1(v)\le\mu$, $h_{i+1}(u)\le h_i(v)+\mu$ for $0\le i<\tau$, and $|h_i(u)-h_i(v)|\le\mu$ for $1\le i\le\tau$, and let $H=h_\tau(v)$. Then
\[
    \beta_{\mathrm{det}}
    \le \frac{8\tau+4}{1+6\ln(5/4)}.
\]
\end{lemma}

\begin{proof}
Write $T=\beta_{\mathrm{det}}$.
The fallback option gives $T\le 2H/\mu$.
If $T\le H/\mu$, the claim follows from
$H/\mu\le2\tau-1$.
We may therefore assume
\[
    \frac{H}{\mu}<T\le\frac{2H}{\mu}.
\]
Set
\[
    q=\frac{2H}{\mu T}\in[1,2),
    \qquad
    z_i=\frac{h_i(u)}{\mu}+i
    \quad (0\le i\le\tau),
    \qquad
    b=\frac{T(2-q)}4.
\]
Since $W=2\mu$, we have $a_i/\mu=4z_i$.

We first show that, whenever $z_i<b$,
\[
    \frac{h_{i+1}(v)-h_i(u)}{\mu}
    \le q+\frac{4z_i}{T}.
\]
For a level that is not considered, and also for $i=0$,
the left-hand side is at most $1$, whereas the
right-hand side is at least $q\ge1$.
For a considered level, $z_i<b$ gives
\[
    4z_i<T,
    \qquad
    q+\frac{4z_i}{T}<2.
\]
Since $\max\{a_i/\mu,(a_i+2H)/t_i\}\ge T$ and $a_i/\mu=4z_i<T$, we get $(a_i+2H)/t_i\ge T$, i.e.\ $t_i/\mu\le q+4z_i/T$, and $t_i/\mu=\min\{2,(h_{i+1}(v)-h_i(u))/\mu\}$.
Hence
\[
    \min\left\{
        2,\frac{h_{i+1}(v)-h_i(u)}{\mu}
    \right\}
    \le q+\frac{4z_i}{T}<2.
\]
This proves the claim.

The height inequalities give, for $0\le i\le\tau-2$,
\[
    z_{i+2}-z_i
    \le
    3+\frac{h_{i+1}(v)-h_i(u)}{\mu}.
\]
Combining this with the preceding claim and the bound
$h_{i+1}(v)-h_i(u)\le3\mu$, we obtain
\[
    z_{i+2}-z_i
    \le
    \begin{cases}
        3+q+4z_i/T, & z_i<b,\\
        6, & z_i\ge b.
    \end{cases}
\]

To sum these bounds, define
\[
    \Psi(z)=
    \int_0^{\min\{z,b\}}
        \frac{2\,ds}{3+q+4s/T}
    +\frac{\max\{z-b,0\}}3,
    \qquad z\ge0.
\]
This function is increasing and concave:
its slope decreases to $2/5$ below $b$,
then drops to $1/3$.
Consequently, the bound on $z_{i+2}-z_i$ gives
\[
    \Psi(z_{i+2})-\Psi(z_i)\le2.
\]
Moreover, $z_0=0$, $z_1\le2$, and the slope of
$\Psi$ is at most $2/(3+q)\le1/2$.
Thus $\Psi(z_0)=0$ and $\Psi(z_1)\le1$.
Induction on the even and odd indices now gives
\[
    \Psi(z_\tau)\le\tau.
\]

Since $H=h_\tau(v)$ and
$\lvert h_\tau(u)-h_\tau(v)\rvert\le\mu$,
\[
    \frac H\mu+\tau\le z_\tau+1.
\]
Also,
\[
    b=\frac{T-H/\mu}{2}
      \le\frac{H}{2\mu}
      \le\tau-\frac12,
\]
so both $z_\tau$ and $H/\mu+\tau$ exceed $b$.
On this interval, $\Psi$ has slope $1/3$. Hence
\[
    \Psi\left(\frac H\mu+\tau\right)
    \le\tau+\frac13.
\]
Evaluating the integral gives
\[
    \Psi\left(\frac H\mu+\tau\right)
    =
    \frac T2\ln\frac5{3+q}
    +\frac{3H/\mu+2\tau-T}{6}.
\]
Substituting $H/\mu=qT/2$ and rearranging, we obtain
\[
    8\tau+4
    \ge
    T\left(3q-2+6\ln\frac5{3+q}\right).
\]
The expression in parentheses is increasing for $q\ge1$,
since its derivative is $3-6/(3+q)>0$.
Its value at $q=1$ is $1+6\ln(5/4)$.
Therefore
\[
    T\le\frac{8\tau+4}{1+6\ln(5/4)},
\]
as required.
\end{proof}

Combining \Cref{lem:weighted-det-opt} with
\eqref{eq:weighted-core} and $c=2$, we obtain
\[
    \hat d(u,v)
    \le
    \left[
        1+(1+\eps)
        \left(
            \frac{8\tau+4}{1+6\ln(5/4)}+10
        \right)
    \right]d(u,v).
\]
Since $\tau\le k-1$, the stretch is at most
\[
    1+(1+\eps)
    \left(
        \frac{8k-4}{1+6\ln(5/4)}+10
    \right).
\]

\begin{theorem}\label{T-weighted-det}
For every integer $k\ge2$ and every $0<\eps<1$, there is a $\bigl(1+(1+\eps)(\frac{8k-4}{1+6\ln(5/4)}+10)\approx(1+\eps)3.42k+O(1)\bigr)$-stretch labeled compact routing scheme for weighted graphs that uses worst-case local storage of $\Ot(\eps^{-1}n^{1/k}\log\Delta)$ words, labels of $\Ot(\eps^{-1}k\log\Delta)$ words, and headers of $\Ot(1)$ words, where $\Delta$ is the aspect ratio of the graph.
\end{theorem}
\begin{proof}
The tables use $\Ot((J+1)n^{1/k})$ words at every vertex and the labels use $\Ot(k(J+1))$ words, where $J+1=O(\eps^{-1}\log\Delta)$. The header carries the routing state of \Cref{S-worst-det} and the identifiers of the selected layer and search structure, that is, $\Ot(1)$ words. The stretch bound is the bound above, by \Cref{lem:weighted-det-opt}.
\end{proof}

\paragraph{Randomized routing.}
Keep $1<c\le2$ as a parameter.
Let $\pi_0$ denote the probability of taking the fallback
route directly, and let $\pi_i$ denote the probability
of visiting a considered level $i$.
The source solves
\[
\begin{aligned}
    \text{minimize}\quad &z\\
    \text{subject to}\quad
    &2H\pi_0+
      \sum_i\pi_i
         \bigl(a_i+2H\,\one[s\ge t_i]\bigr)
      \le zs
      &&\text{for every }
        s\in\{\mu,W\}\cup\{t_i\},\\
    &\pi_0+\sum_i\pi_i=1,\\
    &\pi_0,\pi_i\ge0.
\end{aligned}
\]
The sums range over the considered levels.
The program has $O(k)$ variables and constraints,
and every coefficient is known to the source.

Between consecutive thresholds, the left-hand side
is constant and nonnegative.
The constraints therefore hold for every $s\ge\mu$.
The source samples a route from an optimal solution.
Using \eqref{eq:weighted-core} and
\eqref{eq:weighted-layer-selection}, we obtain
\[
    \EE[\hat d(u,v)]
    \le
    \left[
        1+(1+\eps)(z^*+6+2c)
    \right]d_G(u,v).
\]
The expectation is over the source's choice,
with the preprocessing fixed.
The remaining optimization analysis bounds $z^*$
as a function of $c$ and then chooses $c$.

\begin{lemma}\label{lem:weighted-rand-opt}
For every $1<c\le2$, let $W=c\mu$ and let $z^*$ be the value of the linear program above, over the considered levels $i$ with $a_i=4h_i(u)+2iW$ and $t_i=\min\{W,h_{i+1}(v)-h_i(u)\}$.
Assume $h_1(u),h_1(v)\le\mu$, $h_{i+1}(u)\le h_i(v)+\mu$ for $0\le i<\tau$, and $|h_i(u)-h_i(v)|\le\mu$ for $1\le i\le\tau$, and let $H=h_\tau(v)$. Then
\[
    z^*\le
    \rho(c)\left(\tau+\frac12\right),
    \qquad
    \rho(c)=
    \max\left\{
        \frac{4(c+2)}{c+3},\frac4c
    \right\}.
\]
The minimum of $\rho(c)$ is attained at
\[
    c_*=\frac{\sqrt{13}-1}{2},
    \qquad
    \rho(c_*)=\frac{2(1+\sqrt{13})}{3}.
\]
\end{lemma}

\begin{proof}
We bound every feasible solution of the dual linear program.
Let $y_s\ge0$ be a dual solution, normalized by $\sum_sy_ss=1$, so its value is $\min_{\text{choice}}\sum_sy_s\cdot\mathrm{cost}_s$.
Let $\theta=s/\mu$ for $s$ drawn with probability proportional to $y_s$, and write
$m=\EE[\theta]$.
Thus $1\le\theta\le c$, and the dual value equals
$C/m$, where
\[
    C=
    \min\left\{
        \frac{2H}{\mu},\
        \min_i\left\{
            \frac{a_i}{\mu}
            +\frac{2H}{\mu}
              \Pr\left[\theta\ge\frac{t_i}{\mu}\right]
        \right\}
    \right\}.
\]
The inner minimum is over the considered levels.
It therefore suffices to prove
\[
    \frac Cm\le
    \rho(c)\left(\tau+\frac12\right).
\]

Set
\[
    r=\frac{C\mu}{2H}\in(0,1],
    \qquad
    z_i=\frac{h_i(u)}{\mu}+\frac{ci}{2}
    \quad (0\le i\le\tau).
\]
Then $a_i/\mu=4z_i$.

For $0\le\alpha<1$, let $f(\alpha)$ be the smallest
value in the support of $\theta$ such that
$\Pr[\theta\le f(\alpha)]>\alpha$.
Define
\[
    \widehat f(\alpha)=
    \begin{cases}
        f(\alpha),&f(\alpha)<c,\\
        3,&f(\alpha)=c.
    \end{cases}
\]
The replacement of $c$ by $3$ accounts for the clipping
in $t_i$: a threshold equal to $c\mu$ only tells us that
$h_{i+1}(v)-h_i(u)\le3\mu$.

We claim that, whenever $z_i<C/4$,
\[
    \frac{h_{i+1}(v)-h_i(u)}{\mu}
    \le
    \widehat f\left(1-r+\frac{2\mu z_i}{H}\right).
\]
For a considered level, the definition of $C$ gives
\[
    \Pr\left[\theta\ge\frac{t_i}{\mu}\right]
    \ge r-\frac{2\mu z_i}{H}>0.
\]
By the definition of $f$, this implies
\[
    \frac{t_i}{\mu}
    \le f\left(1-r+\frac{2\mu z_i}{H}\right).
\]
If the right-hand side is below $c$, clipping has no
effect, and the claim follows.
Otherwise, use $h_{i+1}(v)-h_i(u)\le3\mu$.
For a level that is not considered, and for $i=0$,
the claim follows because its left-hand side is at most
$1$, whereas $\widehat f\ge1$.

As in the deterministic proof, we now bound the progress
over two consecutive levels.
Define
\[
\begin{aligned}
    \Psi(z)
    ={}&
    \int_0^{\min\{z,C/4\}}
        \frac{2\,dw}
        {1+c+\widehat f(1-r+2\mu w/H)}
    +\frac{2\max\{z-C/4,0\}}{c+4}.
\end{aligned}
\]
This function is increasing and concave, since
$\widehat f$ is nondecreasing and at most $3$.
The height inequalities give
\[
    z_{i+2}-z_i
    \le
    1+c+\frac{h_{i+1}(v)-h_i(u)}{\mu}.
\]
Below $C/4$, apply the preceding claim.
Above $C/4$, use
$h_{i+1}(v)-h_i(u)\le3\mu$.
Concavity then gives
\[
    \Psi(z_{i+2})-\Psi(z_i)\le2.
\]
Also, $z_0=0$, $z_1\le(c+2)/2$, and the slope of
$\Psi$ is at most $2/(c+2)$.
Induction on the even and odd indices therefore yields
\[
    \Psi(z_\tau)\le\tau.
\]

The endpoint satisfies
\[
    z_\tau
    \ge
    \max\left\{0,\frac H\mu-1\right\}
       +\frac{c\tau}{2}
    >
    \frac{H}{2\mu}
    \ge\frac C4,
\]
where the strict inequality uses $c>1$ and $\tau\ge1$.
Moreover,
\[
    \frac H\mu+\frac{c\tau}{2}\le z_\tau+1.
\]
Since $\Psi$ has slope $2/(c+4)$ above $C/4$,
\[
    \Psi\left(\frac H\mu+\frac{c\tau}{2}\right)
    \le\tau+\frac{2}{c+4}.
\]
Writing
\[
    I=\int_{1-r}^1
        \frac{d\alpha}{1+c+\widehat f(\alpha)}
\]
and evaluating the two parts of $\Psi$, we obtain
\[
    \frac H\mu\bigl((c+4)I+2-r\bigr)
    \le4\tau+2.
\]

It remains to bound $I$ using $m$.
For $1\le t\le c$, replacing $t=c$ by $3$ in
the denominator gives the pointwise bound
\[
    \begin{cases}
        \displaystyle\frac1{1+c+t},&t<c,\\[1mm]
        \displaystyle\frac1{c+4},&t=c
    \end{cases}
    \ \ge\
    \frac1{c+2}
    -\frac{2(t-1)}{(c-1)(c+2)(c+4)}.
\]
At $t=c$, equality holds.
For $t<c$, this follows from the tangent lower bound
for $1/(1+c+t)$ at $t=1$, together with
$2(c+2)\ge(c-1)(c+4)$, which holds as $c\le2$.

Set $p=(m-1)/(c-1)\in[0,1]$.
Since $f$ is the quantile function of $\theta$,
\[
    \int_{1-r}^1(f(\alpha)-1)\,d\alpha
    \le \int_0^1(f(\alpha)-1)\,d\alpha
    =m-1.
\]
Integrating the pointwise bound, and also using
$\widehat f\le3$, gives
\[
    I\ge
    \frac r{c+2}
      -\frac{2p}{(c+2)(c+4)},
    \qquad
    I\ge\frac r{c+4}.
\]

If $p\ge r$, the second bound yields
\[
    \frac Cm
    \le
    \frac{4r}{1+(c-1)p}
        \left(\tau+\frac12\right)
    \le
    \frac4c\left(\tau+\frac12\right).
\]
If $p\le r$, the first bound yields
\[
    \frac Cm
    \le
    \frac{4r}
    {(1+(c-1)p)(1+(r-p)/(c+2))}
    \left(\tau+\frac12\right).
\]
For fixed $p$, the coefficient increases with $r$,
so it is maximized at $r=1$.
Its denominator then becomes
\[
    \frac{(1+(c-1)p)(c+3-p)}{c+2}.
\]
This is a concave quadratic in $p$, so its minimum
on $[0,1]$ occurs at an endpoint.
The two endpoints give respectively
$4(c+2)/(c+3)$ and $4/c$.
Thus every dual solution has value at most
$\rho(c)(\tau+1/2)$, proving the bound on $z^*$.

Finally, $4(c+2)/(c+3)$ is increasing in $c$,
whereas $4/c$ is decreasing.
They coincide when $c^2+c-3=0$, giving the stated
choice of $c_*$.
\end{proof}

Taking $c=c_*$, combining
\Cref{lem:weighted-rand-opt} with
\eqref{eq:weighted-core}, and using $\tau\le k-1$, we obtain
\[
    \EE[\hat d(u,v)]
    \le
    \left[
        1+(1+\eps)
        \left(
            \frac{2(1+\sqrt{13})}{3}
                \left(k-\frac12\right)
            +5+\sqrt{13}
        \right)
    \right]d(u,v).
\]

\begin{theorem}\label{T-weighted-rand}
For every integer $k\ge2$ and every $0<\eps<1$, there is a $\bigl(1+(1+\eps)(\frac{2(1+\sqrt{13})}{3}(k-\frac12)+5+\sqrt{13})\approx(1+\eps)3.07k+O(1)\bigr)$-expected stretch labeled compact routing scheme for weighted graphs that uses worst-case local storage of $\Ot(\eps^{-1}n^{1/k}\log\Delta)$ words, labels of $\Ot(\eps^{-1}k\log\Delta)$ words, and headers of $\Ot(1)$ words, where $\Delta$ is the aspect ratio of the graph.
\end{theorem}
\begin{proof}
The tables use $\Ot((J+1)n^{1/k})$ words at every vertex and the labels use $\Ot(k(J+1))$ words, where $J+1=O(\eps^{-1}\log\Delta)$. The header carries the routing state of \Cref{S-worst-det} and the identifiers of the selected layer and search structure, that is, $\Ot(1)$ words. The expected stretch bound is the bound above, by \Cref{lem:weighted-rand-opt}.
\end{proof}

\subsection{Name-independent routing}

We now consider name-independent routing with worst-case
storage in weighted graphs.
The source knows only the destination's name, so it cannot
use the destination's pivot heights to select a layer or
a search radius.
Instead, it makes increasing guesses for $d(u,v)$.
Each guess determines a layer, a pivot to visit, and
a radius to search.
After an unsuccessful visit, the message returns to the
source and proceeds to the next guess.

\paragraph{Preprocessing.}
Use the same layers $G_j$ and cluster structures as in
the labeled scheme.
Within each connected component of every layer, store
the source-pivot information, the explicit level-$0$
dictionaries, and the distributed search structures for
the cluster trees.
Each record contains the vertex's name, its routing label
in the full cluster tree, and its distance from the root.
As before, the prefix structures support a search at any
requested radius $R$, with all record holders at
distance at most $R$ from the root.

We also keep the source-pivot information and explicit
level-$0$ dictionaries for the original graph $G$.
The storage bound remains
$\Ot((J+1)n^{1/k})$ words per vertex.
The destination supplies no routing label:
the required tree label is retrieved during the search.

\paragraph{The initial lower bound.}
Fix a source $u$ and a destination $v\ne u$.
If $v\in C_0(u)$ in the original graph, route to $v$
along a shortest path using its stored record.
Otherwise, set
\[
    M=\max\left\{1,\frac{h_1(u)}2\right\}.
\]
Then $M\le d(u,v)$.
Indeed, edge weights are at least $1$, and
$d(u,v)<h_1(u)/2$ would imply
\[
    h_1(v)\ge h_1(u)-d(u,v)>d(u,v),
\]
placing $v$ in $C_0(u)$.

\paragraph{A visit with guess $D$.}
For a guess $D\ge M$, consider every layer
with $W_j\ge D$.
The source selects the layer minimizing the visit-cost
bound defined below.
For now, fix one such layer. All pivot heights and clusters
refer to this layer, and we suppress the superscript $j$.
Set $h_0(u)=0$ and $h_k(u)=\infty$, and choose
\[
    i=i(D)
    =
    \min\left\{
        0\le r<k:
        h_{r+1}(u)-h_r(u)>2D
    \right\}.
\]

If $i=0$, inspect the explicit dictionary of $C_0(u)$
locally.
If it contains $v$ at distance at most $D$,
route to $v$.
Otherwise, this guess fails without
sending the message.

If $i\ge1$, route to $p_i(u)$ and search its cluster
with radius
\[
    R=h_i(u)+D.
\]
The positive gap at level $i$ ensures that
$p_i(u)\in A_i\setminus A_{i+1}$ and
$u\in C(p_i(u))$, so the level-$i$ search bound applies.

A search succeeds only if it finds the destination's
record and the stored root distance is at most $R$.
This distance test is necessary because the selected
prefix may also contain records of vertices beyond $R$.
On success, use the retrieved label to reach $v$.
On failure, return to the pivot and then to $u$.

\paragraph{The guarantee of a visit.}
Every guess $D\ge d(u,v)$ succeeds.
To see this, write $d=d(u,v)$.
Since $W_j\ge D\ge d$, a shortest $u$-to-$v$
path is retained in $G_j$, and hence
$d_{G_j}(u,v)=d$.
The choice of $i$ gives
\[
\begin{aligned}
    h_{i+1}(v)
    &\ge h_{i+1}(u)-d\\
    &>h_i(u)+2D-d\\
    &\ge h_i(u)+D
     \ge d_{G_j}(p_i(u),v).
\end{aligned}
\]
Thus $v\in C(p_i(u))$, and its distance from the pivot
is at most the requested radius.
For $i=0$, this is precisely the local dictionary test.
Consequently, every failed guess certifies
\[
    d(u,v)>D.
\]

\paragraph{The cost of a visit.}
For each guess, compute the following bound in every
layer with $W_j\ge D$:
\[
    \operatorname{cost}_j(D)=
    \begin{cases}
        0,
            &i(D)=0,\\
        4h_i(u)+2D+2(i+1)W_j,
            &i=i(D)\ge1.
    \end{cases}
\]
Here $i(D)$ and $h_i(u)$ are computed in $G_j$.
The source chooses a layer attaining
\[
    \operatorname{cost}(D)
    =\min_{j:\,W_j\ge D}
        \operatorname{cost}_j(D).
\]
In the following bounds, $j$ is the selected layer.

A failed visit costs at most
$\operatorname{cost}(D)$.
For $i\ge1$, reaching the pivot and returning to $u$
costs $2h_i(u)$, while the failed search and return
to the pivot cost at most $2R+2(i+1)W_j$.
Their sum is
\[
    2h_i(u)+2R+2(i+1)W_j
    =
    \operatorname{cost}(D).
\]

A successful visit costs at most
$d(u,v)+\operatorname{cost}(D)$.
For $i\ge1$, the search costs at most
$R+2(i+1)W_j$.
Both the record holder and the destination are within
distance $R$ of the pivot, so delivery from the holder
costs at most $2R$.
Including the initial route to the pivot, the total is
at most
\[
    h_i(u)+3R+2(i+1)W_j
    =
    \operatorname{cost}(D)+D.
\]
If $D<d(u,v)$, this proves the claimed bound.
If $D\ge d(u,v)$, shortest paths are preserved
in the selected layer, and the sharper delivery bound
\[
    R+d_{G_j}(p_i(u),v)
    \le R+h_i(u)+d(u,v)
\]
gives the same conclusion.

For $i=0$, a successful dictionary test finds a path
of length at most $D\le W_j$.
This path is globally shortest, since every deleted edge
has weight greater than $W_j$.
Thus the successful route has length $d(u,v)$, consistent
with $\operatorname{cost}(D)=0$.

\paragraph{The available guesses.}
Use the finite set
\[
    \mathcal D=\{D_0,\ldots,D_L\},
    \qquad
    D_\ell=
    \min\{(1+\eps)^\ell M,\Delta\},
\]
where $L$ is the first index for which
$D_L=\Delta$.
A schedule is an increasing sequence of guesses from
$\mathcal D$ ending at $\Delta$.
The message follows this sequence until a search succeeds.

\paragraph{Deterministic routing.}
Consider a schedule
$\sigma=(t_0,\ldots,t_q)$, where $t_q=\Delta$.
For a distance $d\ge M$, let $r$ be the first index
with $t_r\ge d$.
The search succeeds no later than this guess, so
\[
    \hat d(u,v)
    \le
    d+\sum_{a=0}^{r}\operatorname{cost}(t_a).
\]
For $M\le d\le t_0$, the ratio of this extra-cost bound
to $d$ is largest at $M$.
For $t_{r-1}<d\le t_r$, its supremum is attained
as $d$ approaches $t_{r-1}$ from above.
Thus define
\[
    \beta(\sigma)=
    \max\left\{
        \frac{\operatorname{cost}(t_0)}{M},
        \max_{1\le r\le q}
        \frac{\sum_{a=0}^{r}\operatorname{cost}(t_a)}
             {t_{r-1}}
    \right\}.
\]
The source chooses a schedule minimizing this expression:
\[
    \beta_{\mathrm{det}}^*
    =\min_\sigma\beta(\sigma).
\]
The resulting route satisfies
\[
    \hat d(u,v)
    \le(1+\beta_{\mathrm{det}}^*)d(u,v).
\]
Unlike in the unweighted analysis, distances need not
be integral. The denominators are therefore the preceding
guesses themselves, with no additive $1$.

\begin{lemma}\label{lem:weighted-ni-det-opt}
The deterministic optimization satisfies
\[
    \beta_{\mathrm{det}}^*
    \le \frac{128}{5}(1+\eps)^2 k.
\]
Consequently, the deterministic routing scheme has stretch
at most
\[
    1+\frac{128}{5}(1+\eps)^2 k.
\]
\end{lemma}

\begin{proof}
It suffices to exhibit a schedule with the claimed bound.
For each $r\ge0$, round $\min\{2^rM,\Delta\}$ up to
the next available guess:
\[
    t_r=\min\{t\in\mathcal D:
        t\ge\min\{2^rM,\Delta\}\}.
\]
Delete repetitions and stop at the first occurrence of
$\Delta$.
Since consecutive available guesses differ by a factor
of at most $1+\eps$, we have $t_r\le(1+\eps)2^rM$.

We first bound the sum of $\operatorname{cost}(t_r)$ over
at most four consecutive indices $r$.
Let $t$ be the largest unrounded guess in this block,
so its unrounded guesses belong to
$\{t,t/2,t/4,t/8\}$.
Choose the smallest $j$ for which $W_j$ is at least
the largest rounded guess in this block. Then
\[
    W_j\le(1+\eps)^2t.
\]
This layer is eligible for every guess in the block.
Since $\operatorname{cost}$ minimizes over eligible layers,
we may bound all these visits using this one layer.
All heights and levels below refer to this layer.

Visits with $i=0$ have zero cost. For $i\ge1$, write
\[
    \operatorname{cost}_j(D)
    =2D+2W_j
     +\sum_{a=0}^{i-1}
        \bigl(4(h_{a+1}(u)-h_a(u))+2W_j\bigr).
\]
Fix a gap $g=h_{a+1}(u)-h_a(u)$ that contributes to
$\ell$ visits in the block.
The selected level in this fixed layer is nondecreasing
with the guess, so these visits correspond to the $\ell$
largest unrounded guesses, where $1\le\ell\le4$.
At the smallest of them, the minimality of the selected
level and the rounding bound give
\[
    g\le 2(1+\eps)\frac{t}{2^{\ell-1}}.
\]
Thus this gap contributes at most
\[
    \ell(4g+2W_j)
    \le
    (1+\eps)^2t\,
       \ell\left(\frac{8}{2^{\ell-1}}+2\right)
    \le12(1+\eps)^2t.
\]
The last coefficient is $10,12,12,12$ for
$\ell=1,2,3,4$, respectively.
There are at most $k-1$ such gaps.
The remaining terms $2D+2W_j$ contribute at most
\[
    (1+\eps)^2t
    \left[2\left(1+\frac12+\frac14+\frac18\right)+8\right]
    =\frac{47}{4}(1+\eps)^2t
    \le12(1+\eps)^2t.
\]
The total cost of the block is therefore at most
$12k(1+\eps)^2t$.

Now fix $d\in[M,\Delta]$, and let $m$ be the first
index with $2^mM\ge d$.
Then $t_m\ge d$ and $2^mM\le2d$.
Partition the guesses indexed by $0,\ldots,m$ into
blocks of four, starting from index $m$ and working
backwards. The first block in the schedule may be shorter.
Their largest unrounded guesses are
$2^mM,2^{m-4}M,2^{m-8}M,\ldots$.
Hence
\[
\begin{aligned}
    \sum_{r=0}^{m}\operatorname{cost}(t_r)
    &\le
    12k(1+\eps)^2\,2^mM
        \sum_{b\ge0}2^{-4b}\\
    &\le
    \frac{128}{5}(1+\eps)^2k\,d.
\end{aligned}
\]
The search succeeds no later than $t_m$.
Deleting repetitions or stopping earlier only decreases
the accumulated cost.
Thus the exhibited schedule has $\beta(\sigma)$ at most
the claimed bound, and so does the minimizing schedule.
Adding $d(u,v)$ gives the stretch bound.
\end{proof}

\begin{theorem}\label{T-weighted-ni-det}
For every integer $k\ge2$ and every $0<\eps<1$, there is a $(1+25.6(1+\eps)^2k)$-stretch name-independent compact routing scheme for weighted graphs that uses worst-case local storage of $\Ot(\eps^{-1}n^{1/k}\log\Delta)$ words and headers of $\Ot(1)$ words, where $\Delta$ is the aspect ratio of the graph.
\end{theorem}
\begin{proof}
The tables use $\Ot((J+1)n^{1/k})$ words at every vertex, where $J+1=O(\eps^{-1}\log\Delta)$. The header carries the current guess, the selected layer, level, and search structure, and the current tree identifier, target label, and return label, that is, $\Ot(1)$ words. The stretch bound is \Cref{lem:weighted-ni-det-opt}.
\end{proof}

\paragraph{Randomized routing.}
The source now chooses a distribution over schedules.
We express this optimization using transitions between
guesses, avoiding a separate variable for every schedule.

Introduce a start index $-1$.
For each $-1\le a<b\le L$, let $p_{a,b}$ be the
probability that the sampled schedule moves directly
from index $a$ to index $b$.
An edge from $-1$ selects the first guess.
These probabilities satisfy
\[
\begin{aligned}
    &\sum_{b=0}^{L}p_{-1,b}=1,\\
    &\sum_{a=-1}^{r-1}p_{a,r}
      =\sum_{b=r+1}^{L}p_{r,b}
      &&(0\le r<L),\\
    &p_{a,b}\ge0
      &&(-1\le a<b\le L).
\end{aligned}
\]
These are the constraints for one unit of flow from
the start index to $L$.
Every such flow defines a distribution over increasing
schedules ending at $\Delta$.

Fix $0\le r<L$ and a distance
$d\in(D_r,D_{r+1}]$.
A transition contributes to the cost bound precisely
when it selects the first guess or its preceding guess
is at most $D_r$.
The expected extra-cost bound is therefore
\[
    \sum_{b=0}^{L}
        p_{-1,b}\operatorname{cost}(D_b)
    +
    \sum_{a=0}^{r}\sum_{b=a+1}^{L}
        p_{a,b}\operatorname{cost}(D_b).
\]
This expression is constant throughout the interval,
so its ratio to $d$ is largest as $d$ approaches
$D_r$ from above.

The source minimizes $z$, subject to the flow
constraints and
\[
\begin{aligned}
    &\sum_{b=0}^{L}
        p_{-1,b}\operatorname{cost}(D_b)
    \le zM,\\
    &\sum_{b=0}^{L}
        p_{-1,b}\operatorname{cost}(D_b)
    +
    \sum_{a=0}^{r}\sum_{b=a+1}^{L}
        p_{a,b}\operatorname{cost}(D_b)
    \le zD_r
    &&(0\le r<L).
\end{aligned}
\]
The first constraint bounds the cost at $d=M$.
The program has $O((L+1)^2)$ variables and $O(L+1)$
constraints.

Given an optimal flow, the source samples the next
index according to the outgoing flow at its current
index.
It performs the corresponding visit and, on failure,
returns to $u$ and samples the next index.
Every sampled schedule ends at $\Delta$, where success
is guaranteed.
If $z^*$ is the optimum, then
\[
    \EE[\hat d(u,v)]
    \le(1+z^*)d(u,v).
\]
The expectation is over the source's choices,
with the preprocessing fixed.

\begin{lemma}\label{lem:weighted-ni-rand-opt}
The randomized optimization satisfies
\[
    z^*
    \le
    \frac{4e+20\sqrt e}{3}(1+\eps)^2
    \left(k+\frac{\ln5}{4}\right)
\approx (1+\eps)^2(14.62k+5.88).
\]
Consequently, the randomized routing scheme has expected \avi{I guess sleeping is overrated.}
stretch at most
\[
    1+\frac{4e+20\sqrt e}{3}(1+\eps)^2
    \left(k+\frac{\ln5}{4}\right).
\]
\end{lemma}

\begin{proof}
It suffices to exhibit a distribution over schedules.
Choose $U$ uniformly from $[0,1)$ and use the guesses
$Me^{r+U}$, for $r=0,1,\ldots$, rounding each up to
the next member of $\mathcal D$ and capping it at $\Delta$.
Delete repetitions and stop at $\Delta$.

For $t\ge M$, let $f(t)$ be the cost of the rounded and
capped guess, and set $f(t)=0$ for $0<t<M$.
The function $f$ is nonnegative and nondecreasing:
within each layer, the selected level and its cost are
nondecreasing, while the set of eligible layers only shrinks.

Fix $d\in[M,\Delta]$.
The schedule succeeds by the first unrounded guess
at least $d$.
A guess after the first contributes to the resulting
cost bound exactly when its predecessor is less than $d$.
Averaging over $U$ therefore gives
\begin{equation}\label{eq:weighted-ni-random-shift}
    \mathbb E[\text{extra-cost bound at distance }d]
    \le \int_M^{ed}\frac{f(t)}{t}\,dt.
\end{equation}
Indeed, the first guess accounts for $t\in[M,eM)$,
and the subsequent guesses account for $t\in[eM,ed)$.
Deleting repeated guesses or stopping earlier only
decreases the cost.

We first bound a weighted average of $f$.
For every $W>0$, define
\[
    F(W)=\frac{f(W)}{10W}
        +\int_0^W\frac{2f(t)}{(4t+W)^2}\,dt.
\]
We claim that
\begin{equation}\label{eq:weighted-ni-average-cost}
    F(W)\le
    (1+\eps)^2\left(k+\frac{\ln5}{4}\right).
\end{equation}
For $W<M$, this is immediate.
Otherwise, choose the smallest layer covering the
rounded and capped guess corresponding to $W$.
This layer is eligible for all smaller guesses and
satisfies $W_j\le(1+\eps)^2W$.
We upper-bound $f$ using the costs in this one layer.

As in the deterministic proof, a nonzero visit cost is
the sum of $2D+2W_j$ and one term $4g+2W_j$
for each gap $g$ below the selected level.
Fix such a gap, and let $a$ be the infimum of the
unrounded guesses in $[M,W]$ for which it contributes.
By the choice of the level and the rounding rule,
$g\le2(1+\eps)a$.
Its total weight in $F(W)$ is at most
\[
    \frac1{10W}
    +\int_a^W\frac2{(4t+W)^2}\,dt
    =\frac1{8a+2W}.
\]
Since
\[
    4g+2W_j\le(1+\eps)^2(8a+2W),
\]
each gap contributes at most $(1+\eps)^2$.
There are at most $k-1$ gaps.
The remaining terms satisfy
$2D+2W_j\le(1+\eps)^2(2t+2W)$, so their
contribution is at most
\[
\begin{aligned}
    &(1+\eps)^2
    \left[
        \frac25+
        \int_0^W\frac{4(t+W)}{(4t+W)^2}\,dt
    \right]\\
    &\hspace{15mm}
    =(1+\eps)^2\left(1+\frac{\ln5}{4}\right).
\end{aligned}
\]
This proves \eqref{eq:weighted-ni-average-cost}.

To pass from this weighted average to the integral
in \eqref{eq:weighted-ni-random-shift}, set
\[
    \kappa=\frac{20}{\sqrt e}-2.
\]
We use the elementary inequality
\begin{equation}\label{eq:weighted-ni-quadratic-weight}
    \int_x^1\frac{1+(\kappa/2)y^2}{y+4x}\,dy
    \ge \ln\frac1x
    \qquad(0<x\le1).
\end{equation}
For completeness, let $\Gamma(x)$ be the left-hand side
minus the right-hand side.
Integration and differentiation give
\[
    \Gamma(x)-\frac{x}{2}\Gamma'(x)
    =
    \ln\left(\frac{\sqrt e(1+4x)}5\right)
    +\frac5{\sqrt e(1+4x)}-1
    \ge0,
\]
where the last step uses
$\ln y+1/y-1\ge0$ for $y>0$.
Thus $(\Gamma(x)/x^2)'\le0$.
Since $\Gamma(1)=0$, we obtain $\Gamma(x)\ge0$, proving
\eqref{eq:weighted-ni-quadratic-weight}.

It follows that, for every $T>0$,
\[
    \int_0^T\frac{f(t)}t\,dt
    \le
    \int_0^T\left(2+\kappa\frac{W^2}{T^2}\right)F(W)\,dW.
\]
To verify this, use that $f$ is a nonnegative linear
combination of step functions.
For a single step $f(t)=\mathbf1[t\ge a]$, we have
$F(W)=0$ for $W<a$ and
$F(W)=1/(8a+2W)$ for $W\ge a$.
If $a\le T$, the desired inequality is precisely
\eqref{eq:weighted-ni-quadratic-weight} with $x=a/T$,
after substituting $y=W/T$.
If $a>T$, both sides vanish.
Values at the finitely many jump points do not affect
these integrals.

Using \eqref{eq:weighted-ni-average-cost}, we conclude
\[
    \int_0^T\frac{f(t)}t\,dt
    \le
    \left(2+\frac\kappa3\right)(1+\eps)^2
    \left(k+\frac{\ln5}{4}\right)T.
\]
Take $T=ed$ in this inequality and apply
\eqref{eq:weighted-ni-random-shift}.
The expected extra-cost bound is at most
\[
    e\left(2+\frac\kappa3\right)(1+\eps)^2
    \left(k+\frac{\ln5}{4}\right)d
    =
    \frac{4e+20\sqrt e}{3}(1+\eps)^2
    \left(k+\frac{\ln5}{4}\right)d.
\]
This holds for every $d\in[M,\Delta]$.
The exhibited distribution therefore gives a feasible
solution to the randomized optimization with the
claimed value of $z$.
The optimum is no larger, and adding $d(u,v)$ gives
the expected stretch bound.
\end{proof}

\begin{theorem}\label{T-weighted-ni-rand}
For every integer $k\ge2$ and every $0<\eps<1$, there is a $\bigl(1+\frac{4e+20\sqrt e}{3}(1+\eps)^2(k+\frac{\ln5}{4})\approx(1+\eps)^2(14.62k+5.88)\bigr)$-expected stretch name-independent compact routing scheme for weighted graphs that uses worst-case local storage of $\Ot(\eps^{-1}n^{1/k}\log\Delta)$ words and headers of $\Ot(1)$ words, where $\Delta$ is the aspect ratio of the graph.
\end{theorem}
\begin{proof}
The tables use $\Ot((J+1)n^{1/k})$ words at every vertex, where $J+1=O(\eps^{-1}\log\Delta)$. The header carries the current guess, the selected layer, level, and search structure, and the current tree identifier, target label, and return label, that is, $\Ot(1)$ words. The expected stretch bound is \Cref{lem:weighted-ni-rand-opt}.
\end{proof}

\end{document}